\documentclass{article}

\usepackage{kr}
\usepackage{times}
\usepackage{soul}
\usepackage{url}
\usepackage[hidelinks]{hyperref}
\usepackage[utf8]{inputenc}
\usepackage[small]{caption}
\usepackage{graphicx}
\usepackage{amsmath}
\usepackage{amsthm}
\usepackage{booktabs}
\usepackage[activate={true,nocompatibility},final,tracking=true,
kerning=true,spacing=true,factor=1100,stretch=10,shrink=10]{microtype}
\SetTracking{encoding={*}, shape=sc}{40}
\usepackage{ellipsis}         

\usepackage[abbreviations,british]{foreign} 

\usepackage{scalerel}

\usepackage{etoolbox}

\usepackage{xcolor}

\definecolor{darkmidnightblue}{rgb}{0.0, 0.2, 0.4}
\definecolor{persianplum}{rgb}{0.44, 0.11, 0.11}

\hypersetup{
  colorlinks  = true, 
  citecolor   = persianplum, 
  urlcolor    = persianplum, 
  linkcolor   = persianplum
}

\usepackage{tikz}
\usepackage{todonotes}
\usepackage{xspace}
\usepackage{marginnote}
\usepackage{rotating}

\usepackage{amssymb} 
\usepackage{mathtools}
\usepackage[bb=boondox]{mathalfa} 

\usepackage[capitalise, noabbrev]{cleveref}

\newtheorem{thm}{Theorem}[section]
\newtheorem*{thm*}{Theorem}
\newtheorem{definition}[thm]{Definition}

\newtheorem{corollary}[thm]{Corollary}

\newtheorem{lemma}[thm]{Lemma}

\newtheorem{claim}[thm]{Claim}

\usetikzlibrary{arrows, decorations.markings, shapes, calc}
\usepackage{graphics/nicolas-markey-expose}

\usetikzlibrary{shadows}
\tikzset{
diagonal fill/.style 2 args={fill=#2, path picture={
\fill[#1, sharp corners] (path picture bounding box.south west) -|
                         (path picture bounding box.north east) -- cycle;}},
reversed diagonal fill/.style 2 args={fill=#2, path picture={
\fill[#1, sharp corners] (path picture bounding box.north west) |- 
                         (path picture bounding box.south east) -- cycle;}}
}

\usetikzlibrary{hobby,backgrounds,calc,trees}

\usepackage{tikz-cd}
\usetikzlibrary{matrix}

 \makeatletter
 \def\desclabel#1#2{\begingroup
    \def\@currentlabel{#1}%
    #1\label{#2}\endgroup
 }
 \makeatother

\usepackage{float}

\usepackage{braket}

\usepackage[shortlabels]{enumitem} 
\usepackage{multicol}

\usepackage[h]{esvect}

\usepackage[linesnumbered,ruled,vlined]{algorithm2e}

\SetCommentSty{mycommfont}
\SetKwInput{KwData}{Input}

\usepackage{xurl} 
\usepackage{balance}

\newcommand{\orcid}[1]{\href{https://orcid.org/#1}{\includegraphics[width=9pt]{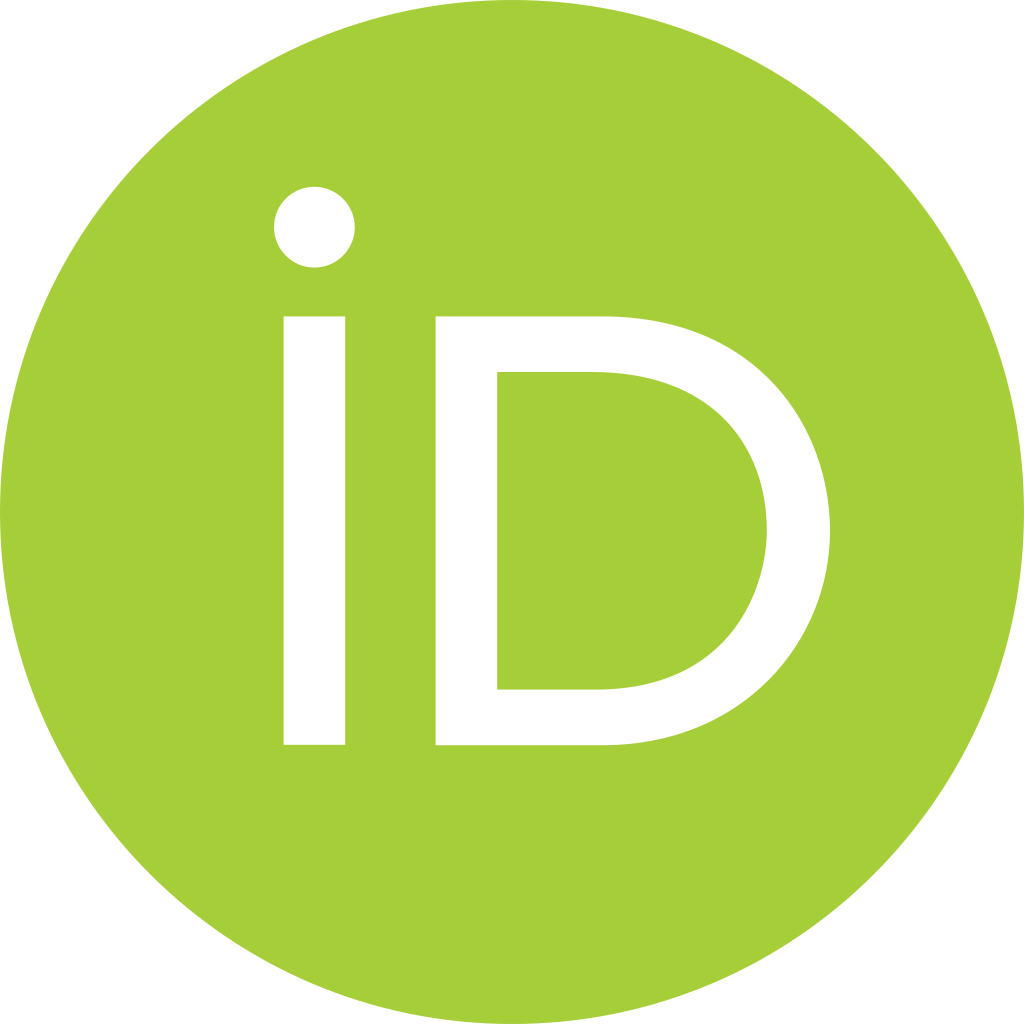}}}

\DeclareMathAlphabet{\mathdutchcal}{U}{dutchcal}{m}{n}
\SetMathAlphabet{\mathdutchcal}{bold}{U}{dutchcal}{b}{n}
\DeclareMathAlphabet{\mathdutchbcal}{U}{dutchcal}{b}{n}

\usepackage[switch, columnwise]{lineno}

\usepackage{fontawesome5}

\definecolor{ao(english)}{rgb}{0.0, 0.5, 0.0}
\definecolor{brickred}{rgb}{0.8, 0.25, 0.33}
\definecolor{ballblue}{rgb}{0.13, 0.67, 0.8}

\newcommand{\Logic}[1]{\ensuremath{\mathrm{#1}}} 
\newcommand{\logicL}{\Logic{L}} 
\newcommand{\logicQ}{\Logic{Q}} 
\newcommand{\FO}{\Logic{FO}}   
\newcommand{\FF}{\Logic{FF}}   
\newcommand{\FL}{\Logic{FL}}   

\newcommand{\GF}{\Logic{GF}}   
\newcommand{\GFt}{\GF^2}   
\newcommand{\FGF}{\Logic{FGF}}   
\newcommand{\FGFt}{\FGF^2}    

\newcommand{\RGF}{\Logic{RGF}}   
\newcommand{\RGFt}{\RGF^2}       
\newcommand{\FRGF}{\Logic{FRGF}}   
\newcommand{\FRGFt}{\FRGF^2}       
\newcommand{\GFF}{\Logic{GFF}}   
\newcommand{\GFL}{\Logic{GFL}}   

\newcommand{\GNFPup}{\Logic{GNFP}^{\mathsf{up}}}   

\newcommand{\UNFOreg}{\Logic{UN}_{\mathsf{reg}}}   

\newcommand{\GFTG}{\GF\hspace{-0.1em}{+}\hspace{-0.1em}\Logic{TG}}   
\newcommand{\GFtTG}{\GFt\hspace{-0.1em}{+}\hspace{-0.1em}\Logic{TG}}   

\newcommand{\PDL}{\Logic{PDL}}   
\newcommand{\ICPDL}{\Logic{ICPDL}}   

\newcommand{\complexityclass}[1]{\textsc{#1}} 
\newcommand{\ExpTime}{\complexityclass{ExpTime}} 
\newcommand{\AExpTime}{\complexityclass{AExpTime}} 
\newcommand{\ExpSpace}{\complexityclass{ExpSpace}} 
\newcommand{\NExpTime}{\complexityclass{NExpTime}} 
\newcommand{\TwoExpTime}{\complexityclass{2ExpTime}} 
\newcommand{\deff}{\coloneqq}

\newcommand{\N}{\mathbb{N}} 
\newcommand{\Z}{\mathbb{Z}} 
\renewcommand{\iff}{\leftrightarrow} 
\newcommand{\cupdot}{\mathbin{\mathaccent\cdot\cup}} 
\newcommand{\restr}{\!\!\restriction\!\!}
\newcommand{\powerset}[1]{\mathcal{P}(#1)}

\makeatletter 
\newcommand{\hathat}[1]{%
\begingroup%
  \let\macc@kerna\z@%
  \let\macc@kernb\z@%
  \let\macc@nucleus\@empty%
  \hat{\mathchoice%
    {\raisebox{.2ex}{\vphantom{\ensuremath{\displaystyle #1}}}}%
    {\raisebox{.2ex}{\vphantom{\ensuremath{\textstyle #1}}}}%
    {\raisebox{.16ex}{\vphantom{\ensuremath{\scriptstyle #1}}}}%
    {\raisebox{.14ex}{\vphantom{\ensuremath{\scriptscriptstyle #1}}}}%
    \smash{\hat{#1}}}%
\endgroup%
}
\makeatother

\def\myrm#1{\expandafter\def\csname rm#1\endcsname{\mathrm{#1}}}
\def\myrmall#1{\ifx#1\myrmall\else\myrm#1\expandafter\myrmall\fi}
\myrmall ABCDEFGHIJKLMNOPQRSTUVWXYZ\myrmall

\newcommand{\str}[1]{\mathfrak{#1}} 
\def\mystr#1{\expandafter\def\csname str#1\endcsname{\str{#1}}} 
\def\mystrall#1{\ifx#1\mystrall\else\mystr#1\expandafter\mystrall\fi}
\mystrall ABCDEFGHIJKLMNOPQRSTUVWXYZ\mystrall

\newcommand{\pred}[1]{\mathrm{#1}} 
\def\mypred#1{\expandafter\def\csname p#1\endcsname{\pred{#1}}}
\def\mypredall#1{\ifx#1\mypredall\else\mypred#1\expandafter\mypredall\fi}
\mypredall ABCDEFGHIJKLMNOPQRSTUVWXYZ\mypredall

\newcommand{\el}[1]{\mathrm{#1}} 
\def\myel#1{\expandafter\def\csname el#1\endcsname{\el{#1}}}
\def\myelall#1{\ifx#1\myelall\else\myel#1\expandafter\myelall\fi}
\myelall abcdefghijklmnopqrstuvwxyz\myelall

\newcommand{\rexp}[1]{\mathdutchcal{#1}} 
\def\myregexp#1{\expandafter\def\csname rexp#1\endcsname{\rexp{#1}}} 
\def\myregexpall#1{\ifx#1\myregexpall\else\myregexp#1\expandafter\myregexpall\fi}
\myregexpall ABCDEFGHIJKLMNOPQRSTUVWXYZ\myregexpall

\newcommand{\aut}[1]{\mathbb{#1}} 
\def\myaut#1{\expandafter\def\csname aut#1\endcsname{\aut{#1}}} 
\def\myautall#1{\ifx#1\myautall\else\myaut#1\expandafter\myautall\fi}
\myautall ABCDEFGHIJKLMNOPQRSTUVWXYZ\myautall

\newcommand{\CQ}{\textrm{CQ}}   
\newcommand{\UCQ}{\textrm{UCQ}}   
\newcommand{\query}[1]{\mathrm{#1}}
\newcommand{\queryq}{\query{q}}

\newcommand{\homo}[1]{\mathfrak{#1}} 
\newcommand{\homof}{\homo{f}} 
\newcommand{\homog}{\homo{g}} 
\newcommand{\homoh}{\homo{h}} 
\newcommand{\injto}{\hookrightarrow} 
\newcommand{\sinjto}{%
  \hookrightarrow\mathrel{\mspace{-15mu}}\rightarrow
}

\newcommand{\myqed}{\hfill{$\blacktriangleleft$}}
\newcommand{\sledom}{\Relbar\joinrel\mathrel{|}}
\newcommand{\sig}{\Sigma}
\newcommand{\sigfo}{\Sigma_{\FO}}
\newcommand{\sigreg}{\Sigma_{\rexpR}}
\newcommand{\AAA}{\mbox{\boldmath $\alpha$}} 
\newcommand{\BBB}{\mbox{\boldmath $\beta$}} 
\newcommand{\mldiamond}[1]{\langle #1 \rangle}
\newcommand{\mlbox}[1]{\left[ #1 \right]}
\newcommand{\IW}{\mathrm{I}} 
\newcommand{\iw}{\mathrm{i}} 
\newcommand{\pointer}{\text{\faHandPointRight}}

\newcommand{\question}[1]{\noindent\pointer\; #1}

\newcommand{\tp}[2]{\mathrm{tp}^{#1}_{#2}}
\newcommand{\prof}[1]{\mathrm{prof}^{#1}}

\newcommand{\forallfo}{\forall^{\FO}}
\newcommand{\forallreg}{\forall^{\rexpR}}
\newcommand{\existsfo}{\exists^{\FO}}
\newcommand{\existsreg}{\exists^{\rexpR}}

\newcommand{\etaiex}{\eta^{\raisebox{-0.4ex}{\scalebox{0.6}{$\exists$}}}_{\raisebox{0.4ex}{\scalebox{0.6}{$i$}}}}

\newcommand{\varthetaiex}{\vartheta^{\raisebox{-0.4ex}{\scalebox{0.6}{$\exists$}}}_{\raisebox{0.4ex}{\scalebox{0.6}{$i$}}}}

\newcommand{\etaifa}{\eta^{\raisebox{-0.4ex}{\scalebox{0.6}{$\forall$}}}_{\raisebox{0.4ex}{\scalebox{0.6}{$i$}}}}
\newcommand{\psiifa}{\psi^{\raisebox{-0.4ex}{\scalebox{0.6}{$\forall$}}}_{\raisebox{0.4ex}{\scalebox{0.6}{$i$}}}}
\newcommand{\phiifa}{\phi^{\raisebox{-0.4ex}{\scalebox{0.6}{$\forall$}}}_{\raisebox{0.4ex}{\scalebox{0.6}{$i$}}}}
\newcommand{\piifa}{\pi^{\raisebox{-0.4ex}{\scalebox{0.6}{$\forall$}}}_{\raisebox{0.4ex}{\scalebox{0.6}{$i$}}}}
\newcommand{\psiiex}{\psi^{\raisebox{-0.4ex}{\scalebox{0.6}{$\exists$}}}_{\raisebox{0.4ex}{\scalebox{0.6}{$i$}}}}
\newcommand{\gammaiex}{\gamma^{\raisebox{-0.4ex}{\scalebox{0.6}{$\exists$}}}_{\raisebox{0.4ex}{\scalebox{0.6}{$i$}}}}
\newcommand{\phiiex}{\phi^{\raisebox{-0.4ex}{\scalebox{0.6}{$\exists$}}}_{\raisebox{0.4ex}{\scalebox{0.6}{$i$}}}}
\newcommand{\piiex}{\pi^{\raisebox{-0.4ex}{\scalebox{0.6}{$\exists$}}}_{\raisebox{0.4ex}{\scalebox{0.6}{$i$}}}}

\newcommand{\SAT}{\mathsf{Sat}}
\newcommand{\FSAT}{\mathsf{FSat}}
\newcommand{\FoptSAT}{\mathsf{(F)Sat}}

\makeatletter
\newcommand{\genericdot@}[2]{
  \mathbin{\mathpalette\genericdot@@{{#1}{#2}{\cdot}}}%
}
\newcommand{\genericdotop@}[2]{
  \DOTSB\mathop{\mathpalette\genericdot@@{{#1}{#2}{\boldsymbol{\cdot}}}}\slimits@
}
\newcommand{\genericdot@@}[2]{\genericdot@@@#1#2}
\newcommand{\genericdot@@@}[4]{%
  \vphantom{#3}%
  \begingroup
  \sbox\z@{$\m@th#1#3$}%
  \ooalign{%
    \usebox{\z@}\cr
    \hidewidth
    \raisebox{#2\ht\z@}[\z@][\z@]{$\m@th#1#4$}%
    \hidewidth\cr
  }%
  \endgroup
}
\newcommand{\veedot}{\genericdot@{0.3}{\vee}}
\newcommand{\capdot}{\genericdot@{-0.2}{\cap}}
\newcommand{\wedgedot}{\genericdot@{-0.2}{\wedge}}

\newcommand{\bigcupdot}{\genericdotop@{0.25}{\bigcup}}
\newcommand{\bigveedot}{\genericdotop@{0.25}{\bigvee}}

\newcommand{\AAAphifo}{\AAA^{\raisebox{-0.4ex}{\scalebox{0.6}{$\FO$}}}_{\raisebox{0.4ex}{\scalebox{0.6}{$\varphi$}}}}
\newcommand{\BBBphifo}{\BBB^{\raisebox{-0.4ex}{\scalebox{0.6}{$\FO$}}}_{\raisebox{0.4ex}{\scalebox{0.6}{$\varphi$}}}}

\newcommand{\octant}{\mathbb{O}}
\newcommand{\tilingsys}{\mathdutchcal{D}}
\newcommand{\tilingH}{\mathdutchcal{H}}
\newcommand{\tilingV}{\mathdutchcal{V}}
\newcommand{\tiles}{\mathdutchcal{T}}

\newcommand{\tile}{\mathdutchcal{t}}

\newcommand{\tilesmap}{\xi}

\usepackage{framed}
\newcommand{\wrong}[1]{{\color{red}\underline{#1}}}

\newcommand{\tab}{\mathdutchcal{T}}
\newcommand{\bg}{\mathrm{bg}}
\newcommand{\pr}{\mathrm{pr}}
\newcommand{\im}{\mathrm{im}}
\newcommand{\varphireg}[1]{\varphi_{\rexpR, #1}}
\newcommand{\varphifo}[1]{\varphi_{\FO, #1}}
\newcommand{\varphiregfull}{\varphi_{\rexpR}}
\newcommand{\varphifofull}{\varphi_{\FO}}
\newcommand{\betaminus}{\beta^-}
\newcommand{\strAminus}{\strA^-}
\newcommand{\strAFO}{\strA_{\FO}}%
\newcommand{\strAReg}{\strA_{\rexpR}}
\newcommand{\ALC}{\mathcal{ALC}}
\usetikzlibrary{decorations.pathreplacing}

\title{Guarded Fragments Meet Dynamic Logic: The Story of Regular Guards\\ (Extended Version)}

\author{%
Bartosz Bednarczyk$^{1,2}$\and
Emanuel Kieroński$^{1}$\and
\affiliations
$^1$Institute of Computer Science, University of Wroc{\l}aw\\
$^2$Knowledge-Based Systems Group, TU Wien
\emails
\{bartek, kiero\}@cs.uni.wroc.pl
}

\begin{document}
\maketitle


\begin{abstract}\label{sec:abstract}
We study the \emph{Guarded Fragment with Regular Guards} $(\RGF)$, which combines the expressive power of the Guarded Fragment $(\GF)$ with Propositional Dynamic Logic with Intersection and Converse ($\ICPDL$).
Our logic generalizes, in a uniform way, many previously-studied extensions of $\GF$, including (conjunctions of) transitive or equivalence guards, transitive or equivalence closure and more.
We prove $\TwoExpTime$-completeness of the satisfiability problem for $\RGF$, showing that $\RGF$ is not harder than~$\ICPDL$~or~$\GF$. 
Shifting to the query entailment problem, we provide undecidability results that significantly strengthen and solidify earlier results along those lines.
We conclude by identifying, in a natural sense, the maximal $\ExpSpace$-complete fragment of $\RGF$.
\end{abstract} 

\section{Introduction}\label{sec:introduction}

The \emph{Guarded Fragment} ($\GF$) of first-order logic ($\FO$), introduced by Andréka et al.~\shortcite{AndrekaNB98}, generalizes modal and description logics (DLs) to higher-arity relational vocabularies. Over the past 25 years, $\GF$ has become the canonical first-order fragment that balances expressive power with attractive model-theoretic properties, such as the finite model property~\cite{Gradel99}, preservation theorems~\cite{Otto12}, and robust decidability under various extensions involving fixed-point operators~\cite{GradelW99} or query languages (Bárány et al.~\citeyear{BGO14}). Since classical (polyadic) multi-modal and description logic formul{\ae} embed naturally into $\GF$ via standard translations, this fragment serves as a versatile logical framework central to both theoretical studies and applications in knowledge representation and databases.

\noindent However, not all widely-used families of modal and description logics (DLs) are expressible within the scope of $\GF$, as it cannot express properties such as transitivity or equivalence of relations. 
Consequently, translating transitive description logics like those from the $\mathcal{S}$ family of DLs or modal logics interpreted over equivalence frames, including $\Logic{S5}$, into the guarded fragment is not directly possible. To overcome this limitation, Ganzinger et al.~\shortcite{GanzingerMV99} initiated the study of \emph{semantically-constrained guards}, an extension of $\GF$ allowing certain relations—--confined to guards---to be interpreted with additional semantic constraints, notably transitivity or equivalence.
This direction spurred intensive research, yielding several positive results, notably the $\TwoExpTime$-completeness of $\GF$ extended with (conjunctions of) transitive guards (consult the works of Szwast\&Tendera~\shortcite{SzwastT04}, Kazakov~\shortcite{KazakovPHD} and Kieroński\&Rudolph~\shortcite{KieronskiR21}), as well as the two-variable fragment of $\GF$ augmented by transitive or equivalence closures of binary guards, established by Michaliszyn and his co-authors~\shortcite{Michaliszyn09,KieronskiPT17}. 
Check Tendera's survey~\shortcite{Tendera17} for a comprehensive overview. 
On the negative side, natural extensions of $\GF$ \emph{with equality} ($\GF_\approx$), intended to capture popular description logics from the $\mathcal{SR}$ family, turned out to be undecidable. 
Examples include~$\GF_{\approx}$ with exponentiation (regular expressions that are compositions of the same letter)~\cite{KubaMgr} or associative compositional axioms~\cite{KazakovPHD}. 
The decidability status of these logics without equality $\approx$ is still open. 
Consequently, there is no known decidable extension of $\GF$ with semantically-constrained guards capturing propositional dynamic logic ($\PDL$) and its generalizations such as the $\mathcal{Z}$ family~(Calvanese et al.~\citeyear{CalvaneseEO09}) of DLs, $\PDL$ with intersection and converse~($\ICPDL$)~(G\"oller et al.~\citeyear{GeollerICPDL}), or higher-arity extensions of DLs such as $\mathcal{DLR}_{\mathsf{reg}}$~(Calvanese et al.~\citeyear{CalvaneseGL08}).

\vspace{-1em}
\paragraph*{Our contributions.}
We introduce and study a novel logic called $\RGF$, which extends (equality- and constant-free) $\GF$ by allowing $\ICPDL$-programs as guards (\cf Definition~\ref{def:RGF}).
Our main result establishes that $\SAT(\RGF)$ is $\TwoExpTime$-complete, matching the complexity of plain $\GF$ and $\ICPDL$. This also lifts decidability to several logics where it was previously known only in their two-variable case.
Our proof employs a \emph{fusion} technique, reducing $\SAT(\RGF)$ to instances of the satisfiability problem in plain $\GF$ and two-variable $\RGF$, 
solving the latter via a careful encoding into $\ICPDL$.
We further address two questions: (i) Is the query entailment problem decidable for $\RGF$? (ii) Is there an expressive fragment of $\RGF$ of complexity lower than $\TwoExpTime$?
We~answer question (i) negatively, showing undecidability of conjunctive query entailment even for two-variable fluted $\GF$ with a single transitive guard, substantially strengthening prior results of Gottlob et al.~\shortcite{GottlobPT13} in three ways: our logic is more restricted (belongs to the so-called fluted fragment), our queries do not use disjunction (we use conjunctive queries rather than the unions thereof), and our proof is also applicable to the finite-model scenario (which remained open). 
This transfers to the DL $\ALC$ extended with unqualified existential restrictions with intersection of the form $\exists{(\pred{r} \cap \pred{s})}.\top$, inclusion axioms of the form $\pred{r} \subseteq \pred{s} \cup \pred{t}$, and a single transitivity statement.
For~(ii), we conduct a thorough case analysis, pinpointing when subfragments of $\RGF$ admit lower complexity than $\TwoExpTime$. 
We conclude that a novel forward variant of $\GF$ extended with transitive closure is the largest (in a natural sense) $\ExpSpace$-complete fragment~of~$\RGF$.

\vspace{-1em}
\paragraph*{Our motivations.}
We explain our motivations behind the study of the $\GF$ with regular guards in the form of a Q\&A.

\question{Why the guarded fragment ($\GF$)?}\\
Because $\GF$ is \emph{the canonical} extension of modal and description logics to the setting of higher-arity relations~\cite{Gradel98}, heavily investigated in the last 25 years. $\GF$ is not only well-behaved both computationally~\cite{Gradel99} and model-theoretically~\cite{Gradel014}, but is also robust under extensions like fixed~points~\cite{GradelW99}~or~semantically-constrained~guards~\cite{Tendera17}.
It~was studied also in the setting of knowledge representation in multiple recent papers (Lutz et al.~\citeyear{LutzM24}, Jung et al.~\citeyear{JungLPW21}, Figueira et al.~\citeyear{FigueiraFB20}, Zheng et al.~\citeyear{ZhengS20}, Bourhis et al.~\citeyear{BourhisMP17}).


\question{Do we generalize any previously studied logics?}\\
Yes, many of them. 
First, as $\GF$ encodes (via the standard translation, see \eg Section 2.6.1 of Baader's textbook~\citeyear{dlbook}) multi-modal and description logics~\cite{Gradel98}, our logic also encodes (via an analogous translation) $\ICPDL$ and its subfragments such as $\ALC_{\mathsf{reg}}$ or $\mathcal{SRI}$~(Horrocks~et~al.~\citeyear{HorrocksKS06}). 
There also exists a natural translation from (counting-free fragment of) $\mathcal{DLR}_{\mathsf{reg}}$~(Calvanese et al.~\citeyear{CalvaneseGL08}) to $\RGF$. 
Second, there is a long tradition of studying~$\GF$ extended with semantically-constrained guards~\cite{Tendera17}, \ie distinguished relations (available only as guards) interpreted as \emph{transitive}~(Ganzinger et al.~\citeyear{GanzingerMV99}) or equivalence~\cite{Kieronski05} relations, or as transitive~\cite{Michaliszyn09} or equivalence~(Kieroński et al.~\citeyear{KieronskiPT17}) closures of another relation (that may also appear only as a guard).
As one can simulate transitive (or equivalence) relation~$\pR$ with $\pS^+$ (resp.~$(\pS\cup\pS^-)^*$) for a fresh relation $\pS$, our logic strictly extends all of the mentioned logics. 
Moreover, the mentioned papers concerning transitive and equivalence closures only focused on the extensions of $\GFt$, and hence our logic lifts them (without $\approx$) to the case of full~$\GF$ and provides the tight complexity bound.
Other ideas concern $\GF$ with exponentiation (regular expressions that are composition of the same letter)~\cite{KubaMgr} or associative compositional axioms~\cite{KazakovPHD} (\ie axioms of the form $\pR \circ \pS \subseteq \pT$ for $\pR, \pS, \pT$ predicates occurring only in guards). Both of them can be easily simulated in our framework.
Finally, $\GF$ with conjunctions of transitive relations in guards~\cite{KazakovPHD} can be expressed in $\RGF$ by employing $\cap$ operator.
All of this makes our logic a desirable object~of~study.

\question{Are there any closely related but incomparable logics?}\\
The closest logic is the Unary Negation Fragment~\cite{SegoufinC13} with regular-path expressions~\cite{JungLM018} $\UNFOreg$, together with its very recent\footnote{The papers by Figueira brothers and by Nakamura (still under review) appeared on arXiv during the preparation of this article.} generalizations with transitive closure operators~\cite{Figueirabrothers25} and guarded negation~\cite{Nakamura25}.
All of these logics share $\TwoExpTime$ complexity of their  satisfiability problem, but their expressive powers are incomparable. 
Indeed, $\RGF$ is not able to express conjunctive queries, while the other logics cannot express that $\pR^\ast$-reachable elements are $\pB$-connected. 
Yet another related logic is $\GNFPup$ by Benedikt et al.~\citeyear{BenediktBB16}, which extends the guarded negation fragment~\cite{BaranyCS15} with fixed-point operators with unguarded parameters. 
The syntax of $\GNFPup$ is complicated, but the logic seems to embed $\ICPDL$. Unfortunately, according to our understanding, such an encoding requires non-constant ``pdepth'' of the resulting formul{\ae}, leading to a non-elementary complexity of the logic.
The expressive powers of $\GNFPup$ and $\RGF$ are again incomparable and the separating examples are as before.

\setlength{\FrameSep}{3pt} 
\colorlet{shadecolor}{black!10}
\begin{shaded}
\noindent Additional examples, complete proofs, omitted standard definitions, and almost all the content from Section~\ref{sec:expspace} are delegated to the extended version of this paper (on arXiv).
\end{shaded} 

\section{Technical Background}\label{sec:preliminaries}

\noindent Throughout the paper, we assume familiarity with the basics of finite model theory, formal languages, and complexity. 

\vspace{-1em}
\noindent \paragraph*{Models.} We work with structures over a countably-infinite \emph{equality-} and \emph{constant-free} relational signature $\sig \deff \sigfo \cupdot \sigreg$, where all predicates in $\sigreg$, called \emph{regular predicates}, are binary. Fraktur letters denote structures; corresponding Roman letters denote their domains. For $\sigma \subseteq \sig$, a~\emph{$\sigma$-structure} interprets predicates in $\sig \setminus \sigma$ as~$\emptyset$.
A~\emph{sentence} is a formula without free variables. For a formula~$\varphi$, we write $\varphi(\bar{x})$ to indicate that its free variables lie in~$\bar{x}$. For a structure $\strA$ and a tuple $\bar{\ela}$, we write $\strA \models \varphi[\bar{\ela}]$ to indicate that $\varphi(\bar{x})$ holds in $\strA$ under the assignment $\bar{x} \mapsto \bar{\ela}$. If $\strA \models \varphi$, we call $\strA$ a \emph{model} of~$\varphi$.
A~sentence is \emph{(finitely) satisfiable} if it has a (finite) model. The \emph{(finite) satisfiability problem} $\FoptSAT(\logicL)$ for a logic~$\logicL$ asks if an input $\logicL$-sentence has a (finite) model. We say that $\logicL$ has the \emph{finite model property} (FMP) if all satisfiable $\logicL$-sentences admit finite models. Note that in this case, the finite and general satisfiability problems for $\logicL$~coincide.

\vspace{-1em}
\noindent \paragraph*{Types.} An \emph{atomic $k$-type} over $\sigma$ is a $\subseteq$-maximal satisfiable set of literals over $\sigma$ with variables in $x_1,\ldots,x_k$. A $k$-tuple~$\bar{\ela}$ \emph{realizes} a $k$-type $\gamma$ in~$\str{A}$ if $\str{A} \models \gamma[\bar{\ela}]$; this unique type is denoted $\tp{\str{A}}{k}(\bar{\ela})$.
The sets of $1$- and $2$-types over the signature of $\varphi$ are denoted $\AAA_{\varphi}$ and $\BBB_{\varphi}$, with sizes exponential and doubly-exponential in $|\varphi|$. 
With superscripts $\cdot^{\strA}$ and $\cdot^{\FO}$ we restrict such sets to types realized in $\strA$ and the ones restricted to the $\sigfo$-atoms.
For a $2$-type~$\beta$, let $\beta^{-1}$ be obtained by swapping its variables, and let $\beta \restr x_i$ be the $1$-type formed from literals involving only $x_i$.
With a $k$-type $\gamma$ over a signature~$\sigma$ we associate its \emph{canonical structure} $\strA_{\gamma}$ over the domain $\{ 1, 2, \ldots, k \}$ with the interpretation of $l$-ary predicates~$\pP$ defined as: $\strA_{\gamma} \models \pP[i_1, \ldots, i_{l}]$ iff $\pP(x_{i_1}, \ldots, x_{i_l}) \in \gamma$.
For a formula $\varphi(\bar{x})$ we write $\gamma \models \varphi$ to indicate that $\strA_{\gamma} \models \forall{\bar{x}}\, \varphi$.

\vspace{-1em}
\noindent \paragraph*{Morphisms.} With $\strA \injto_{\homof} \strB$ we denote the existence of an injective homomorphism $\homof$ between $\strA$ and $\strB$ over a common vocabulary.
If $\homof$ preserves $1$-types, we call it \emph{semi-strong} ($\strA \sinjto_{\homof} \strB$). 
We omit $\homof$ if it is~unimportant.
A~(union of) \emph{conjunctive queries} (U){\CQ} is a (disjunction of) existentially quantified conjunction of atoms.
The \emph{(finite) entailment problem}  for a logic $\logicL$ and a class of queries $\logicQ$, asks whether an input $\logicL$-sentence $\varphi$ (finitely) entails an input $\logicQ$-sentence~$\queryq$.

\vspace{-1em}
\noindent \paragraph*{First-Order Fragments.}
We consider fragments of first-order logic $\FO$ over $\sigfo$.
In the \emph{guarded fragment} $\GF$,
all quantification takes the form $\forall{\bar{x}} (\alpha \rightarrow \psi)$ and $\exists{\bar{x}} (\alpha \land \psi)$, where $\alpha$ (called a~\emph{guard}) is an atom featuring all the variables in~$\bar{x}$ and all the free variables~of~$\psi$. 
E. Gr\"adel~\shortcite{Gradel99} proved:

\begin{thm}\label{thm:deciding-plain-GF}
$\SAT(\GF)$ is $\TwoExpTime$-complete and it has a procedure that given an input $\GF$-sentence
works in time bounded polynomially in the length of the input, exponentially in the size of the signature,
and doubly exponentially in its width (maximal arity among all predicates in the input).~\myqed
\end{thm}

\noindent The \emph{two-variable} fragment~$\GFt$ of $\GF$, allows for sentences featuring only the variables~$x_1, x_2$ (re-quantification is permitted).
We call $\varphi$ \emph{index-normal} if on any branch of its syntax tree, the $i$-th quantifier bounds precisely~$x_i$.
Such a $\varphi$ is \emph{fluted} (\emph{forward}) if for any atom $\alpha(\bar{x})$ that occurs in the scope of a quantifier bounding $x_n$ (but not $x_{n{+}1}$), the sequence $\bar{x}$ is a suffix (infix) of the sequence $x_1, x_2, \ldots, x_n$.
To help the reader, we provide a few examples of forward formul{\ae}:
\vspace{-0.25em}
\begin{align*}
\forall{x_1}\; (\pS(x_1) \to \neg \forall{x_2} \;(\pP(x_2) \to \pA(x_1 x_2))),\\
\forall{x_1}\; \pL(x_1) \to \neg \exists{x_2}[\pP(x_2) \land \forall{x_3}\;(\pS(x_3) \to \pI(x_1 x_2 x_3))].
\end{align*}
Next, we present formul{\ae} that are syntactically not forward.
The \wrong{red highlight} means a mismatch in the variable ordering.
\begin{align*}
  \forall{x_1}\forall{x_2}\forall{x_3}\; \pR(x_1 x_2) \land \pR(x_2 x_3) \to \pR(\wrong{x_1 x_3}),\\
  \forall{x_1}\ \pS(x_1) \to \pR(\wrong{x_1x_1}), \qquad \forall{x_1}\forall{x_2}\ \pR(x_1 x_2) \iff \pR(\wrong{x_2 x_1}).  
\end{align*}

\noindent The forward $\GF$ ($\FGF$), designed to generalize inversion-free DLs, restricts $\GF$ to forward index-normal sentences.

\vspace{-1em}
\noindent \paragraph*{ICPDL.} Given a logic $\logicL$, we define \emph{$\logicL$-programs} $\pi$ with:
\begin{center}
\vspace{-0.1em}
$\pi,\rho \; \Coloneqq \; \pB \mid \bar{\pB} \mid \pi{\circ}\rho \mid \pi{\cup}\rho \mid \pi{\cap}\rho \mid \pi^{\ast} \mid \pi^{+} \mid \varphi{?}$,
\end{center}
\vspace{-0.1em}
where $\pB \in \sigreg$ and $\varphi$ is an $\logicL$-formula with a sole free~variable. 
Their semantics is given below.
We call $\pi$ \emph{simple} if all its subformul{\ae} $\varphi{?}$ (tests) are of the form $\pU{?}$ for~a~unary~$\pU$.

\vspace{-0.75em}
\begin{table}[H]
  \begin{center}
  \hspace{-2em}
  \vspace{-1.7em}
    \scalebox{0.86}{\begin{tabular}{@{}l@{\ \ }c@{\ \ }r@{}}
      \hline\\[-2ex]
      Name & Syntax of $\pi$ & Semantics $\pi^{\strA}$ of $\pi$ in a structure $\strA$ \\\hline\\[-2ex]
      Test / Predicate & $\varphi{?}$ / $\pB$ & $\{ (\ela, \ela) \mid \strA \models \varphi[\ela] \}$ / Binary relation \\ 
      Converse operator & $\bar{\pi}$ & $\{ (\elb, \ela) \mid (\ela,\elb) \in \pi^{\strA}\}$\\ 
      Concatenation & $\pi{\circ}\rho$ & $\{ (\ela, \elc) \mid \exists{\elb}. (\ela,\elb) \in \pi^{\strA} \land (\elb,\elc) \in \rho^{\strA}  \}$\\
      Union / Intersection & $\pi{\cup}\rho$ / $\pi{\cap}\rho$ & $\pi^{\strA} \cup \rho^{\strA}$ / $\pi^{\strA} \cap \rho^{\strA}$\\
      Kleene star/plus & $\pi^{\ast}$ / $\pi^{+}$ & $\textstyle\bigcup_{i=0}^{\infty} (\pi^i)^{\strA}$ / $\textstyle\bigcup_{i=1}^{\infty} (\pi^i)^{\strA}$,\\
      & & where $\pi^0 \deff \top?$ and $\pi^{i{+}1} \deff (\pi^{i}) \circ \pi$.
    \end{tabular}}
  \end{center}
\end{table}

\noindent The~\emph{intersection width} $\iw(\pi)$ of an $\logicL$-program $\pi$ is defined as: 
\begin{center}
$\iw(\pB) \deff \iw(\bar{\pB}) \deff \iw(\varphi?) \deff 1, \; \iw(\pi^\ast) \deff \iw(\pi^+) \deff \iw(\pi)$,\\ 
$\iw(\pi{\cup}\rho) \deff \iw(\pi{\circ}\rho) \deff \max(\iw(\pi), \iw(\rho)),  \iw(\pi{\cap}\rho) \deff \iw(\pi) {+} \iw(\rho)$.
\end{center}
The \emph{iwidth} $\IW(\varphi)$ of a formula $\varphi$ is $1$ if $\varphi$ contains no programs; otherwise it is the maximum of $\iw(\pi)$ over all programs $\pi$ in~$\varphi$. 

\noindent Formul{\ae} of \emph{PDL with Intersection and Converse} ({\ICPDL}) 
are given by
$\varphi, \psi \Coloneqq \top \mid \bot \mid \pU \mid \neg \varphi \mid \varphi \land \psi \mid \mldiamond{\pi}\varphi \mid \mlbox{\pi}\varphi$,
for $\pU \in \sigfo$ and $\ICPDL$-programs $\pi$, with the semantics:
\begin{table}[H]
  \begin{center}
  \vspace{-1em}
    \scalebox{0.8}{\begin{tabular}{@{}l@{\ \ }c@{\ \ }r@{}}
      \hline\\[-2ex]
      Name & Syntax of $\varphi$ & Given $\strA$ and $\ela \in A$, $\strA \models \varphi[\ela]$ if \\\hline\\[-2ex]
      Top / Bot / Predicate & $\top$ / $\bot$ / $\pU$ & always / never / $\ela \in \pU^{\strA}$, \\ 
      Negation / Conjunc. & $\neg\varphi$ / $\varphi{\land}\psi$ & $\strA \not\models \varphi[\ela]$ / $\strA \models \varphi[\ela]$ and $\strA \models \psi[\ela]$, \\ 
      Modalities & $\mldiamond{\pi}\varphi$\,/\,$\mlbox{\pi}\varphi$ & $\strA \models \varphi[\elb]$ for some/all $\elb$ with $(\ela,\elb) \in \pi^{\strA}$. \\ 
    \end{tabular}}
  \end{center}
\end{table}
\vspace{-1.5em}
\noindent $\strA$ is a \emph{model} of $\varphi$ in $\ICPDL$ if~$\strA \models \varphi[\ela]$ for some $\ela \in A$.
S.~G\"oller et al.~(\citeyear[Th.~3.28]{GoellerThesis};\;\citeyear[Th.~4.8]{GeollerICPDL})~proved that:

\begin{thm}\label{thm:ICPDL-upper-bound}
There exists an exponential function $\mathrm{exp}$ such that the satisfiability of each sentence
$\varphi \in {\ICPDL}$ can be decided in time bounded by~$\mathrm{exp}\left( |\varphi|^{\IW(\varphi)} \right)$.~\myqed
\end{thm}

\section{Guarded Fragment with Regular Guards}\label{sec:RGF}
\noindent We now introduce $\RGF$, the \emph{Guarded Fragment with Regular Guards}, a novel and highly-expressive logic extending $\GF$ by allowing for $\ICPDL$-programs as binary guards. Formally:

\begin{definition}[$\RGF$]\label{def:RGF}
  An $\RGF$-\emph{guard} $\vartheta$ for a formula $\varphi$ is either an atom over $\sigfo$ or $\pi(xy)$ for some $\RGF$-program~$\pi$, such that free variables of $\vartheta$ include all free variables of $\varphi$.
  The set $\RGF$ of \emph{$\RGF$-formul{\ae}} is defined with the grammar:
  \begin{center}
  \vspace{-0.25em}
  $\varphi,\varphi' \; \Coloneqq \; \pA(\bar{x}) \mid \neg \varphi \mid \varphi \land \varphi' \mid \exists{x}\, \varphi(x) \mid \exists{\bar{x}}(\vartheta \land \varphi)$,
  \end{center}
  \vspace{-0.25em}
  where $\pA \in \sigfo$ and $\vartheta$ is an $\RGF$-guard for $\varphi$.~\myqed
\end{definition}

\noindent Standard connectives and the universal quantifier are defined as usual.
For each $\rmS \subseteq \{ \cdot^-, \circ, \cup, \cap, \cdot^{\ast}, \cdot^{+}, ? \}$ we let $\RGF[\rmS]$ denote $\RGF$-formulae with operators in programs in guards restricted to $\rmS$. Additionally, $\FRGF[\rmS]$ is the forward restriction of $\RGF[\rmS]$ and $\FRGF^2[\rmS]$ is its two-variable~fragment.\\

\vspace{-0.5em}
\noindent Our logic generalizes  a plethora of extensions of $\GF$ with semantically-constrained guards (consult the introduction for details).
In particular, transitive relations in guards can be simulated in $\RGF[\cdot^+]$ using~$\pR^+$ for a fresh binary relation~$\pR$. Hence, $\GFTG$, the extension of $\GF$ with transitive guards, is a fragment of~$\RGF$.
We explain our design choices below.\\

\vspace{-0.4em}
\question{Why is the signature separated, \ie $\sig \deff \sigfo \cupdot \sigreg$?}\\
This is to ensure that binary predicates from $\sigreg$ appear \emph{only in guards}; otherwise, even the two-variable guarded fragment with transitivity is undecidable~(Ganzinger et al.~\citeyear[Th.~2]{GanzingerMV99}).

\vspace{0.1em}
\question{Why is the equality symbol $\approx$ excluded from $\sig$?}\\
Its inclusion makes our logic undecidable. This holds already for $\GFt$ with assoc. compositional axioms~\cite[Th.~5.3.1]{KazakovPHD}, conjunctions of transitive guards~\cite[Th.~5.3.2]{KazakovPHD}, or exponentiation~\cite[Th.~3.1]{KubaMgr}. 

\vspace{0.1em}
\question{Why are constant symbols excluded from $\sig$?}\\
They are expected to preserve decidability, especially given that \cref{thm:ICPDL-upper-bound} extends to Hybrid {$\ICPDL$}. We leave a full analysis to the journal version of this paper.



\section{Solving the Satisfiability Problem}\label{sec:sat}

We solve $\SAT(\RGF)$ in $\TwoExpTime$ in two steps.
First, \Cref{sub:the_two_variable_case} addresses $\SAT(\RGFt)$ by carefully encoding it into $\ICPDL$, producing an exponentially larger formula with only polynomial iwidth. This translation utilizes models realizing few $2$-types and tables, \ie lists of $2$-types that can be globally realized by pairs connected via any subsets of guards.
Then, \Cref{sub:the_case_of_many_variables} handles the general case via a reduction to instances of $\SAT(\RGFt)$ and~$\SAT(\GF)$, based on a \emph{fusion} --- a generalization of a method by Kieroński\&Rudolph~\shortcite{KieronskiR21} for $\GF$ with transitive guards. 
Its correctness hinges on the restriction that regular predicates are confined~to~guards.

\noindent As usual in the context of the satisfiability problem, we can w.l.o.g. restrict attention to formul{\ae} in a handy normal~form.

\begin{definition}\label{def:normal-form-RGF}
An $\RGF$-sentence $\varphi$ is in \emph{normal form}~(NF) if $\varphi = \forall{x}\lambda(x) \land \textstyle\bigwedge_{i} \varphi_i$ where each $\varphi_i$ has one~of~the~forms:
$\bullet\,  \forall{\bar{x}}\, \etaiex(\bar{x}) \to \exists{\bar{y}}\, \varthetaiex(\bar{x}\bar{y}) \land \psiiex(\bar{x}\bar{y}),\; \bullet\, \forall{\bar{x}}\, \etaifa(\bar{x}) \to \psiifa(\bar{x})$,\\
$\bullet\, \forall{x_1}\, \gammaiex(x_1) \to \exists{x_2}\, \piiex(x_1x_2) \land \phiiex(x_1x_2),\\ 
\bullet\, \forall{x_1}\forall{x_2}\, \piifa(x_1x_2) \to \phiifa(x_1x_2)$, \\
for (decorated) $\sigfo$-atoms $\eta$, $\vartheta$, $\gamma$, simple $\RGF$-programs~$\pi$, and quantifier-free $\psi$, $\phi$, $\lambda$ over $\sigfo$ with~$\bar{x} {\cap} \bar{y}$~$=$~$\emptyset$.~\myqed
\end{definition}


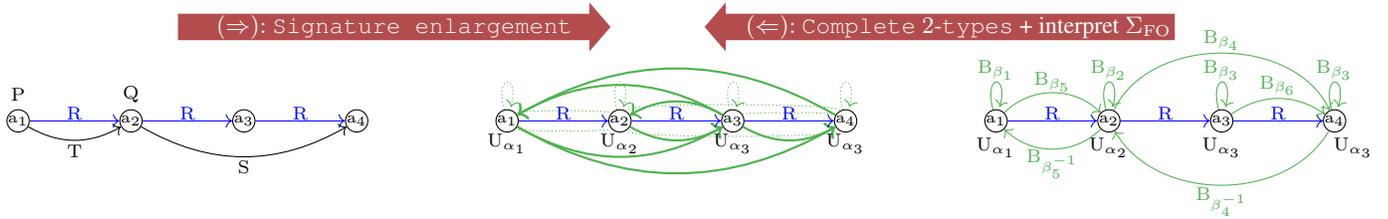
\begin{figure*}
  \centering
  \begin{tikzpicture}[transform shape]


  \draw (-12,0) node[minirondslightlybigger] (A1) {\tiny{$\ela_1$}};
  \draw (-10.5,0) node[minirondslightlybigger] (A2) {\tiny{$\ela_2$}};
  \draw (-9,0) node[minirondslightlybigger] (A3) {\tiny{$\ela_3$}};
  \draw (-7.5,0) node[minirondslightlybigger] (A4) {\tiny{$\ela_4$}};

  \path[->] (A1) edge [blue, ->] node[yshift=0.285em] {\scriptsize{$\pR$}} (A2);
  \path[->] (A2) edge [blue, ->] node[yshift=0.285em] {\scriptsize{$\pR$}} (A3);
  \path[->] (A3) edge [blue, ->] node[yshift=0.285em] {\scriptsize{$\pR$}} (A4);
  \path[->] (A1) edge [black, ->, bend right=30] node[yshift=-0.4em] {\scriptsize{$\pT$}} (A2);
  \path[->] (A2) edge [black, ->, bend right=30] node[yshift=-0.4em] {\scriptsize{$\pS$}} (A4);
  \draw (-10.5,0.35) node (aboveA2) {\scriptsize{$\pQ$}};
  \draw (-12,0.35) node (aboveA1) {\scriptsize{$\pP$}};


  \draw (-5.5,0) node[minirondslightlybigger](B1) {\tiny{$\ela_1$}};
  \draw (-4.0,0)  node[minirondslightlybigger] (B2) {\tiny{$\ela_2$}};
  \draw (-2.5,0)  node[minirondslightlybigger] (B3) {\tiny{$\ela_3$}};
  \draw (-1.0,0)  node[minirondslightlybigger] (B4) {\tiny{$\ela_4$}};


  \path[-] (B1) edge[looseness=0.1, green!40!gray, densely dotted, loop above] node[yshift=-0.25em] {} ();
  \path[-] (B2) edge[looseness=0.1, green!40!gray, densely dotted, loop above] node[yshift=-0.25em] {} ();
  \path[-] (B3) edge[looseness=0.1, green!40!gray, densely dotted, loop above] node[yshift=-0.25em] {} ();
  \path[-] (B4) edge[looseness=0.1, green!40!gray, densely dotted, loop above] node[yshift=-0.25em] {} ();

  \path[-] (B1) edge[looseness=0.1, green!40!gray, densely dotted, bend left=45] node[yshift=0.4em] {} (B2);
  \path[-] (B2) edge[looseness=0.1, green!40!gray, densely dotted, bend left=45] node[yshift=-0.55em] {} (B1);

  \path[-] (B2) edge[looseness=0.1, green!40!gray, densely dotted, bend left=65] node[yshift=0.5em] {} (B4);
  \path[-] (B4) edge[looseness=0.1, green!40!gray, densely dotted, bend left=60] node[yshift=-0.625em] {} (B2);

  \path[-] (B3) edge[looseness=0.1, green!40!gray, densely dotted, bend left=35] node[yshift=0.5em] {} (B4);

  \path[->] (B4) edge [green!40!gray, ->, thick, bend right=30] node[yshift=0.5em] {} (B1);  
  \path[->] (B1) edge [green!40!gray, ->, thick, bend right=30] node[yshift=0.5em] {} (B4);  
  \path[->] (B3) edge [green!40!gray, ->, thick, bend right=30] node[yshift=0.5em] {} (B1);  
  \path[->] (B1) edge [green!40!gray, ->, thick, bend right=30] node[yshift=0.5em] {} (B3);  
  \path[->] (B3) edge [green!40!gray, ->, thick, bend right=30] node[yshift=0.5em] {} (B2);  
  \path[->] (B2) edge [green!40!gray, ->, thick, bend right=30] node[yshift=0.5em] {} (B3);  
  \path[->] (B3) edge [green!40!gray, ->, thick, bend right=30] node[yshift=0.5em] {} (B4);  

  \path[->] (B1) edge [blue, ->] node[yshift=0.285em] {\scriptsize{$\pR$}} (B2);
  \path[->] (B2) edge [blue, ->] node[yshift=0.285em] {\scriptsize{$\pR$}} (B3);
  \path[->] (B3) edge [blue, ->] node[yshift=0.285em] {\scriptsize{$\pR$}} (B4);

  \draw (-5.5,-0.3) node (BelowB1) {\scriptsize{$\pU_{\alpha_{1}}$}};
  \draw (-4.0,-0.3) node (BelowB2) {\scriptsize{$\pU_{\alpha_{2}}$}};
  \draw (-2.5,-0.3) node (BelowB3) {\scriptsize{$\pU_{\alpha_{3}}$}};
  \draw (-1.0,-0.3) node (BelowB4) {\scriptsize{$\pU_{\alpha_{3}}$}};


  \draw (1.0,0)  node[minirondslightlybigger] (C1) {\tiny{$\ela_1$}};
  \draw (2.5,0) node[minirondslightlybigger] (C2) {\tiny{$\ela_2$}};
  \draw (4.0,0) node[minirondslightlybigger] (C3) {\tiny{$\ela_3$}};
  \draw (5.5,0) node[minirondslightlybigger] (C4) {\tiny{$\ela_4$}};

  \path[->] (C1) edge [blue, ->] node[yshift=0.285em] {\scriptsize{$\pR$}} (C2);
  \path[->] (C2) edge [blue, ->] node[yshift=0.285em] {\scriptsize{$\pR$}} (C3);
  \path[->] (C3) edge [blue, ->] node[yshift=0.285em] {\scriptsize{$\pR$}} (C4);

  \draw (1.0,-0.355) node (BelowC1) {\scriptsize{$\pU_{\alpha_{1}}$}};
  \draw (2.5,-0.355) node (BelowC2) {\scriptsize{$\pU_{\alpha_{2}}$}};
  \draw (4.0,-0.355) node (BelowC3) {\scriptsize{$\pU_{\alpha_{3}}$}};
  \draw (5.75,-0.355) node (BelowC4) {\scriptsize{$\pU_{\alpha_{3}}$}};

  \path[-] (C1) edge[green!40!gray, ->, loop above] node[yshift=-0.25em] {\scriptsize{$\pB_{\beta_{1}}$}} ();
  \path[-] (C2) edge[green!40!gray, ->, loop above] node[yshift=-0.25em] {\scriptsize{$\pB_{\beta_{2}}$}} ();
  \path[-] (C3) edge[green!40!gray, ->, loop above] node[yshift=-0.25em] {\scriptsize{$\pB_{\beta_{3}}$}} ();
  \path[-] (C4) edge[green!40!gray, ->, loop above] node[yshift=-0.25em] {\scriptsize{$\pB_{\beta_{3}}$}} ();

  \path[-] (C1) edge[green!40!gray, ->, bend left=45] node[yshift=0.4em] {\scriptsize{$\pB_{\beta_{5}}$}} (C2);
  \path[-] (C2) edge[green!40!gray, ->, bend left=45] node[yshift=-0.55em] {\scriptsize{$\pB_{\beta_{5}^{-1}}$}} (C1);

  \path[-] (C2) edge[green!40!gray, ->, bend left=65] node[yshift=0.5em] {\scriptsize{$\pB_{\beta_{4}}$}} (C4);
  \path[-] (C4) edge[green!40!gray, ->, bend left=60] node[yshift=-0.625em] {\scriptsize{$\pB_{\beta_{4}^{-1}}$}} (C2);

  \path[-] (C3) edge[green!40!gray, ->, bend left=35] node[yshift=0.5em] {\scriptsize{$\pB_{\beta_{6}}$}} (C4);

  \draw (-10,1.25) node[] (X0) {};
  \draw (-4,1.25) node[] (X1) {};

  \draw (3.5,1.25) node[] (X4) {};
  \draw (-3,1.25) node[] (X3) {};

  \draw[-{Triangle[width=18pt,length=8pt]}, line width=10pt, red!40!gray] (X0) -- (X1);
  \draw[-{Triangle[width=18pt,length=8pt]}, line width=10pt, red!40!gray] (X4) -- (X3);

  \draw (-7,1.25) node[] (T1) {\textcolor{white}{\small{$(\Rightarrow){:}$ \texttt{Signature enlargement}}}};
  \draw (0.5,1.25) node[] (T1) {\textcolor{white}{\small{$(\Leftarrow){:}$ \texttt{Complete} $2${-}\texttt{types} + interpret $\sigfo$}}};

  \end{tikzpicture}%
  \vspace{-0.9em}
  \caption{Structures $\strA$ (left), $\strA^+$ (middle), and $\strA^*$ (right) from our example illustrating the proof of \Cref{thm:RGFt-correctness}.}
  \label{f:example-two-variables}
\end{figure*}

\noindent We refer to above types of conjuncts as $\existsfo$-, $\forallfo$-, $\existsreg$-, and $\forallreg$-conjuncts.
We call a $k$-type $\gamma$ $\FO$-\emph{compatible} with $\varphi$ if $\gamma \models \lambda$ and for each $\forallfo$-conjunct: $\gamma \models \etaifa$ implies $\gamma \models \psiifa$.\\
\noindent 

\vspace{-1em}
\begin{lemma}
Consider $\FRGF[\cdot^+] \subseteq \logicL \subseteq \RGF$. Then if the satisfiability problem for $\logicL$-sentences in NF is decidable in $\TwoExpTime$ (resp. $\ExpSpace$) then so is $\SAT(\logicL)$.~\myqed 
\end{lemma}

\subsection{The Two-Variable Case}\label{sub:the_two_variable_case}

Consider $\varphi \in \RGFt$ in NF over a signature $\sigma$ with its usual components (\cf Def.~\ref{def:normal-form-RGF}). 
Let $\bg(\varphi) \deff \{ \vartheta_1, \ldots, \vartheta_K \}$ collect $2$-variable guards of $\varphi$, and $\pr(\varphi)$ be its restriction to~programs.
We focus on \emph{sparse models} of $\varphi$,  
\ie models realizing only exponentially-many~($\leq 2^{12|\varphi|}$)~$2$-types~from~$\BBB_{\varphi}^{\FO}$. 
Their presence helps keep the size of our translation in control.%

\begin{lemma} \label{lemma:RGF2-small-number-of-2-types}
Suppose $\varphi$ is (finitely) satisfiable, 
and let $\strA$ be its (finite) model $\varphi$. 
Then $\varphi$ also has a (finite) sparse model~$\strB$
such that: (i) $\strA$ and $\strB$ realise precisely 
the same $1$-types, and (ii) all $2$-types realised 
in $\strB$ are also realised in $\strA$.~\myqed
\end{lemma}

\begin{proof}[Sketch]
We use a filtration-based argument.
Let $\strA \models \varphi$.
Consider $\sigma^* \deff \sigma \cupdot \{ \pG_1, \ldots, \pG_{K} \}$ for fresh binary~$\pG_i$s, and let~$\strA^*$ be the $\sigma^*$-expansion of $\strA$ interpreting each $\pG_i$ in $\strA$ as the set of these pairs $(\ela, \elb)$ with $\strA \models \vartheta_i[\ela,\elb]$. 
Denoting by $\BBB^*$ the set of all $2$-types over $(\sigma^* \cap \sigfo)$ \emph{realized} in $\strA^*$, we define an equivalence relation $\sim$ on it as follows:~$\beta \sim \beta'$~if
\begin{itemize}[itemsep=0em, leftmargin=*]
\vspace{-0.25em}
\item $\beta$ and $\beta'$ are equal when restricted to literals involving either only a single variable or a binary predicate; and
\item for all conjuncts $\forall{x_1}\, \eta(x_1) {\to} \exists{x_2}\, \vartheta_i(x_1x_2) {\land} \psi_i(x_1x_2)$~of~$\varphi$:\\ 
$\beta \models \pG_i(x_1x_2) \land \psi_i$ if and only if $\beta' \models \pG_i(x_1x_2) \land \psi_i$\\
holds, and similarly for $\beta^{-1}$ and $\beta'^{-1}$ in place of $\beta$ and~$\beta'$.
\end{itemize}
\vspace{-0.25em}
Fix a representative for each $\sim$-class (of $\leq 2^{12|\varphi|}$ many) and denote the chosen member in $[\beta]_{\sim}$ by $\beta^\star$.
A sparse model is obtained by replacing all $2$-types $\beta$ between any unordered pair of elements in $\strA^*$ with the~corresponding $2$-type~$\beta^\star$.~\qedhere
\end{proof}

\noindent We reduce the satisfiability of $\varphi$ to the existence of a \emph{$\varphi$-table} $\tab \colon \left(\AAAphifo \times \mathcal{P}(\pr(\varphi)) \times \AAAphifo \right) \to \mathcal{P}(\BBBphifo)$ governing realizable $2$-types in an intended~sparse model of $\varphi$ and satisfiability of $\varphi_{\tab}$ in $\ICPDL$ governing its intended shape.~We~use unary $\pU_{\alpha}$ and binary $\pB_{\beta}$ indexed by types $\alpha \in \AAAphifo$, $\beta \in \BBBphifo$, intended to represent types of (pairs of) elements.
Based on them, we produce $\varphi_{\tab}$ incorporating obvious properties of sparse models, \eg that every element has its $1$-type and its witnesses (realizing some $2$-type and its inverse) for all $\existsfo$- and $\existsreg$-conjuncts of $\varphi$.
The main issue is that $\ICPDL$ cannot express that any two elements are connected by some binary relation, hence models of $\varphi_{\tab}$ have only \emph{partially} assigned $2$-types. 
To lift them to the actual models of $\varphi$ by ``completing'' the missing $2$-types, we employ $\varphi$-tables $\tab$.
Given its input $(\alpha_{\ela}, \rmS, \alpha_{\elb})$, a map $\tab$ outputs possible $2$-types $\beta$ realizable by pairs $(\ela, \elb)$ with $1$-types $\alpha_{\ela}$ and $\alpha_{\elb}$ in a hypothetical model of~$\varphi$, provided that $\ela$ and $\elb$ are connected by (at least) all the guards~in~$\rmS$. 
To qualify as a $\varphi$-table, $\tab$ must satisfy the four conditions listed below, met by any sparse model of~$\varphi$. 
\begin{description}[itemsep=0em, leftmargin=*]
    \item[\desclabel{(Size)}{tab:size}] The image $\im(\tab)$ of $\tab$ is bounded by $2^{12|\varphi|}$.
    \item[\desclabel{(Clo)}{tab:clo}] If $\beta \in \im(\tab)$ then $\im(\tab)$ contains $\beta^{-1}$ and the \emph{symmetric $2$-types} $\beta_{\alpha}$ \emph{induced by} each $1$-type $\alpha \in \{ \beta \restr x_1, \beta \restr x_2 \}$, namely the unique $2$-types $\beta_{\alpha}$ satisfying $\beta_{\alpha}[x_2 \mapsto x_1] \equiv \alpha$.
    %
    %
    \item[\desclabel{(Com)}{tab:com}] If $\beta \in \tab(\alpha_1,\rmS,\alpha_2)$ then $\beta \restr x_1 = \alpha_1$ and $\beta \restr x_2 = \alpha_2$ and the types $\alpha_1$, $\alpha_2$, $\beta$ are $\FO$-compatible with $\varphi$.
    \item[\desclabel{(Sat)}{tab:sat}] For all $\forallreg$-conjuncts $\forall{x_1x_2}\, \piifa(x_1x_2) \to \phiifa$~from~$\varphi$ and all $\beta \in \tab(\alpha_1,\rmS,\alpha_2)$ we have $\beta \models \phiifa$ if $\piifa(x_1x_2) \in \rmS$.
\end{description} 
\noindent Every $\strA$ naturally defines its table $\tab_{\strA}$, where $\tab_{\strA}(\alpha_1, \rmS, \alpha_2)$ consists of all $\tp{\strA}{2}(\ela_1, \ela_2)$ over all $\ela_1,\ela_2$ from $\strA$ witnessing $\tp{\strA}{1}(\ela_i) = \alpha_i$ and $\strA \models \vartheta[\ela_1,\ela_2]$ for all guards~$\vartheta \in \rmS$. 
An $\strA$ satisfies~$\tab$ ($\strA \models \tab$) if $\tab_{\strA} \subseteq \tab$.
Clearly, if $\strA \models \varphi$ then $\tab_{\strA}$ is a $\varphi$-table but not all $\strA$ satisfying $\varphi$-tables~are~models~of~$\varphi$.\\ 


\vspace{-0.75em}
\noindent We can now translate $\varphi$ and $\tab$ into $\varphi_{\tab} \deff \mlbox{\star}(\varphi_\tab^1 \land \ldots \land \varphi_\tab^7)$ in $\ICPDL$, where the formul{\ae} $\varphi_\tab^i$ with informal explainations are given below and $\star$ is the ``universal modality'', \ie 
\begin{center}$
\star \deff \textstyle( \bigcup_{\beta \in \im(\tab)}\bigcup_{\pR \in \sigma \cap \sigreg} \pB_{\beta} \cup \overline{\pB_{\beta}} \cup \pR \cup \overline{\pR})^\ast.
$\end{center}
\vspace{-0.15em}
\noindent For a program $\pi$, let $\pi_{\tab}$ be the result of replacing each unary $\pU$ in $\pi$ with the union of $\pU_\alpha$ over all $\alpha \models \pU(x_1)$ from $\AAAphifo$.

\noindent\textbf{1.} Each element has a unique $1$-type, and each pair of elements shares at \emph{most one} $2$-type (existence of a $2$-type is not expressible, justifying the use of $\varphi$-tables~by~our~reduction).
\begin{center}$
\varphi_\tab^1 \deff \textstyle\bigveedot_{\alpha \in \AAAphifo}\pU_{\alpha} \land \textstyle\bigwedge_{\beta\neq\beta' \in \im(\tab)}  \mlbox{\pB_{\beta} \cap \pB_{\beta'}}\bot
$\end{center}
\vspace{-0.15em}
\noindent\textbf{2.} Each element \emph{self}-realize its unique (symmetric) $2$-type.
\begin{center}$
\varphi_\tab^2 \deff \textstyle\bigwedge_{\alpha \in \AAAphifo} \left( \pU_{\alpha} \to \mldiamond{\pB_{\beta_{\alpha}} \cap \overline{\pB_{\beta_{\alpha}}} \cap \top?}\top \right)
$\end{center}
\vspace{-0.15em}
\noindent\textbf{3.} Compatibility of $2$-types with their inverses and induced $1$-types.
The big conjunction is over $\beta,\beta' \in \im(\tab)$: $\beta' \neq \beta^{-1}$.
\begin{center}$
\varphi_\tab^3 \deff \bigwedge \mlbox{\pB_{\beta} {\cap} \overline{\pB_{\beta'}}\;}\hspace{-0.2em}\bot 
\land 
\mlbox{\neg\pU_{\beta \, \restr\, x_1}{?}{\circ}\pB_{\beta} \cup \pB_{\beta}{\circ}\neg\pU_{\beta\, \restr\, x_2}{?}}\hspace{-0.2em}\bot 
$\end{center}

\noindent\textbf{4\&5.} Ensuring witnesses for all $\existsfo$- and $\existsreg$-conjuncts of~$\varphi$.
\begin{center}$
\varphi_\tab^4 \deff \textstyle\bigwedge_{i, \alpha \models \etaiex} \pU_{\alpha} \to \bigvee_{\beta \in \im(\tab), \beta \models \varthetaiex, \beta \models \psiiex} \mldiamond{\pB_{\beta} \cap \overline{\pB_{\beta^{-1}}}\,}\top
$\end{center}
\vspace{-1em}
\begin{center}$
\varphi_\tab^5\hspace{-0.15em}\deff\hspace{-0.15em}\textstyle\bigwedge_{i, \alpha \models \gammaiex} \hspace{-0.1em}\pU_{\alpha} \to \bigvee_{\hspace{-0.15em}\beta \in \im(\tab), \beta \models \phiiex} \mldiamond{(\piiex)_{\tab} {\cap} \pB_{\beta} {\cap} \overline{\pB_{\beta^{-1}}}\,}\hspace{-0.15em}\top
$\end{center}
\vspace{-0.5em}
\noindent\textbf{6.} No pair $(\ela, \elb)$ of elements in the intended model are $\pB_{\beta}$-connected for a $2$-type $\beta$ with either $\beta$ or $\beta^{-1}$ violating $\tab$. 
\begin{center}$
\varphi_\tab^6 \deff\hspace{-0.1em}\textstyle\bigwedge\hspace{-0.1em}\mlbox{ 
\hspace{-0.1em}\left( 
\pU_{\alpha}{?} {\circ}\hspace{-0.1em}\left( \bigcup \pB_{\beta} \right) {\circ} \pU_{\alpha'}{?}\hspace{-0.1em}\right)\hspace{-0.1em} \cap \bigcap_{\pi \in \rmS}\hspace{-0.1em}\pi_{\tab} \hspace{-0.1em} \cap \bigcap_{\pi \in \rmS'}\hspace{-0.1em}\overline{\pi_{\tab}}
}\hspace{-0.2em}\bot
$\end{center}
\vspace{-0.25em}
Above, $\textstyle\bigwedge$ is over $\alpha, \alpha' \in \AAAphifo$ ($1$-types of $\ela$, $\elb$), $\rmS, \bar{\rmS} \subseteq \pr(\varphi)$ (guards from $\ela$ to $\elb$ and from $\elb$ to $\ela$), and 
$\textstyle\bigcup$ ranges over $\beta \in \im(\tab)$ satisfying $\beta \not\in \tab(\alpha, \rmS, \alpha')$ or $\beta^{-1} \not\in \tab(\alpha', \rmS', \alpha)$.

\vspace{0.35em}
\noindent\textbf{7.} There should always be an available $2$-type and its inverse to assign to each pair of elements in the intended model.
\begin{center}$
\varphi_\tab^7 \deff \textstyle\bigwedge
\mlbox{ 
\hspace{-0.1em}\left( 
\pU_{\alpha}{?} {\circ}\hspace{-0.1em} \star {\circ} \pU_{\alpha'}{?}\hspace{-0.1em}\right) \cap \bigcap_{\pi \in \rmS}\hspace{-0.1em}\pi_{\tab} \cap \bigcap_{\pi \in \rmS'}\hspace{-0.1em}\overline{\pi_{\tab}}
}\hspace{-0.2em}\bot
$\end{center}
\vspace{-0.25em}
As in the 6th case, $\textstyle\bigwedge$ iterate $\alpha,\alpha', \rmS, \rmS'$ with $\tab(\alpha, \rmS, \alpha')$ either empty or containing no $\beta$ with $\beta^{-1} \in \tab(\alpha', \rmS', \alpha)$.\myqed

\vspace{0.35em}
\noindent Observe that $|\varphi_{\tab}|$ is exponential in $|\varphi|$, but the iwidth~of~$\varphi_{\tab}$ is bounded by $2|\pr(\varphi)|{+}2$ (hence only polynomially in~$|\varphi|$). 
Thus, by \Cref{thm:ICPDL-upper-bound}, satisfiability of each $\varphi_{\tab}$ can be tested in doubly-exponential time in~$|\varphi|$. By enumerating possible $\varphi$-tables~$\tab$ and testing each $\varphi_{\tab}$ for satisfiability yields the optimal algorithm (in $\TwoExpTime$) for satisfiability of $\varphi$. 

\begin{thm}\label{thm:RGFt-correctness}
$\varphi$ is satisfiable iff there is a $\varphi$-table $\tab$ with a satisfiable $\varphi_{\tab}$.
Thus, $\SAT(\RGFt)$ is $\TwoExpTime$-compl.~\myqed
\end{thm}
\vspace{-0.75em}
\begin{proof}[Sketch.]
For a (sparse) $\strA \models \varphi$, let $\strA^{+}$ be its expansion that interpret each~$\pU_{\alpha}$ (resp.$\pB_{\beta}$) as the set of elements (resp. pairs) satisfying $\alpha$ (resp. $\beta$). Clearly $\strA^{+} \models \varphi_{\tab}$ for $\tab \deff \tab_{\strA}$.
Conversely, take a connected~$\strA^* \models \varphi_{\tab}$ for some $\tab$. We ``complete''~$\strA^*$ by resolving all typeless pairs, \ie $(\ela_1,\ela_2)$ such that $\strA \not\models \pB_{\beta}[\ela_1,\ela_2]$ for any $\beta$. For each such pair, let $\alpha_i$ be the unique $1$-type with $\strA^* \models \pU_{\alpha_i}[\ela_i]$.
If some $\beta$ satisfies $\strA^* \models \pB_{\beta}[\elb,\ela]$, we add $(\ela,\elb)$ to~$(\pB_{\beta^{-1}})^{\strA^*}$. Otherwise, fix a $\beta \in \tab(\alpha_1, \rmS, \alpha_2)$, where $\rmS \subseteq \pr(\varphi)$ collects all guards $\vartheta$ satisfied by $(\ela,\elb)$, and then include~$(\ela,\elb)$~in~$(\pB_{\beta})^{\strA^*}$.
Repeat until no typeless pair remain. $\strA^*$ becomes a model $\strA^+$ of $\varphi$ after interpreting $\sigfo$ minimally to satisfy $\strA^+ \models \pB_{\beta}[\ela,\elb]$ iff $\strA^* \models \beta[\ela,\elb]$ for all $\beta \in \im(\tab)$ and all $\ela, \elb \in A^*$.
\end{proof}

\noindent To illustrate the proof of \Cref{thm:RGFt-correctness}, consider $\varphi$ (over the signature $\sigma \deff \{\pP, \pQ, \pS, \pT, \pR \}$ with $\pR$ being the only regular~predicate) composed of the following four conjuncts:\\
$\bullet\, \forall{x_1}\, \pQ(x_1) \rightarrow \exists{x_2}\, \pR^+(x_1x_2) \land \pS(x_1x_2),$\\
$\bullet\, \forall{x_1}\, \pQ(x_1) \rightarrow \exists{x_2}\, \bar{\pR}(x_1x_2) \land \pP(x_2),$\\
$\bullet\, \forall{x_1}\, \pP(x_1) \rightarrow \exists{x_2}\, \pT(x_1x_2) \land \pQ(x_2), $\\
$\bullet\, \forall{x_1}\forall{x_2}\, \pR(x_1x_2) \to \neg \pS(x_1x_2)$.

\noindent When starting from 
$\strA \models \varphi$ (left part of Fig.~\ref{f:example-two-variables}), 
we construct its expansion $\strA^+ \models \varphi_{\tab_{\strA}}$, in which  
(i) each~$\ela_i$ satisfies $\pU_{\alpha}$, where $\alpha \deff \tp{\strA}{1}(\ela_i)$, and  
(ii) for each pair $(\ela_i, \ela_j)$, the tuple $(\ela_i, \ela_j)$ satisfies $\pB_{\beta}$ and  
$(\ela_j, \ela_i)$ satisfies $\pB_{\beta^{-1}}$, where $\beta \deff \tp{\strA}{2}(\ela_i, \ela_j)$ (dotted and solid green arrows).
For the other direction, we start with~$\strA^* \models \varphi_{\tab}$~(right~part~of~Fig.~\ref{f:example-two-variables}).~Here:

\begin{itemize}[itemsep=0em, leftmargin=*]
  \item Some $\pU_{\alpha}$ are assigned to the $\ela_i$ due to $\varphi_\tab^1$: 
    $\alpha_1$ contains $\pP(x_1)$, 
    $\alpha_2$ contains $\pQ(x_1)$ (both contain no other positive atoms), 
    and $\alpha_3$ contains no positive atoms.
  \item Some $\pB_{\beta}$ are assigned to the pairs $(\ela_i, \ela_i)$ due to $\varphi_\tab^2$:~$\beta_1$, $\beta_2$, $\beta_3$ are the trivial $2$-types without positive binary atoms, 
    compatible with the $1$-types of the corresponding~$\ela_i$.
  \item $\beta_4$ contains $\pS(x_1, x_2)$ (and no other positive binary atoms);  
    $\pB_{\beta_4}$ is assigned to $(\ela_2, \ela_4)$ and  
    $\pB_{\beta_4^{-1}}$ to $(\ela_4, \ela_2)$ due to $\varphi_\tab^4$, providing the first witness for $\ela_2$.
  \item $\beta_5$ contains $\pT(x_1, x_2)$ (and no other positive binary atoms);  
    $\pB_{\beta_5}$ is assigned to $(\ela_1, \ela_2)$ and  
    $\pB_{\beta_5^{-1}}$ to $(\ela_2, \ela_1)$ due to $\varphi_\tab^5$, providing the witness for $\ela_1$.
  \item The second witness for $\ela_2$ is provided by $\varphi_\tab^4$, 
    which requires some $\pB_\beta$ to be assigned to $(\ela_2, \ela_1)$ and  
    $\pB_{\beta^{-1}}$ to $(\ela_1, \ela_2)$.  
    By $\varphi_\tab^3$, this $\beta$ must be equal to ${\beta_5^{-1}}$.
  \item $\beta_6$ is the $2$-type, compatible with the $1$-type $\alpha_3$ of its endpoints,  
    containing no positive binary atoms. Assigning $\pB_{\beta_6}$ to $(\ela_3, \ela_4)$ is not enforced by $\varphi_\tab$, but it is not forbidden.
\end{itemize}

\noindent We first append $(\ela_4, \ela_3)$ to (the interpretation of) $\textstyle\pB_{\beta_6^{-1}}$.  
Then, for each pair not yet contained in any $\pB_\beta$, take a~$\beta \in \im(\tab)$ compatible with its $1$-types,  
and assign $\pB_\beta$ to this pair; for instance, one may choose $2$-types containing no positive atoms.  
(The newly added connections are represented by solid green arrows in Fig.~\ref{f:example-two-variables}).  
This structure becomes a model $\strA^+ \models \varphi$ after assigning the $1$-types and $2$-types over the original signature $\sigma$ of $\varphi$
in accordance with the $\pU_\alpha$ and $\pB_\beta$.

\subsection{The Multi-Variable Case}\label{sub:the_case_of_many_variables}

\noindent To lift the result from the previous section to the general case, we generalize the approach of Kieroński\&Rudolph~\shortcite{KieronskiR21}, originally developed for the finite satisfiability problem in $\GFTG$.
Fix an $\RGF$-sentence $\varphi$ in NF with its usual components (\cf Def.~\ref{def:normal-form-RGF}).
The core idea is to decompose $\varphi$ into two parts: a sentence $\varphifofull$ in the guarded fragment and a sentence $\varphiregfull$ in $\RGFt$, augmented with some additional synchronizing information concerning the set of realizable  $1$-types $\AAA_0$ and (first-order) $2$-types $\BBB_0$.
Both formul{\ae} express that precisely all $1$-types from $\AAA_0$ are realized. However, they differ in how they handle the realization of $2$-types from $\BBB_0$: $\varphiregfull$ asserts that no $2$-type outside of $\BBB_0$ is realized, whereas $\varphifofull$ requires the realization of all $2$-types in $\BBB_0$ (and possibly others as well).
We then verify the satisfiability of $\varphifofull$ and $\varphiregfull$ separately and, by carefully combining multiple copies of models of these two sentences, we construct a model of $\varphi$. 
Our approach heavily relies on the fact that regular predicates occur only~in~guards (otherwise our logic is undecidable).

\noindent For $\beta \in \BBB_{\varphi}$, we define $\betaminus$ (the \emph{regular-free reduction} of $\beta$) as the set of formul{\ae} obtained by simultaneously removing all $\sigreg$-literals involving both variables $x_1, x_2$ (preserving those involving only a single variable). For a structure~$\strA$, we define $\strAminus$ as its substructure obtained by restricting the interpretation of all regular predicates to equal-element tuples.

\vspace{-1em}
\paragraph*{The formul{\ae}.}
We first construct the formul{\ae} $\varphifofull$ and $\varphiregfull$, parametrized by ($\FO$-compatible with $\varphi$, parametrization in names omitted due to space reasons) sets of types $\AAA_0 \subseteq \AAA_{\varphi}$ and $\BBB_0 \subseteq \BBB_{\varphi}$.
Define $\varphifofull \deff \varphifo{1} \land \varphifo{2} \land \varphifo{3}$, where\\
\textbf{1.} Our $\varphifo{1}$ comprises all $\forallfo$- and $\existsfo$-conjuncts of $\varphi$.\\
\textbf{2.} The formula $\varphifo{2}$ enforces that precisely the $1$-types in $\AAA_0$ are realized in the intended model:
\[
  \varphifo{2} \deff \textstyle\bigwedge_{\alpha \in \AAA_0} \exists{x_1}\, \alpha(x_1) \;\; \land \;\; \forall{x_1}\,\textstyle\bigvee_{\alpha \in \AAA_0}  \alpha(x_1). \label{gf5}
\]
\textbf{3.} The formula $\varphifo{3}$ ensures realizations of regular-free reductions of (at least) $2$-types $\beta \in \BBB_0$ in the intended~model:
\[
  \varphifo{3} \deff \textstyle\bigwedge_{\beta \in \BBB_0} \exists{x_1}\exists{x_2}.\, \pred{Aux}(x_1,x_2) \land \betaminus(x_1,x_2), \label{gf4}
\]
where $\pred{Aux}$ is an auxiliary binary predicate included solely for syntactic correctness.
Treating all $\sigreg$-predicates in $\varphifofull$ as standard predicates, we conclude~that~$\varphifofull$~is~in~$\GF$.

\noindent Next, we define $\varphiregfull \deff \varphireg{1} \land \varphireg{2} \land \varphifo{2}$ as follows.\\
\textbf{1.} Our $\varphireg{1}$ comprises all $\forallreg$- and $\existsreg$-conjuncts of $\varphi$.\\
\textbf{2.} Our $\varphireg{2}$ ensures that $2$-types realized in the intended models belong to $\BBB_0$.
More specifically, it has the form: 
  \[
    \varphireg{2} \deff  \textstyle\bigwedge_{\pG}\textstyle\bigwedge_{\bar{x}} \forall{x_1}\forall{x_2}\, \pG(\bar{x}) \to \textstyle\bigvee\nolimits_{\beta \in \BBB_{0}} \betaminus(x_1x_2),  \label{rgf1}
  \]
  where $\pG$ is any guard from $\varphi$ (a predicate over the signature of $\varphi$ or program from $\varphi$) and $\bar{x}$ is a variable sequence, composed solely of $x_1$ and $x_2$, of length matching the arity~of~$\pG$.

  \noindent We stress that $\varphiregfull$, as written, is not in $\RGF$ due to ocurrences outside guard of literals like $\pm \pP(x_1x_1)$ for regular predicates $\pP \in \sigreg$ in~$\betaminus$. 
  This can be remedied by replacing each such literal with (an equivalent guarded) formula $\exists{x_2} \pP(x_1x_1) \land \top$.

\noindent The following lemma is the main ingredient of our reduction.

\begin{lemma} \label{lemma:fusion_main}
$\varphi$ is (finitely) satisfiable iff for some (bounded exponentially in $|\varphi|$) $\FO$-compatible sets of types $\AAA_0$ and $\BBB_0$ both formul{\ae} $\varphifofull$ and $\varphiregfull$ are (finitely) satisfiable.~\myqed
\end{lemma}

\noindent \Cref{lemma:fusion_main} tells us that to check the (finite) satisfiability of~$\varphi$, it suffices to iterate over possible sets of $1$-types $\AAA_0 \subseteq \AAA_{\varphi}$ and ``sparse'' sets $\BBB_0 \subseteq \BBB_{\varphi}$---a process that can be performed in doubly-exponential time as done in the previous section---then construct  $\varphifofull$ and $\varphiregfull$ and apply the appropriate algorithms to test their satisfiability. Since the constructed formul{\ae} are exponential in $|\varphi|$, and the satisfiability problems of the target logics are $\TwoExpTime$-complete, this naively appears to yield a triply-exponential-time algorithm. 
However, the second part of Thm.~\ref{thm:deciding-plain-GF} shows that the satisfiability of $\varphifofull$ can be tested in time doubly-exponential in~$|\varphi|$.
The same holds for $\varphiregfull$ by examining the proof of Thm.~\ref{thm:RGFt-correctness} (by incorporating $\AAA_0$ and $\BBB_0$ directly into the $\varphi$-tables instead of explicitly including $\varphireg{2}$ and $\varphifo{2}$ in $\varphiregfull$).~We~conclude:

\begin{thm} \label{theorem:RGF-is-two-exp-compl}
$\SAT(\RGF)$ is $\TwoExpTime$-complete.\myqed
\end{thm}

\noindent We stress that the finite satisfiability problem for $\ICPDL$ is a notorious open question, and the decidability status even for its tiny fragment \Logic{LoopPDL}~\cite{DaneckiLoopPDL84} is open for more than 40 years.
Our result however, can be used for conditional lifting the decidability of~$\FSAT$~from~$\ICPDL$~to~$\RGF$.
\begin{corollary}\label{corr:transfer-to-finsat}
If the finite satisfiability for $\ICPDL$ is decidable and $\TwoExpTime$-complete, then so is $\FSAT(\RGF)$.~\myqed 
\end{corollary}


\noindent In the remainder of this section, we prove~\Cref{lemma:fusion_main}.

The left-to-right implication is easy.
Let $\strA \models \varphi$, and define $\AAA_{1}$ and $\BBB_{1}$ as the sets of $1$-types and $2$-types realised in $\strA$.  
Since $\BBB_{1}$ may be doubly exponential in $|\varphi|$, we seek a smaller subset $\BBB_0 \subseteq \BBB_1$ to complete the proof.
First, construct an auxiliary $\varphiregfull$ based on $\AAA_1$ and $\BBB_1$. Note that $\strA \models \varphiregfull$.  
Second, applying the construction from \Cref{lemma:RGF2-small-number-of-2-types} we obtain a model $\strB \models \varphiregfull$ that realize only exponentially (w.r.t.~$|\varphi|$) many $2$-types.  
Take $\AAA_0$ and $\BBB_0$ to be the sets of $1$- and $2$-types realised in~$\strB$. Observe that $\AAA_0 = \AAA_1$ and $\BBB_0 \subseteq \BBB_1$.
Now construct $\varphifofull$ and $\varphiregfull$ for $\AAA_0$ and $\BBB_0$.  
It can be readily verified that $\strB \models \varphiregfull$, and that the expansion of $\strA$ (interpreting $\pred{Aux}$ as the full relation) satisfies~$\varphifofull$.

For the other direction, fix $\AAA_0$ and $\BBB_0$ as in the lemma's statement, and consider models $\strAFO \models \varphifofull$ and~$\strAReg \models \varphiregfull$.
W.l.o.g. we impose some extra assumptions on $\strAFO$~and~$\strAReg$.
\begin{description}[itemsep=0em, leftmargin=*]

\item[\desclabel{(C1)}{fuzja:1}] As $\varphiregfull$ is a two-variable sentence, no relation in $\strAReg$ contains tuples composed of more than two distinct elements.

\item[\desclabel{(C2)}{fuzja:2}] $\strAFO$ contains no pair of distinct elements $\ela, \elb$ with $\strAFO \models \pT[\ela,\elb]$ for a regular $\pT \in \sigreg$ (all atoms involving regular predicates from $\varphifofull$ employ only a sole~variable).
\item[\desclabel{(C3)}{fuzja:3}] For convenience,  all conjuncts of $\varphifo{3}$ are witnessed in $\strAFO$ by pairs of distinct elements (\ie $\strAFO \models \betaminus[\ela\elb]$ for some $\ela \not\approx \elb$). It is easy to rebuild $\strAFO$ if this is not the~case.
\item[\desclabel{(C4)}{fuzja:4}] The domains of $\strAFO$ and $\strAReg$ are of equal cardinality (finite or~$\aleph_0$). Each $1$-type is realized in $\strAFO$ and~$\strAReg$ an equal number of~times.
In particular, if $\strAReg$ is finite, then the model of $\varphi$ produced by our construction will be finite as well.
This follows from $\varphifo{2}$, the FMP for $\GF$~\cite{Gradel99}, downward L\"owenheim-Skolem property for $\mathcal{L}_{\omega_1\omega}$, and some standard ``model surgeries'' used in the context of guarded logics.
\end{description}

We now combine $\strAFO$ and $\strAReg$ into a model of~$\varphi$ by constructing a \emph{fusion structure} $\strF$.
From a bird's eye view, $\strF$ is a (potentially infinite) two-dimensional square grid in which each row is an isomorphic copy of $\strAFO$, each column is an isomorphic copy of $\strAReg$, and
the interpretation of regular predicates is confined to the columns of $\strF$.
This structure is nearly a model of $\varphi$; the only remaining issue is that some ``vertical'' pairs of elements may lack witnesses for certain $\existsfo$-conjuncts in $\varphi$.
We resolve this using a ``circular witnessing scheme'' akin to the one by Gr\"adel\&Kolaitis\&Vardi~\shortcite{GKV97}.
\input{sections/fig-fusion-BBE}

\vspace{-1em}
\noindent \paragraph*{Step I: Domain of $\strF$ and $1$-types.} 
Let $\rmK$ denote the cardinality of the domains of $\strAFO$ and $\strAReg$ (equal due to \ref{fuzja:4}).
If $\rmK = \aleph_0$ then, after a bijective renaming, we may assume that the domains of $\strAFO$ and $\strAReg$ are both equal to $\Z$.
We then let $\rmF \deff \Z \times \Z$ and we interpret predicates from $\sig$ minimally to fulfil $\tp{\strF}{1}(k,l) = \tp{\strAFO}{1}(k{+}l)$ for all $k,l \in \Z$.
Otherwise, after a bijective renaming, the domains of $\strAFO$ and $\strAReg$ are both equal to $\Z_{\rmK} \deff \{ 0, 1, \ldots, \rmK{-}1\}$.
We let $\rmF \deff \Z_{\rmK} \times \Z_{\rmK}$ and interpret predicates from $\sig$ minimally to fulfil $\tp{\strF}{1}(k,l) = \tp{\strAFO}{1}(k{+}l \bmod \rmK)$ for all $k,l \in \Z_{\rmK}$.

\vspace{-1em}
\noindent \paragraph*{Step II: Isomorphic copies.} 

As already said, we naturally view $\str{F}$ as a two-dimensional square table and speak about \emph{columns} (resp. \emph{rows}) of $\strF$, meaning the sets of all elements from $\strF$ sharing the same 1st (resp. 2nd) coordinate.
We also speak about \emph{vertical} (resp. \emph{horizontal}) tuples from $\strF$, meaning the (guarded, \ie contained in some relation) tuples having all its elements in the same column (resp. row) of $\strF$.
By construction, each row and column of $\strF$ contains the same number of realizations of each $1$-type $\alpha$.
This implies that from every row there exists a $1$-type-preserving bijection to $\strAFO$, and similarly for columns and $\strAReg$.
Thus, we may define the interpretation of relations in $\strF$ minimally to ensure that every row of $\strF$ becomes an isomorphic copy of $\strAFO$ and that every column of $\strF$ becomes an isomorphic copy of $\strAReg$. 

\vspace{0.5em}
\noindent A few important observations about~$\strF$.
First, note that by condition~\ref{fuzja:2}, no program connect elements from different columns of $\strF$.
Second, all guarded tuples from $\strF$ consisting of a single element are both vertical and horizontal.
Third and last, every guarded tuple in $\strF$ is either vertical or horizontal. 

\vspace{-1em}
\noindent \paragraph*{Step III: ``Almost'' satisfaction.} 
\hspace{-0.5em}We show that $\strF$ is ``almost'' a model of $\varphi$, \ie it satisfies all the conjuncts (with their usual components from Def.~\ref{def:normal-form-RGF}) from $\varphi$ except for $\existsfo$-ones. We deal with each conjunct separately, proving~its~satisfaction.\\
\noindent $\bullet$ $\strF \models \forall{x_1}\ \lambda(x_1)$ and $\strF \models \forall{\bar{x}}\, \etaifa(\bar{x}) \to \psiifa(\bar{x})$ (for all $i$).\\
\noindent It suffices to show that every $1$-type and $2$-type realized in $\strF$ lies in $\AAA_0 \cup \BBB_0$, which consists of types $\FO$-compatible types with~$\varphi$ only. For $1$-types, those realized in $\strF$ also appear in $\strAReg$ and $\strAFO$ (by construction), and thus belong to $\AAA_0$ by $\strAReg, \strAFO \models \varphifo{2}$. For $2$-types, any guarded tuple $\bar{\ela}$ is either horizontal or vertical, and so appears in $\strAFO$ or~$\strAReg$, respectively. In the first case, we invoke $\strAFO \models \varphifo{1}$; Otherwise, as $\bar{\ela}$ uses at most $2$ elements, we apply~$\strAReg \models \varphireg{2}$.\\
\noindent $\bullet$ $\strF \models \forall{x_1x_2}\, \piifa(x_1x_2) \to \phiifa(x_1x_2)$ (for all $i$).\\
Let $\strF \models \piifa[\ela, \elb]$ for some $\ela, \elb$.
By \ref{fuzja:2}, we infer that $(\ela, \elb)$ is vertical (otherwise $\ela = \elb$, so it is vertical as well),
and that the path witnessing $(\ela, \elb) \in (\piifa)^{\strF}$ is entirely contained within the column of $\strF$ determined by $\ela$ (or~$\elb$).
Since each column satisfies~$\varphireg{1}$,~we~are~done.

\noindent $\bullet$ $\strF \models \forall{x_1}\, \gammaiex(x_1) \to \exists{x_2}\, \piiex(x_1x_2) \land \phiiex(x_1x_2)$ (for all $i$).

\noindent Suppose $\strF \models \gammaiex[\ela]$.
By construction, the column of $\ela$ is isomorphic to $\strAReg$ and it satisfies $\varphireg{1}$.
Hence, the element $\ela$ finds its witness for the satisfaction of $\piiex(x_1x_2) \land \phiiex(x_1x_2)$.

\noindent We now address the satisfaction of $\existsfo$-conjuncts of the form $\forall{\bar{x}}, \etaiex(\bar{x}) \to \exists{\bar{y}}, \varthetaiex(\bar{x}\bar{y}) \land \psiiex(\bar{x}\bar{y})$ in $\strF$.
Observe that all horizontal guarded tuples $\bar{\ela}$ with $\strF \models \etaiex[\bar{\ela}]$ have the required witnesses within their rows—this follows directly from the satisfaction of $\varphifo{1}$ by every row of $\strF$.
Thus, the only obstruction to $\strF$ being a model of $\varphi$ is the potential lack of witnesses for some vertical guarded tuples.
By \ref{fuzja:1}, such vertical tuples consist of exactly two elements (since single-element tuples already have witnesses in their rows).
To resolve this, we take multiple copies of $\strF$ and, for each vertical guarded pair lacking witnesses, embed it into a row of a different copy of $\strF$, thereby ensuring all required $\existsfo$-witnesses~are~present.

\vspace{-1em}
\noindent \paragraph*{Step IV: Providing missing $\existsfo$-witnesses.} 
\hspace{-0.5em} Since the case of $\rmK = \aleph_0$ can be handled analogously, let us focus only on the case when $\rmK$ is a natural number.
We define a structure $\strG$ as the disjoint union of $3\rmK$ isomorphic~copies~of~$\strF$,~\ie:\\ 
\noindent $\bullet$ The domain $G$ of $\strG$ is equal to $\{ 0, 1, 2 \} \times \Z_{\rmK} \times F$.\\\noindent $\bullet$ The interpretation of all predicates is minimal to make $\strG$ restricted to $\{ i \} \times \{ j\} \times F$, for each $i$ and $j$, isomorphic~to~$\strF$.\\

\vspace{-0.5em}
\noindent For convenience, let $G_i \deff \{ i \} \times \Z_{\rmK} \times F$ for each $i \in \{0,1,2\}$, and let $\strG_i$ be the restriction of $\strG$ to $G_i$. 
When referring to rows of $\str{G}$ or $\str{G}_i$ we mean rows in their $\str{F}$-components, that is sets of the form $\{i \} \times \{j\} \times E$, for some $i,j$, where
$E$ forms a row in $\str{F}$.
Analogously for columns and vertical/horizontal tuples. 
Consider a vertical guarded pair $(\elb_1, \elb_2)$ of distinct elements in $\strG$.
If the pair $(b_1, b_2)$ lacks some  $\existsfo$-witnesses, that is, there is at least one $i$ such that
$\strG \models \etaiex[\bar{\elc}]$ for some tuple $\bar{\elc}$ consisting solely of $\elb_1$ and $\elb_2$ (each occurring~at~least~once),
then we provide a witnesses by ``connecting'' $(\elb_1, \elb_2)$ to (to be selected) row~$\strE$~as~follows.

%
\noindent As $\strG \models \varphireg{2}$, we infer $\strG \models \betaminus[\elb_1, \elb_2]$ for some $\beta \in \BBB_0$.  
By $\strE \models \varphifo{3}$, there are $\ela_1, \ela_2 \in E$ with $\strE \models \betaminus[\ela_1, \ela_2]$.
By condition \ref{fuzja:3}, we can assume that $\ela_1 \neq \ela_2$.  
For any tuple $\bar{\ela}$ containing at least one of $\elb_1$, $\elb_2$, some elements from $E \setminus \{\ela_1, \ela_2\}$, and a relation $\pR \in \sigfo$ of arity $|\bar{\ela}|$, we add $\pR(\bar{\ela})$ to $\strG$ if and only if $\strE \models \pR[\homoh(\bar{\ela})]$, where $\homoh$ maps $\elb_s \mapsto \ela_s$ (for all $s \in \{1,2\}$) and acts as the identity on $E$.  
Thus, $\homoh$ defines an isomorphism between $(E \cup \{\elb_1, \elb_2\}) \setminus \{\ela_1, \ela_2\}$ and $E$, if regular predicates are ignored.  
No other facts are added—specifically, no fresh regular atoms. 

\noindent Note that since $\strE \models \varphifo{1}$, all tuples built out of elements of $E$ have in $\str{E}$ all the $\existsfo$-witnesses required by~$\varphi$.
After the above described connection of $\elb_1, \elb_2$ to $\str{E}$ this becomes true also for all tuples $\bar{\ela}$ built out of elements from  $\{\elb_1, \elb_2 \} \cup E \setminus \{\ela_1, \ela_2\}$ (because  $\bar{\ela}$ starts to mimic the behaviour of the tuple $\homoh(\ela)$ from $E$, if regular predicates~are~ignored).

\noindent We repeat the process for all vertical guarded $(\elb_1, \elb_2)$. 
To avoid conflicts, if $\elb_1, \elb_2$ are in $G_i$, we select a row $\strE$ from the $G_{(i{+}1) \bmod 3}$ that is not yet used by any other pair from their column. Since each $G_i$ has $\rmK$ copies of $\strF$, each with $\rmK$ rows, and the number of such pairs is less than $\rmK^2$, there are enough rows to complete the process.  
This circular strategy, using three disjoint sets $G_i$, guarantees that if an element $\elb$ is connected to a row $\strE$, no element of $\strE$ is ever connected back to $\elb$'s row.  
One can check that this scheme preserves all $\forallfo$-, $\forallreg$-, and $\existsreg$-conjuncts, yielding~a~model~of~$\varphi$.

\subsection{A Slightly Extended Setting}\label{sub:slightly-extended-setting}

In description logics and ontology-based data access, the focus often shifts from satisfiability to the more general problem of query entailment over knowledge bases (see the survey by~\citeauthor{OrtizS12} for extra background). 
We show how this general entailment problem reduces to plain satisfiability, assuming that the queries are acyclic. Without this restriction, the problem becomes undecidable~(see~Sec.~\ref{sec:undecidability-querying}).

\noindent An $\RGF$-knowledge base~(KB) is a pair $\mathcal{K} \deff (\mathcal{A}, \mathcal{T})$, where $\mathcal{A}$ (ABox) is a finite set of ground facts and $\mathcal{T}$ (TBox) a finite set of $\RGF$-sentences. A structure $\strA$ satisfies $\mathcal{K}$ if it satisfies all elements of $\mathcal{A} \cup \mathcal{T}$. A query $\queryq$ is entailed by $\mathcal{K}$ (written: $\mathcal{K} \models \queryq$) if it holds in all models of $\mathcal{K}$.
A {\CQ} $\queryq$ is \emph{tree-shaped} if it contains only atoms of arity at most two, and its variables can be bijectively mapped to the nodes of some tree in a way that each binary atom $\pR(x,y)$ satisfies one of: (i) $x = y$, (ii)~$x$ is a child of $y$, or (iii) $x$ is the parent of $y$.
In the (tree-shaped) {\CQ} entailment problem over $\RGF$-KBs we ask if an input (tree-shaped) {\CQ} is entailed by an input $\RGF$-KB.

\begin{thm}
The entailment problem of tree-shaped {\CQ} over $\RGF$-KBs is $\TwoExpTime$-complete.~\myqed
\end{thm}
\begin{proof}
We reduce the problem to formula satisfiability in polynomial time.  
Given a tree-shaped {\CQ} $\queryq$, the well-known ``rolling-up'' technique~\cite{HorrocksT00} yields an $\RGF$-sentence $\varphi_{\queryq}$ such that, for all models $\strA$ of $\mathcal{K}$, we have that 
$\strA \models \queryq$ iff $\strA \models \varphi_{\queryq}$ (cf.\ Section~3.2 of \citeauthor{BednarczykPHD}'s PhD thesis).  
It remains to encode the KB as a single formula.  
Let $\rmN$ be the number of individual names in $\mathcal{A}$. Introduce a fresh $\rmN$-ary predicate $\pred{aux}$, and let $\mathcal{A}^{\dagger}$ be obtained from $\mathcal{A}$ by bijectively renaming these names to $x_1, \dots, x_{\rmN}$.  
Then 
\[
  \left( \exists x_1\ldots x_{\rmN}\; \pred{aux}(x_1,\ldots,x_{\rmN}) \land \bigwedge \mathcal{A}^{\dagger} \right) \; \land \; \bigwedge \mathcal{T} \; \land \; \neg \varphi_{\queryq}.
\]
is unsatisfiable iff $\mathcal{K} \not\models \queryq$, concluding the proof.
\end{proof}
 

\newcommand{\snk}{\mathbb{S}}
\newcommand{\snkN}{\snk_{\rmN}}
\newcommand{\strSnk}{\str{S}}
\newcommand{\lesssnk}{<_{\snk}}
\newcommand{\lesseqsnk}{\leq_{\snk}}

\newcommand{\pHnot}{\overline{\pred{H}}}
\newcommand{\phisnake}{\varphi_{\strSnk}}
\newcommand{\phisnakeprop}{\varphi_{\textrm{pro}}}
\newcommand{\phisnakepropeven}{\varphi_{\textrm{pro}}^{\textrm{even}}}
\newcommand{\phisnakepropodd}{\varphi_{\textrm{pro}}^{\textrm{even}}}
\newcommand{\phisnakecons}{\varphi_{\textrm{con}}}
\newcommand{\phisnakeinit}{\varphi_{\textrm{ini}}}
\newcommand{\phisnakegrid}{\varphi_{\textrm{grd}}}
\newcommand{\querysnake}{\mathrm{q}_{\textrm{clo}}}

\newcommand{\CogOOne}{(\text{\faCog}_{\mathrm{O}}^{1})}
\newcommand{\CogOTwo}{(\text{\faCog}_{\mathrm{O}}^{2})}
\newcommand{\CogEOne}{(\text{\faCog}_{\mathrm{E}}^{1})}
\newcommand{\CogETwo}{(\text{\faCog}_{\mathrm{E}}^{2})}

\newcommand{\phitiling}{\varphi_{\tilingsys}}

\section{Undecidability of Querying}\label{sec:undecidability-querying}

\noindent We prove undecidability of (both general and finite) {\CQ}-entailment problems for $\RGF$, already for its very restricted fragment: $\FGFt$ with a \emph{single} transitive guard.
This significantly tightens the undecidability of the unrestricted \UCQ-entailment over $\GFtTG$ by Gottlob et al.~\shortcite[Th.~2]{GottlobPT13} in three essential~ways: (i) our logic is stricter [fluting],  (ii) we utilize only~a~single {\CQ} [rather than a {\UCQ}], and (iii) our undecidability applies also to the finite case [which was open].

\noindent We first introduce ``triangular grid-like'' $\sigma$-structures dubbed \emph{snakes}, where $\sigma \deff \{\pB, \pT_0, \pT_1, \pT_2, \pE, \pH, \pV, \pP\}$ and all but the last three predicates in $\sigma$ (which are binary) are~unary.
In snakes, the predicates $\pH$ and $\pV$ denote the \emph{horizontal} and the \emph{vertical} successors in the intended grid. 
By ``triangular'' we mean that the $i$-th column of a snake contains $i{+}1$ (resp. $i{+}2$) elements if $i$ is even (resp. odd).
Additionally, $\pE$ labels \emph{even} columns, $\pB$ labels the \emph{bottom} of the grid, $\pT_i$s label the elements exactly~$i$ steps from the \emph{top} of the grid, and (a transitive) $\pP$ relates the elements reachable via a directed \emph{path} in the underlying graph.
Quite unusually, the direction of $\pV$ alternates between columns. 
Consult \Cref{fig:snake} for intuitions.


\begin{definition}\label{def:snake}
  Fix $\rmN \in \N{\cup}\{\infty\}$. The set $\snkN \subseteq \N^2$ consists of all pairs $(n,m)$ with $m \leq n \leq \rmN$ and $m{-}1 \leq n \leq \rmN$~if~$n$ is even and odd, respectively. 
  The \emph{snake ordering} $\lesssnk$ (also known as boustrophedon) on $\snkN$ is defined by cases: $(n,m) \lesssnk (n',m')$ if $n < n'$ or if $n = n'$ and either (i) $n$ is odd and $m < m'$, or (ii) $n$ is even and~$m' < m$.\\
  The \emph{$\rmN$-snake} $\strSnk \deff (\snkN, \cdot^{\strSnk})$ is a structure defined~as:
  \begin{itemize}[itemsep=0em, leftmargin=*]
    \item $\pE^{\strSnk} \deff \{ (n,m) \in \snkN \mid n \ \text{even} \}$; $\pB^{\strSnk} \deff \{ (0,m) \in \snkN \}$;
    \item $\pT_i^{\strSnk} {\deff} \{ (n,m) \in \snkN \hspace{-0.1em} \mid \hspace{-0.1em} (n,m{+}i{+}1) {\not\in} \snkN, (n,m{+}i) {\in} \snkN \}$;
    \item $\pV^{\strSnk} \deff \pV_{\textrm{even}}^{\strSnk} \cup \pV_{\textrm{odd}}^{\strSnk}$, where: 
      \begin{itemize}\itemsep0em
        \item $\pV_{\textrm{even}}^{\strSnk} \deff \{ \left( (n,m), (n,m{-}1) \right) \in \snkN^2 \mid n \ \text{is even} \}$,
        \item $\pV_{\textrm{odd}}^{\strSnk} \deff \{ \left( (n,m), (n,m{+}1) \right) \in \snkN^2 \mid n \ \text{is odd} \}$;
      \end{itemize}
    \item $\pH^{\strSnk} \deff \{ \left( (n,m), (n{+}1,m) \right) {\in} \snkN^2 \}$; and $\pP^{\strSnk}$ is $\lesssnk$ on $\snkN$.
  \end{itemize}
  Given $(a,b) \in \N^2$, let $\strSnk_{(a,b)}$ denote the restriction of the~$\infty$-snake to $\{ (n,m) \in \snk_{\infty} \mid (n,m) \lesseqsnk (a,b) \}$.
  The pairs $(\rmN, 0)$ and $(\rmN, \rmN{+}1)$ for an even and odd $\rmN$, respectively, are called \emph{final}. 
  Let $\strSnk_{\rmN}$ denote $\strSnk_{(\rmN,\rmM)}$ for the final $(\rmN,\rmM)$.~\myqed   
\end{definition}  
\vspace{-1.25em}
\begin{figure}[H]
  \begin{center}
          \begin{tikzpicture}[transform shape, scale=1.03]

      \draw (0,0) node[minirond] (V00) {\tiny{a}};
      \draw (1.75,0) node[minirond] (V10) {\tiny{b}};
      \draw (1.75,1.25) node[minirond] (V11) {\tiny{c}};
      \draw (1.75,2.5) node[minirond] (V12) {\tiny{d}};

      \draw (3.5,0) node[minirond] (V20) {\tiny{g}};
      \draw (3.5,1.25) node[minirond] (V21) {\tiny{f}};
      \draw (3.5,2.5) node[minirond] (V22) {\tiny{e}};

      \draw (5.25,0) node[minirond] (V30) {\tiny{h}};
      \draw (5.25,1.25) node[minirond] (V31) {\tiny{i}};
      \draw (5.25,2.5) node[minirond] (V32) {\tiny{j}};      
      \draw (5.25,3.75) node[minirond] (V33) {\tiny{k}};      
      \draw (5.25,5) node[minirond] (V34) {\tiny{l}};

      \draw (7,0) node[minirond] (V40) {\tiny{q}};
      \draw (7,1.25) node[minirond] (V41) {\tiny{p}};
      \draw (7,2.5) node[minirond] (V42) {\tiny{o}};      
      \draw (7,3.75) node[minirond] (V43) {\tiny{n}};      
      \draw (7,5) node[minirond] (V44) {\tiny{m}};
      
      \path[->] (V00) edge [red, -stealth] node[fill=white] {\scriptsize{$\pH$}} (V10);
      \path[->] (V10) edge [red, -stealth, densely dashdotted] node[fill=white] {\scriptsize{$\pH$}} (V20);
      \path[->] (V20) edge [red, -stealth] node[fill=white] {\scriptsize{$\pH$}} (V30);
      \path[->] (V30) edge [red, -stealth, densely dashdotted] node[fill=white] {\scriptsize{$\pH$}} (V40);
      \path[->] (V11) edge [red, -stealth, densely dashdotted] node[fill=white] {\scriptsize{$\pH$}} (V21);
      \path[->] (V21) edge [red, -stealth, densely dashdotted] node[fill=white] {\scriptsize{$\pH$}} (V31);
      \path[->] (V31) edge [red, -stealth, densely dashdotted] node[fill=white] {\scriptsize{$\pH$}} (V41);
      \path[->] (V12) edge [red, -stealth] node[fill=white] {\scriptsize{$\pH$}} (V22);
      \path[->] (V22) edge [red, -stealth, densely dashdotted] node[fill=white] {\scriptsize{$\pH$}} (V32);
      \path[->] (V32) edge [red, -stealth, densely dashdotted] node[fill=white] {\scriptsize{$\pH$}} (V42);
      \path[->] (V33) edge [red, -stealth, densely dashdotted] node[fill=white] {\scriptsize{$\pH$}} (V43);
      \path[->] (V34) edge [red, -stealth] node[fill=white] {\scriptsize{$\pH$}} (V44);

      \path[->] (V10) edge [blue, -stealth] node[fill=white] {\scriptsize{$\pV$}} (V11);
      \path[->] (V11) edge [blue, -stealth] node[fill=white] {\scriptsize{$\pV$}} (V12);
      \path[->] (V30) edge [blue, -stealth] node[fill=white] {\scriptsize{$\pV$}} (V31);
      \path[->] (V31) edge [blue, -stealth] node[fill=white] {\scriptsize{$\pV$}} (V32);      
      \path[->] (V32) edge [blue, -stealth] node[fill=white] {\scriptsize{$\pV$}} (V33);      
      \path[->] (V33) edge [blue, -stealth] node[fill=white] {\scriptsize{$\pV$}} (V34);      

      \path[->] (V22) edge [blue, -stealth] node[fill=white] {\scriptsize{$\pV$}} (V21);
      \path[->] (V21) edge [blue, -stealth] node[fill=white] {\scriptsize{$\pV$}} (V20);
      \path[->] (V44) edge [blue, -stealth] node[fill=white] {\scriptsize{$\pV$}} (V43);
      \path[->] (V43) edge [blue, -stealth] node[fill=white] {\scriptsize{$\pV$}} (V42);
      \path[->] (V42) edge [blue, -stealth] node[fill=white] {\scriptsize{$\pV$}} (V41);
      \path[->] (V41) edge [blue, -stealth] node[fill=white] {\scriptsize{$\pV$}} (V40);

      \node[below=0.01em of V00] {\scriptsize{$\pB\, \pE\, \pT_0$}};
      \node[below=0.01em of V10] {\scriptsize{$\pB\, \pT_2$}};
      \node[below=0.01em of V20] {\scriptsize{$\pB\, \pE\, \pT_2$}};
      \node[below=0.01em of V30] {\scriptsize{$\pB$}};
      \node[below=0.01em of V40] {\scriptsize{$\pB\, \pE$}};

      \node [below right=-0.2em and -0.2em of V11] {\scriptsize{$\pT_1$}};
      \node [below right=-0.2em and -0.2em of V12] {\scriptsize{$\pT_0$}};
      \node [below right=-0.2em and -0.2em of V21] {\scriptsize{$\pT_1$}};
      \node [below right=-0.2em and -0.2em of V22] {\scriptsize{$\pT_0$}};
      \node [above right=-0.2em and -0.2em of V21] {\scriptsize{$\pE$}};
      \node [above right=-0.2em and -0.2em of V22] {\scriptsize{$\pE$}};

      \node [below right=-0.2em and -0.2em of V32] {\scriptsize{$\pT_2$}};
      \node [below right=-0.2em and -0.2em of V33] {\scriptsize{$\pT_1$}};
      \node [below right=-0.2em and -0.2em of V34] {\scriptsize{$\pT_0$}};

      \node [below right=-0.2em and -0.2em of V42] {\scriptsize{$\pT_2$}};
      \node [below right=-0.2em and -0.2em of V43] {\scriptsize{$\pT_1$}};
      \node [below right=-0.2em and -0.2em of V44] {\scriptsize{$\pT_0$}};
      \node [above right=-0.2em and -0.2em of V41] {\scriptsize{$\pE$}};
      \node [above right=-0.2em and -0.2em of V42] {\scriptsize{$\pE$}};
      \node [above right=-0.2em and -0.2em of V43] {\scriptsize{$\pE$}};
      \node [above right=-0.2em and -0.2em of V44] {\scriptsize{$\pE$}};

      \path[-] (V00) edge[looseness=0.1, black!30, densely dotted] node[] {} (V10);
      \path[-] (V00) edge[looseness=0.1, black!30, densely dotted] node[] {} (V11);
      \path[-] (V00) edge[looseness=0.1, black!30, densely dotted] node[fill=white] {\scriptsize{$\pP$}} (V12);
      \path[-] (V00) edge[looseness=0.1, black!30, densely dotted] node[] {} (V20);
      \path[-] (V00) edge[looseness=0.1, black!30, densely dotted] node[] {} (V21);
      \path[-] (V00) edge[looseness=0.1, black!30, densely dotted] node[] {} (V22);
      \path[-] (V00) edge[looseness=0.1, black!30, densely dotted] node[] {} (V30);
      \path[-] (V00) edge[looseness=0.1, black!30, densely dotted] node[] {} (V31);
      \path[-] (V00) edge[looseness=0.1, black!30, densely dotted] node[] {} (V32);
      \path[-] (V00) edge[looseness=0.1, black!30, densely dotted] node[] {} (V33);
      \path[-] (V00) edge[looseness=0.1, black!30, densely dotted] node[fill=white, xshift=3em, yshift=2.75em] {\scriptsize{$\pP$}} (V34);
      \path[-] (V00) edge[looseness=0.1, black!30, densely dotted] node[] {} (V40);
      \path[-] (V00) edge[looseness=0.1, black!30, densely dotted] node[] {} (V41);
      \path[-] (V00) edge[looseness=0.1, black!30, densely dotted] node[] {} (V42);
      \path[-] (V00) edge[looseness=0.1, black!30, densely dotted] node[] {} (V43);
      \path[-] (V00) edge[looseness=0.1, black!30, densely dotted] node[] {} (V44);

      \path[-] (V10) edge[looseness=0.1, black!08, densely dotted] node[] {} (V11);
      \path[-] (V10) edge[looseness=0.1, black!08, densely dotted] node[] {} (V12);
      \path[-] (V10) edge[looseness=0.1, black!08, densely dotted] node[] {} (V20);
      \path[-] (V10) edge[looseness=0.1, black!08, densely dotted] node[] {} (V21);
      \path[-] (V10) edge[looseness=0.1, black!08, densely dotted] node[] {} (V22);
      \path[-] (V10) edge[looseness=0.1, black!08, densely dotted] node[] {} (V30);
      \path[-] (V10) edge[looseness=0.1, black!08, densely dotted] node[] {} (V31);
      \path[-] (V10) edge[looseness=0.1, black!08, densely dotted] node[] {} (V32);
      \path[-] (V10) edge[looseness=0.1, black!08, densely dotted] node[] {} (V33);
      \path[-] (V10) edge[looseness=0.1, black!08, densely dotted] node[] {} (V34);
      \path[-] (V10) edge[looseness=0.1, black!08, densely dotted] node[] {} (V40);
      \path[-] (V10) edge[looseness=0.1, black!08, densely dotted] node[] {} (V41);
      \path[-] (V10) edge[looseness=0.1, black!08, densely dotted] node[] {} (V42);
      \path[-] (V10) edge[looseness=0.1, black!08, densely dotted] node[] {} (V43);
      \path[-] (V10) edge[looseness=0.1, black!08, densely dotted] node[] {} (V44);

      \path[-] (V11) edge[looseness=0.1, black!08, densely dotted] node[] {} (V12);
      \path[-] (V11) edge[looseness=0.1, black!08, densely dotted] node[] {} (V20);
      \path[-] (V11) edge[looseness=0.1, black!08, densely dotted] node[] {} (V21);
      \path[-] (V11) edge[looseness=0.1, black!08, densely dotted] node[] {} (V22);
      \path[-] (V11) edge[looseness=0.1, black!08, densely dotted] node[] {} (V30);
      \path[-] (V11) edge[looseness=0.1, black!08, densely dotted] node[] {} (V31);
      \path[-] (V11) edge[looseness=0.1, black!08, densely dotted] node[] {} (V32);
      \path[-] (V11) edge[looseness=0.1, black!08, densely dotted] node[] {} (V33);
      \path[-] (V11) edge[looseness=0.1, black!08, densely dotted] node[] {} (V34);
      \path[-] (V11) edge[looseness=0.1, black!08, densely dotted] node[] {} (V40);
      \path[-] (V11) edge[looseness=0.1, black!08, densely dotted] node[] {} (V41);
      \path[-] (V11) edge[looseness=0.1, black!08, densely dotted] node[] {} (V42);
      \path[-] (V11) edge[looseness=0.1, black!08, densely dotted] node[] {} (V43);
      \path[-] (V11) edge[looseness=0.1, black!08, densely dotted] node[] {} (V44);

      \path[-] (V12) edge[looseness=0.1, black!08, densely dotted] node[] {} (V20);
      \path[-] (V12) edge[looseness=0.1, black!08, densely dotted] node[] {} (V21);
      \path[-] (V12) edge[looseness=0.1, black!08, densely dotted] node[] {} (V22);
      \path[-] (V12) edge[looseness=0.1, black!08, densely dotted] node[] {} (V30);
      \path[-] (V12) edge[looseness=0.1, black!08, densely dotted] node[] {} (V31);
      \path[-] (V12) edge[looseness=0.1, black!08, densely dotted] node[] {} (V32);
      \path[-] (V12) edge[looseness=0.1, black!08, densely dotted] node[] {} (V33);
      \path[-] (V12) edge[looseness=0.1, black!08, densely dotted] node[] {} (V34);
      \path[-] (V12) edge[looseness=0.1, black!08, densely dotted] node[] {} (V40);
      \path[-] (V12) edge[looseness=0.1, black!08, densely dotted] node[] {} (V41);
      \path[-] (V12) edge[looseness=0.1, black!08, densely dotted] node[] {} (V42);
      \path[-] (V12) edge[looseness=0.1, black!08, densely dotted] node[] {} (V43);
      \path[-] (V12) edge[looseness=0.1, black!08, densely dotted] node[] {} (V44);

      \path[-] (V20) edge[looseness=0.1, black!08, densely dotted] node[] {} (V21);
      \path[-] (V20) edge[looseness=0.1, black!08, densely dotted] node[] {} (V22);
      \path[-] (V20) edge[looseness=0.1, black!08, densely dotted] node[] {} (V30);
      \path[-] (V20) edge[looseness=0.1, black!08, densely dotted] node[] {} (V31);
      \path[-] (V20) edge[looseness=0.1, black!08, densely dotted] node[] {} (V32);
      \path[-] (V20) edge[looseness=0.1, black!08, densely dotted] node[] {} (V33);
      \path[-] (V20) edge[looseness=0.1, black!08, densely dotted] node[] {} (V34);
      \path[-] (V20) edge[looseness=0.1, black!08, densely dotted] node[] {} (V40);
      \path[-] (V20) edge[looseness=0.1, black!08, densely dotted] node[] {} (V41);
      \path[-] (V20) edge[looseness=0.1, black!08, densely dotted] node[] {} (V42);
      \path[-] (V20) edge[looseness=0.1, black!08, densely dotted] node[] {} (V43);
      \path[-] (V20) edge[looseness=0.1, black!08, densely dotted] node[] {} (V44);

      \path[-] (V21) edge[looseness=0.1, black!08, densely dotted] node[] {} (V22);
      \path[-] (V21) edge[looseness=0.1, black!08, densely dotted] node[] {} (V30);
      \path[-] (V21) edge[looseness=0.1, black!08, densely dotted] node[] {} (V31);
      \path[-] (V21) edge[looseness=0.1, black!08, densely dotted] node[] {} (V32);
      \path[-] (V21) edge[looseness=0.1, black!08, densely dotted] node[] {} (V33);
      \path[-] (V21) edge[looseness=0.1, black!08, densely dotted] node[] {} (V34);
      \path[-] (V21) edge[looseness=0.1, black!08, densely dotted] node[] {} (V40);
      \path[-] (V21) edge[looseness=0.1, black!08, densely dotted] node[] {} (V41);
      \path[-] (V21) edge[looseness=0.1, black!08, densely dotted] node[] {} (V42);
      \path[-] (V21) edge[looseness=0.1, black!08, densely dotted] node[] {} (V43);
      \path[-] (V21) edge[looseness=0.1, black!08, densely dotted] node[] {} (V44);

      \path[-] (V22) edge[looseness=0.1, black!08, densely dotted] node[] {} (V30);
      \path[-] (V22) edge[looseness=0.1, black!08, densely dotted] node[] {} (V31);
      \path[-] (V22) edge[looseness=0.1, black!08, densely dotted] node[] {} (V32);
      \path[-] (V22) edge[looseness=0.1, black!08, densely dotted] node[] {} (V33);
      \path[-] (V22) edge[looseness=0.1, black!08, densely dotted] node[] {} (V34);
      \path[-] (V22) edge[looseness=0.1, black!08, densely dotted] node[] {} (V40);
      \path[-] (V22) edge[looseness=0.1, black!08, densely dotted] node[] {} (V41);
      \path[-] (V22) edge[looseness=0.1, black!08, densely dotted] node[] {} (V42);
      \path[-] (V22) edge[looseness=0.1, black!08, densely dotted] node[] {} (V43);
      \path[-] (V22) edge[looseness=0.1, black!08, densely dotted] node[] {} (V44);

      \path[-] (V30) edge[looseness=0.1, black!08, densely dotted] node[] {} (V31);
      \path[-] (V30) edge[looseness=0.1, black!08, densely dotted] node[] {} (V32);
      \path[-] (V30) edge[looseness=0.1, black!08, densely dotted] node[] {} (V33);
      \path[-] (V30) edge[looseness=0.1, black!08, densely dotted] node[] {} (V34);
      \path[-] (V30) edge[looseness=0.1, black!08, densely dotted] node[] {} (V40);
      \path[-] (V30) edge[looseness=0.1, black!08, densely dotted] node[] {} (V41);
      \path[-] (V30) edge[looseness=0.1, black!08, densely dotted] node[] {} (V42);
      \path[-] (V30) edge[looseness=0.1, black!08, densely dotted] node[] {} (V43);
      \path[-] (V30) edge[looseness=0.1, black!08, densely dotted] node[] {} (V44);

      \path[-] (V31) edge[looseness=0.1, black!08, densely dotted] node[] {} (V32);
      \path[-] (V31) edge[looseness=0.1, black!08, densely dotted] node[] {} (V33);
      \path[-] (V31) edge[looseness=0.1, black!08, densely dotted] node[] {} (V34);
      \path[-] (V31) edge[looseness=0.1, black!08, densely dotted] node[] {} (V40);
      \path[-] (V31) edge[looseness=0.1, black!08, densely dotted] node[] {} (V41);
      \path[-] (V31) edge[looseness=0.1, black!08, densely dotted] node[] {} (V42);
      \path[-] (V31) edge[looseness=0.1, black!08, densely dotted] node[] {} (V43);
      \path[-] (V31) edge[looseness=0.1, black!08, densely dotted] node[] {} (V44);

      \path[-] (V32) edge[looseness=0.1, black!08, densely dotted] node[] {} (V33);
      \path[-] (V32) edge[looseness=0.1, black!08, densely dotted] node[] {} (V34);
      \path[-] (V32) edge[looseness=0.1, black!08, densely dotted] node[] {} (V40);
      \path[-] (V32) edge[looseness=0.1, black!08, densely dotted] node[] {} (V41);
      \path[-] (V32) edge[looseness=0.1, black!08, densely dotted] node[] {} (V42);
      \path[-] (V32) edge[looseness=0.1, black!08, densely dotted] node[] {} (V43);
      \path[-] (V32) edge[looseness=0.1, black!08, densely dotted] node[] {} (V44);

      \path[-] (V33) edge[looseness=0.1, black!08, densely dotted] node[] {} (V34);
      \path[-] (V33) edge[looseness=0.1, black!08, densely dotted] node[] {} (V40);
      \path[-] (V33) edge[looseness=0.1, black!08, densely dotted] node[] {} (V41);
      \path[-] (V33) edge[looseness=0.1, black!08, densely dotted] node[] {} (V42);
      \path[-] (V33) edge[looseness=0.1, black!08, densely dotted] node[] {} (V43);
      \path[-] (V33) edge[looseness=0.1, black!08, densely dotted] node[] {} (V44);

      \path[-] (V34) edge[looseness=0.1, black!08, densely dotted] node[] {} (V40);
      \path[-] (V34) edge[looseness=0.1, black!08, densely dotted] node[] {} (V41);
      \path[-] (V34) edge[looseness=0.1, black!08, densely dotted] node[] {} (V42);
      \path[-] (V34) edge[looseness=0.1, black!08, densely dotted] node[] {} (V43);
      \path[-] (V34) edge[looseness=0.1, black!08, densely dotted] node[] {} (V44);

      \path[-] (V40) edge[looseness=0.1, black!08, densely dotted] node[] {} (V41);
      \path[-] (V40) edge[looseness=0.1, black!08, densely dotted] node[] {} (V42);
      \path[-] (V40) edge[looseness=0.1, black!08, densely dotted] node[] {} (V43);
      \path[-] (V40) edge[looseness=0.1, black!08, densely dotted] node[] {} (V44);

      \path[-] (V41) edge[looseness=0.1, black!08, densely dotted] node[] {} (V42);
      \path[-] (V41) edge[looseness=0.1, black!08, densely dotted] node[] {} (V43);
      \path[-] (V41) edge[looseness=0.1, black!08, densely dotted] node[] {} (V44);
      
      \path[-] (V42) edge[looseness=0.1, black!08, densely dotted] node[] {} (V43);
      \path[-] (V42) edge[looseness=0.1, black!08, densely dotted] node[] {} (V44);

      \path[-] (V43) edge[looseness=0.1, black!08, densely dotted] node[] {} (V44);

      \draw (0.75,4.75) node[minirond] (Vx) {\tiny{x}};
      \draw (2.00,4.75) node[minirond] (Vt) {\tiny{t}};
      \draw (0.75,3.5) node[minirond] (Vy) {\tiny{y}};
      \draw (2.00,3.5) node[minirond] (Vz) {\tiny{z}};

      \path[->] (Vx) edge [blue, -stealth] node[fill=white] {\scriptsize{$\pV$}} (Vy);
      \path[->] (Vz) edge [blue, -stealth] node[fill=white] {\scriptsize{$\pV$}} (Vt);
      \path[->] (Vy) edge [red, -stealth] node[fill=white] {\scriptsize{$\pH$}} (Vz);
      \path[->] (Vx) edge [red, -stealth, densely dashdotted] node[fill=white] {\scriptsize{$\pHnot$}} (Vt);

      \draw (2.75,4.0) node (models) {\Huge{$\not\sledom$}};
    \end{tikzpicture}%
  \end{center}
  \vspace{-1.25em}
  \caption{Above depicts the $4$-snake $\strSnk_4$ violating the conjunctive query $\querysnake$. The solid arrows are the ones enforced by the formula $\phisnakegrid$, while the dashed arrows are the ones implied by (the violation of) $\querysnake$. The dotted arrows represent the predicate~$\pP$. The snake ordering $\lesssnk$ coincide with the alphabetic order of elements.}\label{fig:snake}
\end{figure}
\vspace{-1em}

\noindent We employ snakes in our reduction from the (finite) query non-entailment~problem to the undecidable tiling problem of a (finite) octant~\cite[p.~699]{BresolinMGMS10}.
This is achieved through $\phisnake \deff \phisnakeprop \land \phisnakecons \land \phisnakeinit \land \phisnakegrid \land \neg \querysnake$ describing natural properties of snakes.
Here, $\phisnakeprop$ \emph{pro}pagates unary atoms via the grid relations; $\phisnakecons$ enforces basic \emph{con}sistency conditions (\eg disjointness of all~$\pT_i$s); $\phisnakeinit$ captures the \emph{ini}tial element of $\strSnk$; $\phisnakegrid$ builds~a~skeleton of a \emph{gr}i\emph{d} (\cf solid arrows from Fig.~\ref{fig:snake}); and $\querysnake$ is a {\CQ} that ``\emph{clo}ses'' the grid (\cf dashed arrows appearing in Fig.~\ref{fig:snake}).
To manufacture~$\phisnake$ we rely on two ``helper'' predicates: a unary~$\pL$ (denoting the \emph{last} column of a finite snake), and a binary~$\pHnot$ (denoting the relative \emph{complement of $\pH$} w.r.t.~$\pP$).
The second predicate is vital for the query design as {\CQ}s do not contain negation.
It is not difficult to provide the full definitions of the first four conjuncts of~$\phisnake$.
As an example, the formula $\forall{x_1} \neg\pL(x_1) {\to} \exists{x_2} [ \pP(x_1x_2) {\land} \pH(x_1x_2) ]$~captures existence of $\pH$-successors in grids ($\pV$-successors~are treated analogously).
In our construction the grid relations always follow $\pP$. 
Such a ``$\pP$-relativised'' quantification used for constructing grids makes~$\pP$, by its transitivity, resemble the ``universal relation''.
Importantly, with the formula $\forall{x_1}\forall{x_2}\ \pP(x_1x_2) \to \left( \pH(x_1x_2) \leftrightarrow \neg\pHnot(x_1x_2)] \right)$ one can correctly axiomatize that $\pHnot$ is interpreted as a (relative to~$\pP$) complement of~$\pH$.
Finally, the role of (negated) query $\querysnake$ is to identify the cases of missing $\pH$-connections between two consecutive grid cells. 
The conjunctive query $\querysnake \deff \exists{x}\exists{y}\exists{z}\exists{t}\, \pV(xy) {\land} \pH(yz) {\land} \pV(zt) {\land} \pHnot(xt)$~then~does~the~job.

\vspace{0.5em}
\noindent Our main technical (inductive) lemma is stated below.
\begin{lemma}\label{lemma:injecting-snake-main}
  Consider $\strA \models \phisnake$, a pair $(\rmN, \rmM) \in \snk_{\infty}$, and a semi-strong homomorphism $\homof$ witnessing $\strSnk_{(\rmN, \rmM)} \sinjto_{\homof} \strA$. 
  Assume that either $(\rmN, \rmM)$ is not final or  $\homof(\rmN, \rmM) \not\in \pL^{\strA}$.
  Then $\homof$ extends to a homomorphism $\homog$ witnessing $\strSnk_{(\rmN', \rmM')} \sinjto_{\homog} \strA$, where $(\rmN', \rmM')$ is the successor~of~$(\rmN, \rmM)$~w.r.t.~$\lesssnk$.~\myqed
\end{lemma}

\noindent Based on \Cref{lemma:injecting-snake-main} we show correctness of our encoding, \ie that models of $\phisnake$ indeed ``encode'' snakes. Formally: 

\begin{lemma}\label{lemma:injecting-snake-summary}
(A) All (finite) snakes expand to (finite) models of $\phisnake$.
(B) If $\strA$ satisfies $\phisnake \land \forall{x_1}\neg\pL(x_1)$ then $\strSnk_{\infty} \sinjto \strA$.
(C) For any finite $\strA \models \phisnake$ we have $\strSnk_{\rmN} \sinjto_{\homof} \strA$ for some $\rmN$ and a map $\homof$ with $\homof(\rmN, \rmM) \in \pL^{\strA} \cap \pT_0^{\strA}$ for the final~$(\rmN,\rmM)$.~\myqed
\end{lemma}

\noindent Given an $\rmN \in \N{\cup}\{\infty\}$, by the $\rmN$-octant (a.k.a. the lower triangle of $\N \times \N$) we mean the set $\octant_{\rmN}$ of pairs $(n,m) \in \N^2$ satisfying $m \leq n \leq \rmN$.
A \emph{tiling system} $\tilingsys \deff (\tiles, \tilingV, \tilingH)$ is composed of a finite set $\tiles$ of \emph{tiles}, and binary relations~$\tilingV$,~$\tilingH$~on~$\tiles$. 
A~system~$\tilingsys$ \emph{covers} $\octant$ if there is a map $\tilesmap \colon \octant^2 \to \tiles$ satisfying $(\tilesmap(n,m), \tilesmap(n{+}1,m)) \in \tilingH$ and $(\tilesmap(n,m), \tilesmap(n,m{+}1)) \in \tilingV$ for all pairs $(n,m), (n,m{+}1) \in \octant^2$.
The (finite) octant tiling problem asks if $\tilingsys$ covers some (finite) octant.
By \Cref{lemma:injecting-snake-summary}, which encodes grids in models of $\phisnake$, the reduction is standard. For all $\tile \in \tiles$ we introduce fresh unary~$\pT_{\tile}$, and construct a formula $\phitiling$ asserting that all elements from snakes (except for the top odd ones, which are irrelevant to the reduction) are labelled by exactly one $\pT_{\tile}$, and that $\rmH$- and $\rmV$-connected elements respect $\tilingH$ and $\tilingV$. We show that $\phisnake \land \phitiling$ has a (finite) model iff $\tilingsys$ covers some (finite) octant. Thus:

\begin{thm}
Both finite and general {\CQ} entailment problems are undecidable for $\RGF$, already for its fluted two-variable fragment with a single transitive guard.\myqed
\end{thm}

\noindent We conclude by translating our results into the setting of description logics (see the standard textbook for any missing definitions).
Let $\ALC^{\cap_{\top}}$ extend $\ALC$ with concept expressions of the form $\exists(\pred{r} \cap \pred{s}).\top$, denoting elements that have at least one $\pred{r}$-filler also serving as an $\pred{s}$-filler.
We also consider the axiom $\mathsf{tr}(\pred{r})$, asserting that $\pred{r}$ is transitive, and inclusion axioms of the form $\pred{r} \subseteq \pred{s} \cup \pred{t}$, interpreted by $\mathcal{I}$~as~$\pred{r}^{\mathcal{I}} \subseteq \pred{s}^{\mathcal{I}} \cup \pred{t}^{\mathcal{I}}$.
With a routine rewriting of first-order formul{\ae} into a (slightly extended) syntax of $\ALC$, we prove the following theorem.

\begin{corollary}
Fix role names $\pred{p}, \pred{h}, \bar{\pred{h}}$.
Both finite and general {\CQ} entailment problems are undecidable for TBoxes of the form $\mathcal{T} \cup \{ \mathsf{tr}(p), \pred{p} \subseteq \pred{h} \cup \bar{\pred{h}} \}$, where $\mathcal{T}$ is an $\ALC^{\cap_{\top}}$-TBox, already for queries without the (transitive) role $\pred{p}$.~\myqed
\end{corollary} 

\section{Decreasing Computational Complexity}\label{sec:expspace}

We now examine whether $\RGF$ can be restricted to a reasonable subfragment with lower computational complexity (modulo standard complexity-theoretic assumptions). 
Our guiding requirements are twofold: first we aim to retain some form of recursion--specifically, by preserving operators such as $\cdot^*$ or $\cdot^+$; second, we require the ability to express non-trivial properties of relations with unbounded~arity.
There are three main sources of $\TwoExpTime$-hardness~of~$\RGF$:

\begin{itemize}[itemsep=0em, leftmargin=*]

\item The first one is the full $\GF$ itself~\cite{Gradel99}. Fortunately, $\SAT(\GF^k)$ is only $\ExpTime$-complete each $k > 1$.
\item The second one is $\ICPDL$~\cite{LangeL05}. Fortunately, when the input formul{\ae} are of bounded iwidth, the satisfiability problem for $\ICPDL$ is $\ExpTime$-complete.
\item The last one is $\GFtTG$~\cite{Kieronski06}, namely the two-variable~$\GF$ with transitive guards, a strict sublogic of $\RGF^2[\cdot^+]$. 
Fortunately, $\GF^2$ with \emph{one-way transitive guards} (\ie imposing that all atoms involving a transitive $\pT$ have the form $\pT(x_1x_1)$, $\pT(x_2x_2)$, or $\pT(x_1x_2)$) is $\ExpSpace$-complete with hardness holding already for its~fluted~sublogic~\cite{PrattHT23}.
\end{itemize}
\noindent The references make it clear that we cannot hope for a nice fragment of $\RGF$ with complexity below $\ExpSpace$. To identify a suitable $\ExpSpace$-complete logic, we must: (i) restrict ourselves to a proper subfragment of $\GF$, (ii) limit the interaction between operators in $\ICPDL$-programs, and (iii) enforce a one-way behavior for regular guards, in particular by disallowing the converse operator.
A natural base logic satisfying all three criteria is the forward $\GF$ (denoted $\FGF$), which has an $\ExpTime$-complete satisfiability problem~\cite{Bednarczyk21} and fits well with the notion of one-way guards introduced by Kieroński~\shortcite{Kieronski06}.
The main result of this section (available only in the extended version of this paper) establishes $\ExpSpace$-completeness of $\FRGFt[\cdot^+, \cdot^*, ?]$. We also show that adding any further operators to just Kleene's plus results in a higher complexity. 

\begin{thm}
The satisfiability problem for $\FRGF[\cdot^+, \cdot^*, ?]$ is $\ExpSpace$-complete.~\myqed
\end{thm}
\begin{thm}
Both finite and general satisfiability problems for
$\FRGFt[\cdot^{+}, \circ]$,
$\FRGFt[\cdot^{+}, \cup]$, and
$\FRGFt[\cdot^{+}, \cap]$ are already $\TwoExpTime$-hard.\myqed
\end{thm}


\section{Conclusions and Future Work}\label{sec:conclusions}

We introduced $\RGF$, a novel and expressive logic combining $\ICPDL$ with $\GF$, and proved it to be $\TwoExpTime$-complete. Through a detailed analysis, we showed that the forward fragment of $\RGF[\cdot^+, \cdot^*, ?]$ is $\ExpSpace$-complete, while additional operators---even when tests are excluded and only $\cdot^*$ or $\cdot^+$ is allowed---render its two-variable fragment $\TwoExpTime$-hard.
Future work may proceed along two directions. 
One promising path is to extend the $\ICPDL$-based guards to more expressive formalisms, such as linear Datalog or non-binary transitive closure operators. This would help eliminate the asymmetry between the binary nature of regular guards and the higher-arity relations allowed in $\GF$.
The alternative path is to tackle the finite satisfiability problem for $\RGF$. 
As previously noted, this problem remains challenging---even for minimal fragments of $\ICPDL$ like $\Logic{LoopPDL}$---and has resisted resolution for over 40 years, indicating that significant breakthroughs and new techniques will be required.
A more pragmatic direction is to study fragments of $\RGF$ with the FMP. While formul{\ae} like $\forall{x_1}\exists{x_2} \pR(x_1x_2)$ combined with either $\neg \exists{x_1} \pR^+(x_1x_1)$ or $\forall{x_1}{x_2} (\pR^+(x_1x_1) \to \pB(x_1x_2) \land \neg \pB(x_2x_1))$ destroy the FMP by enforcing infinite, non-loopable $\pR$-chains, one might hope that $\FRGF$ retains it. Unfortunately, with similar counter-examples, we show:

\begin{lemma}\label{lemma:negative-results-about-the-FMP}
None of the logics $\FRGF[\cdot^+, ?]$, $\FRGF[\cdot^+, \cdot^*]$, $\FRGF[\cdot^+, \bar{\cdot}]$, and $\FRGF[\cdot^+, \circ]$ has the finite model property.
\end{lemma}

\noindent We can already prove that $\FRGF[\cdot^+]$ has the FMP and we are currently trying to extend our approach~to~$\FRGF[\cdot^+, \cap, \cup]$.

\section*{Acknowledgements}
\noindent Work supported by NCN grants 2024/53/N/ST6/01390 (B. Bednarczyk)
and 2021/41/B/ST6/00996 (E. Kieroński).
\bibliographystyle{kr}
\bibliography{references}

\tableofcontents
\appendix

\section{Appendix to \Cref{sec:preliminaries}}\label{appendix:preliminaries}
Here, a more detailed exposition of Preliminaries is given.

\subsection{Omitted Definitions}\label{app:prelim-omited-def}
  \begin{enumerate}\itemsep0em
    \item For a detailed exposition of syntax and semantics of first-order logic ($\FO$) consult Libkin's book \shortcite[Sec.~2.1]{Libkin2004}. 
    Similarly, for an additional background on complexity theory and formal languages, consult Sipser's textbook~\shortcite[Chs. 1\&4--5\&10.3]{sipser13}.

    \item Some standard mathematical notation. We use $\N$ to denote non-negative integers, $\N_{\omega}$ for $\N \cup \{ \infty \}$, $\cupdot$ for disjoint union of sets, and $\powerset{\rmS}$ for the power-set of a set $\rmS$.

    \item We use $\bar{\cdot}$ to denote tuples of objects, \eg $\bar{x}$ often denotes a tuple of variables.
    We write $\bar{x} \cap \bar{y} = \emptyset$ to denote that the sets of elements from tuples $\bar{x}$ and $\bar{y}$ are disjoint.

    \item The \emph{size} of a formula $\varphi$, denoted $|\varphi|$, is the total number of bits required to write $\varphi$ in some reasonable encoding.
    \item Given a signature $\sigma \subseteq \sig$ and a structure $\strA$, the \emph{$\sigma$-reduct} of $\strA$ is the unique $\sigma$-structure $\strB$ that has the same domain as $\strA$ and for all predicates $\pR \in \sigma$ we have $\pR^{\strA} = \pR^{\strB}$.
    The notion of \emph{expansion} is defined dually.

    \item A \emph{homomorphism} between structures $\strA$ and $\strB$ is any map $\homof\colon A \to B$ such that for every predicate $\pR \in \sig$ and any tuple $\bar{\ela} \deff \ela_1 \ela_2 \ldots \ela_k$ from $\strA$ (where $k$ is the arity of $\pR$), $\strA \models \pR[\bar{\ela}]$ implies $\strB \models \pR[\homof(\ela_1), \homof(\ela_2), \ldots, \homof(\ela_k)]$ (in this case we say that $\homof$ \emph{preserves} $\pR$).


  \end{enumerate}

\subsection{Guarded Fragment}\label{app:guarded_fragment}
  We provide a full definition of the guarded fragment of $\FO$ by Andr{\'{e}}ka, N{\'{e}}meti and van Benthem~\shortcite{AndrekaNB98}.
  \begin{definition}\label{def:GF}
    The set of $\GF$-formul{\ae} over $\sigfo$ is the smallest set of $\FO$-formul{\ae} over $\sigfo$ that:
    \begin{itemize}\itemsep0em
    \item contains all atomic formul{\ae} over $\sigfo \cup \{ \approx \}$;
    \item is closed under the usual connectives $\lor, \land, \neg, \to, \iff$;
    \item and such that if $\psi(\bar{x}, \bar{y})$ is a $\GF$-formula and $\alpha(\bar{x}\bar{y})$ is an atom for $\alpha \in \sigfo$ involving all free variables of $\psi$, then both $\forall{\bar{x}}.\left( \alpha \to \psi \right)$ and $\exists{\bar{x}}.\left( \alpha \land \psi \right)$ are $\GF$-formul{\ae}.
    \end{itemize}
    We call the atom $\alpha$ from the last item above, a \emph{guard}.~\myqed
  \end{definition}

  Note that \Cref{def:RGF} from the main body of the paper is less liberal in a sense that it allows for unguarded formul{\ae} with a single free variable.
  This can be easily simulated with a syntax above, as one can always resort to using a dummy guard $x \approx x$.
  Example formul{\ae} in $\GF$ are the ones expressing reflexivity or symmetry of a binary relation, the usual translations of modal and description logic to $\FO$, as well as
  \[
  \forall{x}\; \pA(x) \to \exists{yz}\; \left( \pB(xxzxy) \land \left[\pC(yz) \iff \neg \pD(zx)\right] \right), \]
  which has no particular intended meaning.
  In contrast, the following four formul{\ae} are in $\FO$ but not in $\GF$. 
  \begin{itemize}\itemsep0em
  \item $\forall{xy}\ \pR(xy)$; 
  \item $\exists{x}\ \pP(x) \land \forall{yz} \left( \pR(xy) \to \pR(xz) \right)$;
  \item $\forall{xyz} \left( [\pR(xy) \land \pR(yz)] \to R(xz) \right)$; 
  \item $\forall{xy}\left( [\pP(x) \land \pP(y)] \to \pR(xy) \right)$.
  \end{itemize}
  Again, there is no intended meaning behind~these~formul{\ae}.

\subsection{Forward Guarded Fragments}\label{app:fluted_fragment}

We introduce the \emph{fluted fragment} $\FL$ by Quine~\shortcite{Quine76} and its recent generalization called the forward fragment $\FF$ by Bednarczyk~\shortcite{Bednarczyk21}. 
In what follows, $\bar{x}_{n \ldots m}$ denotes the sequence of variables $x_n, x_{n{+}1}, \ldots, x_m$.
We call $\bar{x}$ an \emph{infix} of $\bar{x}_{n \ldots m}$ if $\bar{x}$ has the form $\bar{x}_{i \ldots j}$ for some $n \leq i \leq j \leq m$. Moreover, if $m = j$, we call $\bar{x}$ a \emph{suffix} of~$\bar{x}_{n \ldots m}$.
\begin{definition}\label{def:FF}
We define $\FF^{[n]}$, for all $n \in \N$, as the smallest fragment of $\FO$ satisfying:
\begin{itemize}\itemsep0em
    \item $\FF^{[n]}$ contains all atoms $\alpha(\bar{x})$ for $\alpha \in \sigfo$ and $\bar{x}$ being an infix of $\bar{x}_{1\ldots n}$;
    \item $\FF^{[n]}$ is closed under boolean connectives $\land, \lor, \neg, \to, \iff$;
    \item If $\varphi(\bar{x}_{1\ldots n+1})$ is in $\FF^{[n{+}1]}$ then $\exists{x_{n{+}1}} \; \varphi(\bar{x}_{1\ldots n{+}1})$ and $\forall{x_{n{+}1}} \; \varphi(\bar{x}_{1\ldots n{+}1})$ are in $\FF^{[n]}$.
\end{itemize}
We let $\FF$ be the set $\FF^{[0]}$, which is composed exclusively of sentences.
Similarly, the logic $\FL$ is defined by replacing the word ``infix'' by the word ``suffix'' in the definition of~$\FF$.~\myqed
\end{definition}

We next provide examples and coexamples of $\FF$-formul{\ae}.
We begin with the positive ones:
\begin{align*}
\forall{x_1}\; (\pS(x_1) \to \neg \forall{x_2} \;(\pP(x_2) \to \pA(x_1 x_2))),\\
\forall{x_1}\; \pL(x_1) \to \neg \exists{x_2}[\pP(x_2) \land \forall{x_3}\;(\pS(x_3) \to \pI(x_1 x_2 x_3))].
\end{align*}

Now comes formul{\ae} that are syntactically not in $\FF$. 
The \wrong{red highlight} indicates a mismatch in the variable ordering.
\begin{align*}
  \forall{x_1}\forall{x_2}\forall{x_3}\; \pR(x_1 x_2) \land \pR(x_2 x_3) \to \pR(\wrong{x_1 x_3}),\\
  \forall{x_1}\ \pS(x_1) \to \pR(\wrong{x_1x_1}),\\
  \forall{x_1}\forall{x_2}\ \pR(x_1 x_2) \iff \pR(\wrong{x_2 x_1}).  
\end{align*}

\begin{definition}\label{def:guardedFF}
We define the \emph{fluted guarded fragment} $\GFL$ as the intersection of $\FL$ and $\GF$.
We define the \emph{forward guarded fragment} $\GFF$ analogously.~\myqed 
\end{definition}

Despite similarities, it turns out that $\GFF$ is strictly more expressive than $\GFL$~\cite[Thm.~4c]{BednarczykJ22}. 
Moreover, the logic $\GFF$ encodes (modulo variable renaming) the standard translation from polyadic modal logics to $\FO$ ~\cite[Sec. 1.6]{GorankoO07}.

\subsection{Turing $\RGF$-formul{\ae} into Normal Form}\label{app:scott-normal-form-proof}

\begin{lemma}\label{lemma:scott-normal-form} 
There exists a nondeterministic polynomial-time procedure that transforms any $\RGF$-sentence $\varphi$ into a normal form $\RGF$-sentence $\psi$ such that $\varphi$ is (finitely) satisfiable if and only if there exists a run of the procedure producing a (finitely) satisfiable $\psi$. Moreover, the procedure preserves the number of variables, the set of operators used in $\RGF$-programs, and the forwardness of the sentence.\myqed
\end{lemma}


\begin{proof}[Proof sketch.]

Here we outline the proof. It goes in a rather standard fashion by employing the renaming technique.
We remark that similar normal forms for \GF-based logics can be found in some other papers. In some of them, transformation to normal form 
is made in deterministic polynomial time. This however requires either increasing the number of variables (as done by Gr\"adel~\shortcite[Lemma~3.1]{Gradel99}), or allowing
for relation symbols of arity $0$ (as done by Kieroński and Rudolph~\shortcite[Lemma 1]{KieronskiR21}). 
An approach similar in spirit to ours can be found in the work of Szwast and Tendera~\shortcite[Lemma 2]{SzwastT04},
where the reduction is deterministic, but it produces an exponentially large disjunction of normal form sentences.

Let $\varphi$ be an \RGF-sentence. W.l.o.g.~we assume that all its quantifiers are existential. We simplify $\varphi$ in a bottom-up manner by replacing its subformul{\ae} with auxiliary atoms, and we simultaneously construct a conjunction $\varphi'$ of sentences that axiomatize those new atoms.

More precisely, we take an innermost subformula 
$\psi$ of $\varphi$ starting with a maximal block of quantifiers and proceed as follows.

If $\psi$ contains at least one free variable, \ie it is of the form 
$\exists \bar{y} \eta (\bar{x} \bar{y}) \wedge \psi'(\bar{x} \bar{y})$, 
for some non-empty $\bar{x}$ then we replace it by $P_{\psi}(\bar{x})$ for a fresh $P_{\psi} \in \sigfo$ of arity equal to the number of variables in $\bar{x}$ and add to $\varphi'$ two normal form conjuncts axiomatising $\pP_\psi$: 
$\forall \bar{x} (\pP_\psi(\bar{x}) \rightarrow \exists \bar{y} (\eta(\bar{x} \bar{y}) \wedge \psi'(\bar{x} \bar{y})))$ and
$\forall \bar{x} \bar{y} (\eta(\bar{x} \bar{y}) \rightarrow \neg \psi'(\bar{x}\bar{y}) \vee \pP_\psi(\bar{x}))$.

 If $\psi$ is a subsentence, i.e., it is of the form $\exists \bar{y} \psi'(\bar{y})$ then we 
nondeterministically guess its truth value, replace it
by $\top$ or $\bot$ depending on this guess and add $\psi$ or, resp., $\neg \psi$ 
as a new conjunct of $\varphi'$.

Moving up the original sentence $\varphi$, we repeat this procedure for subformul{\ae} that are now innermost, and so on.
 This process eventually converts the original $\varphi$ into a boolean combination of $\top$ and $\bot$ and produces $\varphi'$
that is a conjunction of normal form sentences and some purely existential sentences of the form $\exists \bar{y} (\eta(\bar{y}) \wedge \psi(\bar{y}))$.

If the boolean combination of $\top$ and $\bot$ evaluates to $\bot$ then we return any unsatisfiable normal form sentence, \eg $\forall x (\pP(x) \wedge \neg \pP(x))$.
Otherwise we add to $\varphi'$ the conjunct $\forall x \pC(x)$ for some fresh $\pC \in \sigfo$ and replace every purely existential conjunct  $\exists \bar{y} \psi(\bar{y})$
with the normal form conjunct $\forall x (\pC(x) \rightarrow \exists \bar{y} \psi (\bar{y}))$. We return the modified sentence $\varphi'$ in this case.

Now, if the initial formula $\varphi$ is satisfiable, then we take its model $\str{A}  \models \varphi$.  After each step of renaming we expand $\str{A}$ by interpreting the freshly introduced
$\pP_\psi$ as required by its axiomatizing conjuncts, and we make guesses $\top/\bot$ in accordance with the truth-value of the considered subsentences in the current structure. 
Finally, we expand the obtained structure by interpreting $\pC$ to be true everywhere. It should be clear that the normal form sentence obtained with the above-described nondeterministic choices
is satisfied in the so-obtained structure.

In the opposite direction, if a run of the procedure produces $\varphi'$ satisfied in a structure $\str{A}'$, then the original sentence $\varphi$ is satisfied in the restriction 
of $\str{A}'$ to the original signature. 
\end{proof} 

\section{Appendix to \Cref{sec:sat}}\label{appendix:sat}


\subsection{Normal Form for Two-Variable Logics}\label{appendix:normal-form-for-2-variables}

For the reader's convenience we rewrite \Cref{def:normal-form-RGF} in the context of the two-variable logic.
This should make the resulting formul{\ae} more readable. Indeed:
\begin{definition}\label{def:normal-form-two-variable-RGF}
An $\RGFt$-sentence $\varphi$ is in \emph{normal form}~(NF) if $\varphi = \forall{x_1}\lambda(x_1) \land \textstyle\bigwedge_{i} \varphi_i$ where each $\varphi_i$ has one~of~the~forms:
\begin{itemize}[itemsep=0em, leftmargin=*]
\item $\forall{x_1}\, \etaiex(x_1) \to \exists{x_2}\, \varthetaiex(x_1x_2) \land \psiiex(x_1x_2)$
\item $\forall{x_1}\forall{x_2}\, \etaifa(x_1x_2) \to \psiifa(x_1x_2)$
\item $\forall{x_1}\, \gammaiex(x_1) \to \exists{x_2}\, \piiex(x_1x_2) \land \phiiex(x_1x_2)$
\item $\forall{x_1}\forall{x_2}\, \piifa(x_1x_2) \to \phiifa(x_1x_2)$
\end{itemize}
for (decorated) $\sigfo$-atoms $\eta$, $\vartheta$, $\gamma$, simple $\RGF$-programs~$\pi$, and quantifier-free $\psi$, $\phi$, $\lambda$ over $\sigfo$.~\myqed
\end{definition}

It follows from \Cref{lemma:scott-normal-form} that every $\RGFt$-sentence can be reduced to the above form.

\subsection{Extended proof of \Cref{lemma:RGF2-small-number-of-2-types}}\label{appendix:lemma:RGF2-small-number-of-2-types}

\begin{proof}
Read the proof from the main body of the paper, up to the definition of $\sim$.
The class of $\beta$ in $\sim$ can be characterized by the $1$-types induced by $\beta$, the restriction of $\beta$ to binary atoms, and all formul{\ae} $\psi_i$s from the $\forall\exists$-conjuncts of $\varphi$ entailed by $\beta$ or $\beta^{-1}$.
Thus, we bound the index of $\sim$ by $2^{2|\sigma|} \cdot 2^{4|\sigma^*|} \cdot 2^{2K} \leq 2^{12|\varphi|}$, \ie exponentially in~$|\varphi|$.
We fix a representative for each $\sim$-class and denote the chosen member in $[\beta]_{\sim}$ by $\beta^\star$.
Note that $\beta$ and $\beta^\star$ differ only on atoms with two distinct variables and predicates of arity $>2$.
Consider any unordered pair of distinct elements $\ela$, $\elb$ in $A$ with $2$-type $\beta \deff \tp{\strA}{2}(\ela, \elb)$.
If neither $\beta$ nor $\beta^{-1}$ is distinguished, we modify the interpretation of predicates of arity $>2$ to ensure $(\ela, \elb)$ realizes $\beta^\star$.
Since $\beta \sim \beta'$ iff $\beta^{-1} \sim \beta'^{-1}$, we need not alter the $2$-type~of~$(\elb, \ela)$.
This process is clash-free as $2$-types involve at most two elements.
Let $\strB$ be the $\sigma$-reduct of the resulting structure. By construction, $\strA$ and $\strB$ share the same domain, and $\strB$ realizes only exponentially many $2$-types.
Moreover, every domain element has the same $1$-type in both structures, and all binary predicates (and thus all $\ICPDL$-programs in~$\varphi$) are interpreted identically.
From the definition of $\sim$ (2nd bullet), it follows that each element in $\strB$ has the same witnesses for the satisfaction of $\forall\exists$-conjuncts of $\varphi$ as it had in $\strA$.
Finally, as all $2$-types realized in $\strB$ originate from $\strA \models \varphi$, $\strB$ satisfies the $\forallfo$- and $\forallreg$-conjuncts of~$\varphi$.
Thus, $\strB$ is a desired sparse model of $\varphi$.~\qedhere
\end{proof}

\subsection{Enumerating $\varphi$-tables}\label{appendix:lemma:about-varphi-tables}

\begin{lemma}\label{lemma:about-varphi-tables}
There exists an algorithm that enumerates all $\varphi$-tables in time doubly-exponential w.r.t. $|\varphi|$.\myqed
\end{lemma}
\begin{proof}[Sketch.]
We outline the general idea of the algorithm, omitting all implementation details such as the use of Gray codes or backtracking. Specifically, we present an $\NExpTime$ procedure that produces a $\varphi$-table, which can later be transformed into an enumeration algorithm via determinisation.
We begin by constructing the set $\AAAphifo$ of all possible $\sigfo$-atoms that appear in $\varphi$. This step requires exponential time in the size of $\varphi$.
Next, we compute the set $\pr(\varphi)$, which can be done in polynomial time in $|\varphi|$.
We then iterate over indices $i$ from $0$ to $2^{12|\varphi|}$ (the size of $\im(\tab)$). For each $i$, we guess a $2$-type $\beta \in \BBBphifo$ and append it to $\im(\tab)$. This step also requires exponential time in $|\varphi|$.
The guessing of $\beta$ can be implemented by nondeterministically choosing, for each $(\sigma \cap \sigfo)$-atom with free variables $x_1, x_2$ (of which there are exponentially many), whether or not to include it in $\beta$. Notably, this does not require the explicit construction of $\BBBphifo$.
For every pair $\alpha_1, \alpha_2 \in \AAAphifo$ and subset $\rmS \subseteq \pr(\varphi)$, we guess the value of $\tab(\alpha_1, \rmS, \alpha_2)$. This step also operates in time exponential in $|\varphi|$.
We then verify that the guessed table $\tab$ satisfies all required conditions.
The satisfaction of \ref{tab:size} follows directly from our construction.
The first part of the condition \ref{tab:clo} can be verified by exhaustively enumerating all $\beta \in \im(\tab)$ and comparing them with the table content. For the second part (checking $\FO$-compatibility) the main challenge here is to verify whether a given type $\gamma$ entails a formula $\psi$ (either a guard or a subformula of one of the conjuncts of $\varphi$). This can be implemented in exponential time using a standard textbook model-checking algorithm (we construct the canonical structure $\strA_{\gamma}$ in polynomial time w.r.t. $\gamma$ and then employ a suitable first-order model-checker to see if $\strA_{\gamma} \models \forall{\bar{x}}\,\psi$).
The verification of \ref{tab:sat} proceeds similarly.
This concludes the proof sketch.  
\end{proof}

\subsection{Proof of \Cref{thm:RGFt-correctness}}\label{appendix:thm:RGFt-correctness}

For the reader's convenience, we divide the proof of \Cref{thm:RGFt-correctness} into several lemmas.
We start from the first one. 

\begin{lemma}\label{appendix:lemma:satisfiable-iff-translation-is-part-I}
If $\varphi$ is satisfiable then there exists a $\varphi$-table $\tab$ with a satisfiable $\varphi_{\tab}$.~\myqed
\end{lemma}
\begin{proof}
Suppose that $\varphi$ is satisfiable. Then, by \Cref{lemma:RGF2-small-number-of-2-types}, it admits a sparse model $\strA$. Consider the table $\tab_{\strA}$ associated with $\varphi$. Analyzing the definition of $\varphi$-tables, we observe that:
\begin{itemize}[itemsep=0em, leftmargin=*]
    \item $\tab_{\strA}$ satisfies \ref{tab:size} due to the sparsity of $\strA$;
    \item $\tab_{\strA}$ satisfies \ref{tab:clo} by construction;
    \item $\tab_{\strA}$ satisfies  the first part of \ref{tab:com} by construction, while the second part follows from $\strA \models \varphi$;
    \item $\tab_{\strA}$ satisfies \ref{tab:sat} since $\strA$ is a model of $\varphi$.
\end{itemize}

\noindent Hence, $\tab_{\strA}$ is indeed a $\varphi$-table.
We now show that $\varphi_{\tab_{\strA}}$ is satisfiable. Let $\strA^*$ be the expansion of $\strA$ that interprets each predicate $\pU_{\alpha}$ as the set of elements satisfying the $1$-type $\alpha$, and each $\pB_{\beta}$ as the set of element pairs satisfying the $2$-type~$\beta$.
Thus $(\heartsuit){:}$ $\strA^* \models \pU_{\alpha}[\ela]$ iff $\strA^* \models \alpha[\ela]$ and $\strA^* \models \pB_{\beta}[\ela, \elb]$ iff $\strA^* \models \beta[\ela, \elb]$ for all elements $\ela, \elb$, $1$-types $\alpha$, and $2$-types~$\beta$.
We claim that $\strA^* \models \varphi_{\tab_{\strA}}$, which we verify by showing that each conjunct $\varphi_{\tab_{\strA}}^1, \ldots, \varphi_{\tab_{\strA}}^7$ is satisfied by all elements of~$\strA^*$. 
This verification is straightforward and relies only on a clear understanding of the semantics of $\ICPDL$. Indeed:
\begin{itemize}[itemsep=0em, leftmargin=*]
    \item From $(\heartsuit)$, the fact that each element (resp. pair of elements) realizes a unique $1$-type (resp. $2$-type), we conclude that all elements in $\strA^*$ satisfy $\varphi_{\tab_{\strA}}^1$, $\varphi_{\tab_{\strA}}^2$, and $\varphi_{\tab_{\strA}}^3$. 
    \item By $(\heartsuit)$, the interpretation of any program $\pi$ in $\varphi$ coincides with that of $\pi_{\tab_{\strA}}$ in $\strA^*$. Thus, by the construction of $\tab_{\strA}$, we have that all elements in $\strA^*$ satisfy $\varphi_{\tab_{\strA}}^6$ and $\varphi_{\tab_{\strA}}^7$.
    \item Finally, applying $(\heartsuit)$ again, we observe that if an element $\ela$ has a witness $\elb$ for the satisfaction of the $i$-th $\existsfo$-conjunct of $\varphi$, then $\elb$ also serves as a witness for the $i$-th conjunct of $\varphi_{\tab_{\strA}}^4$. A similar argument holds for $\existsreg$-conjuncts and $\varphi_{\tab_{\strA}}^5$.
\end{itemize}
Therefore, $\strA^* \models \varphi_{\tab_{\strA}}$, as desired.
\end{proof}

\noindent Before proving the reverse implication of the above lemma, we introduce a bunch of handy notions.
We say that $\strA \models \varphi_{\tab}$ is \emph{connected} if $\star^{\strA}$ is the full binary relation on $A$.
It is routine to check that if $\strA \models \varphi_{\tab}[\ela]$ for some $\ela \in A$, then the substructure of $\strA$ induced by all the $\star$-reachable elements from $\ela$ is also a model of $\varphi$.
Thus, we can always restrict ourselves to connected models.
We say that $\strA \models \varphi_{\tab}$ is \emph{complete} if it contains no \emph{typeless pairs}, \ie those $(\ela_1,\ela_2)$ such that $\strA \not\models \pB_{\beta}[\ela_1,\ela_2]$ for any $\beta \in \BBBphifo$.
We prove that also in this case we can restrict attention to complete models.

\begin{lemma}\label{appendix:lemma:satisfiable-iff-translation-is-part-II}
Let $\tab$ be a $\varphi$-table such that $\varphi_{\tab}$ is satisfiable. Then $\varphi_{\tab}$ admits a connected and complete model.\myqed
\end{lemma}
\begin{proof}
Let $\strA$ be a model of $\varphi_{\tab}$. Without loss of generality, assume that $\strA$ is connected. We construct a complete model by eliminating all \emph{typeless} pairs from $\strA$.
Consider such a pair $(\ela_1, \ela_2)$. Let $\alpha_i \in \AAAphifo$ be the unique $1$-type such that $\strA \models \pU_{\alpha_i}[\ela_i]$ (existence and uniqueness follow from $\strA \models \varphi_{\tab}^1$). Note that $\ela_1 \neq \ela_2$, since all $2$-types involving equal elements are assigned by $\varphi_{\tab}^2$.
We distinguish two cases:
\begin{enumerate}[itemsep=0em, leftmargin=*]
\item $\strA \models \pB_{\beta}[\ela_2, \ela_1]$ for some $2$-type $\beta \in \im(\tab)$. 
Note that such a $\beta$ is unique by $\strA \models \varphi_{\tab}^1$. Define $\strA^*$ by adding $(\ela_1, \ela_2)$ to the interpretation of $\pB_{\beta^{-1}}$. Note that $\beta^{-1} \in \im(\tab)$ by property~\ref{tab:clo}. Since the interpretation of all regular predicates remains unchanged, we have $(\pi_{\tab})^{\strA} = (\pi_{\tab})^{\strA^*}$ for all $\ICPDL$ programs $\pi \in \pr(\varphi)$. Thus, all components of $\varphi_{\tab}$ except possibly $\varphi_{\tab}^6$ are still satisfied. Suppose for contradiction that $\varphi_{\tab}^6$ is violated in $\strA^*$. Then the pair $(\ela_2, \ela_1)$ satisfies the inner formula of some conjunct of $\varphi_{\tab}^6$, say for indices $\alpha, \alpha', \rmS, \rmS'$. But then $(\ela_1, \ela_2)$ satisfies the symmetric formula with indices $\alpha', \alpha, \rmS', \rmS$, contradicting the satisfaction of $\varphi_{\tab}^6$ in $\strA$.

\item Both $(\ela_1, \ela_2)$ and $(\ela_2, \ela_1)$ are typeless. Let $\rmS \subseteq \pr(\varphi)$ be the set of guards satisfied by $(\ela_1, \ela_2)$, and let $\rmS' \subseteq \pr(\varphi)$ be those satisfied by $(\ela_2, \ela_1)$. Set $\alpha \deff \alpha_1$, $\alpha' \deff \alpha_2$. Since no conjunct in $\varphi_{\tab}^7$ corresponds to $(\alpha, \alpha', \rmS, \rmS')$ (otherwise $\strA \not\models \varphi_{\tab}^6$), there exists $\beta \in \tab(\alpha_1, \rmS, \alpha_2)$ such that $\beta^{-1} \in \tab(\alpha_2, \rmS', \alpha_1)$. Define $\strA^*$ by adding $(\ela_1, \ela_2)$ to the interpretation of $\pB_{\beta}$ and $(\ela_2, \ela_1)$ to that of $\pB_{\beta^{-1}}$. As in the previous case, one can easily show that this preserves the satisfaction of $\varphi_{\tab}$.
\end{enumerate}
In both cases, the number of typeless pairs strictly decreases, and the resulting structure remains a model of $\varphi_{\tab}$. Repeating this process finitely many times yields a connected and complete model of $\varphi_{\tab}$.
\end{proof}

We now establish the reverse implication of \Cref{appendix:lemma:satisfiable-iff-translation-is-part-I}.

\begin{lemma}\label{appendix:lemma:satisfiable-iff-translation-is-part-III}
If there exists a $\varphi$-table $\tab$ with a satisfiable $\varphi_{\tab}$ then $\varphi$ is satisfiable.~\myqed
\end{lemma}
\begin{proof}
Let $\strA \models \varphi_{\tab}$ be connected and complete, as guaranteed by \Cref{appendix:lemma:satisfiable-iff-translation-is-part-II}.
Consider the expansion $\strA^*$ of $\strA$ that for all $k$-ary predicates $\pP \in (\sigma \cap \sigfo)$, all (possibly equal) elements $\ela_1, \ela_2$ from $\strA$ and all $k$-tuples $\bar{\ela}$ consisting solely of $\ela_1$, $\ela_2$ includes $\bar{\ela}$ in $\pP^{\strA^*}$ if and only if $\pP(\bar{x}) \in \beta$, where~$\beta$ is the unique $2$-type from $\BBBphifo$ for which $\strA \models \pB_{\beta}[\ela_1, \ela_2]$ and $\bar{x}$ is the tuple of variables obtained from $\bar{\ela}$ by renaming each $\ela_i$ to $x_i$.
The existence of such $\beta$ is guaranteed by completeness of $\strA$, while its uniqueness follows from $\strA \models \varphi_{\tab}^1$.
It is not difficult to verify that the construction of $\strA^*$ is clash-free, which is a consequence of the satisfaction of $\varphi_{\tab}^1$, $\varphi_{\tab}^2$ and $\varphi_{\tab}^3$ by $\strA$.
We have now secured the following properties $(\heartsuit)$:  
\begin{align*}
\strA \models \pU_{\alpha}[\ela_1] & \qquad \text{if and only if} & \strA^* \models \alpha[\ela_1],\\
\strA \models \pB_{\beta}[\ela_1,\ela_2] & \qquad \text{if and only if} &  \strA^* \models \beta[\ela_1,\ela_2],
\end{align*}
for all types $\alpha \in \AAAphifo$, $\beta \in \im(\tab)$ and elements $\ela_1, \ela_2 \in A$.
We claim that $\strA^* \models \varphi$.
Note that all $2$-types (and thus also $1$-types induced by them) in $\strA^*$ come from $\im(\tab)$.
Hence, by design of $\varphi$-tables, such types are $\FO$-compatible. This means that $\strA^*$ satisfies the $\forall{x_1}\,\lambda(x_1)$-part of $\varphi$ as well as its $\forallfo$-conjuncts.
What is more, our construction does not alter the interpretation of regular predicates, and thus they are equally interpreted in both $\strA$ and $\strA^*$.
This implies, with a similar reasoning employing the satisfaction of $\varphi_{\tab}^6$ by $\strA^*$ and \ref{tab:sat}, that $\strA^*$ satisfies all the $\forallreg$-conjuncts of $\varphi$ as well.
Finally, the satisfaction of $\existsfo$- and $\existsreg$-conjuncts is immediate by $(\heartsuit)$ and their direct translation to $\ICPDL$ expressed by both $\varphi_{\tab}^4$ and $\varphi_{\tab}^5$ (which are satisfied by $\strA^*$).
So $\strA^* \models \varphi$ as desired.
\end{proof}

\noindent We can now establish $\TwoExpTime$-completeness of $\RGFt$.

\begin{thm}\label{thm:RGFt-upper-bound}
$\SAT(\RGFt)$ is $\TwoExpTime$-complete.~\myqed
\end{thm}
\begin{proof}
The lower bound is inherited from $\ICPDL$ due to Lange\&Lutz~\shortcite{LangeL05} and $\GFtTG$~\cite{Kieronski06}.
For the upper bound we first enumerate all $\varphi$-tables $\tab$ (which can be done in doubly-exponential time w.r.t. $|\varphi|$ by \Cref{lemma:about-varphi-tables}). Then, for each of them, construct $\varphi_{\tab}$ and verify its satisfiability with an algorithm by G\"oller et al. from \Cref{thm:ICPDL-upper-bound}. 
As $|\varphi_{\tab}|$ is exponential in $|\varphi|$, but the iwidth~of~$\varphi_{\tab}$ is bounded by $|\pr(\varphi)|{+}2$ (hence only polynomially in~$|\varphi|$), this is feasible in time doubly-exponential w.r.t. $|\varphi|$.
If some of the constructed formul{\ae} is satisfiable, we conclude that $\varphi$ is satisfiable (\Cref{appendix:lemma:satisfiable-iff-translation-is-part-III}).
Otherwise, we output that $\varphi$ is not satisfiable (\Cref{appendix:lemma:satisfiable-iff-translation-is-part-I}).
The whole procedure can be implemented in $\TwoExpTime$, which concludes the proof.
\end{proof}

\subsection{Ensuring Condition \ref{fuzja:3}} \label{appendix:doubling}
Condition \ref{fuzja:3} is achieved by passing from a model $\strAFO \models \varphifofull$ to its ``doubling'' $2\strAFO$.

For a structure $\str{B}$, its \emph{doubling} $2\str{B}$ is defined as follows.
\begin{itemize}[itemsep=0em, leftmargin=*]
\item The domain of $2\str{B}$ is $B \times \{0, 1\}$.
\item For every $\pR \in \sigfo$ we have that $2\str{B} \models \pR[\bar{\elb}]$ iff $\str{B} \models \pR(\bar{\elb}_\downarrow)$, where $\bar{\elb}_\downarrow$ is the projection of $\bar{\elb}$ onto $B$.
\item For every $\pT \in \sigreg$ we have that $2\str{B} \models \pT[(\elb_1, i_1), (\elb_2, i_2)]$ iff $\elb_1=\elb_2$, $i_1=i_2$ and $\str{B} \models \pT[\elb_1, \elb_2]$. 
\end{itemize}

\noindent One can readily verify that whenever $\strAFO \models \varphifofull$ then $2\strAFO \models \varphifofull$. 
In particular, if both $\strAFO \models \pred{Aux}[\ela,\ela]$ and $\strAFO \models \betaminus[\ela,\ela]$ hold, then $2\strAFO \models \pred{Aux}[(\ela,0),(\ela,1)]$ and  $2\strAFO \models \betaminus[(\ela,0),(\ela,1)]$.
This concludes the proof.

\subsection{Ensuring Condition \ref{fuzja:4}}  \label{appendix:equalizing}
First, we may assume that both $\strAFO$ and $\strAReg$ are countable. 
This follows from the downward L\"owenheim-Skolem property~\cite{keisler1971model} of the infinitary logic $\mathcal{L}_{\omega_1\omega}$, which contains $\RGF$ as a fragment.
If $\strAReg$ is finite, we can assume that $\strAFO$ is finite as well due to the fact that $\GF$ possesses the finite model property~\cite{Gradel99}.
Otherwise, if $\strAReg$ is infinite but $\strAFO$ is not, we replace $\strAFO$ with the result of its doubling (defined in \Cref{appendix:doubling}) applied $\omega$ times.

From the satisfaction of $\varphifo{2}$ by both $\strAFO$ and $\strAReg$, we know that the sets of $1$-types realized in $\strAFO$ and $\strAReg$ are equal (more precisely, they are equal to $\AAA_0$). 
We now construct models  $\strAFO^* \models \varphifofull$ and $\strAReg^* \models \varphiregfull$ such that each $1$-type $\alpha \in \AAA_0$ is realized equally often in both structures.
Take $\rmM$ to be the maximal number of realizations of a $1$-type $\alpha$ in $\strAFO$ (over all $\alpha \in \AAA_0$).
Let $\strAReg^*$ be the disjoint union of $\rmM$ isomorphic copies of $\strAReg$ (by \emph{disjoint} we mean that there are no tuples in relations in $\strAReg^*$ involving elements from two or more distinct copies). 
As these copies are disjoint we have that $\strAReg^* \models \varphiregfull$.
Note that for every $\alpha$, the number of realizations of $\alpha$ in $\strAFO$ is less than or equal to the number of its realizations in $\strAReg^*$. 
To make these numbers equal we successively adjoin additional realizations of the appropriate $1$-types to $\strAFO$.
This is straightforward: to adjoin a realization $\elb$ of a $1$-type $\alpha$, we proceed as follows:  (i) choose a ``pattern element'' $\ela$ of type $\alpha$ in $\strAFO$, then (ii) append a fresh element $\elb$ to the domain of $\strAFO$, and finally, (iii) expand the interpretation of each predicate $\pR$ with tuples $\bar{\elc}'$ that are obtained from each $\bar{\elc}$ in the interpretation of $\pR$ by replacing every occurrence of $\ela$ in $\bar{\elc}$ with $\elb$.
It is routine to verify that after each such step the resulting structure is still a model of $\varphifofull$.
This way we derive the desired $\strAFO^*$ (possibly after taking the natural limit if infinitely many elements need to be added).


\section{Appendix to \Cref{sec:undecidability-querying}}\label{appendix:undecidability-querying}

\subsection{Definition of $\phisnake$.}\label{app:definition_of_phisnake}


\begin{definition}\label{def:phisnake}
  Let $\phisnake \deff \phisnakeprop \land \phisnakecons \land \phisnakeinit \land \phisnakegrid \land \neg \querysnake \land \mathrm{trans}(\pP)$ be defined as the conjunction of:

 \noindent 1. Propagation of atoms: even columns. (part I of $\phisnakeprop$)
  \begin{description}[itemsep=0em, leftmargin=*]
    \item[\desclabel{(PE1)}{snk:PE1}] $\forall{x_1} \pE(x_1) \to \forall{x_2} \left[ \pH(x_1x_2) \to \neg\pE(x_2) \right]$.
    \item[\desclabel{(PE2)}{snk:PE2}] $\forall{x_1} \pE(x_1) \to \pT_0(x_1) \to \forall{x_2} \left[ \pH(x_1x_2) \to \pT_2(x_2) \right]$.
    \item[\desclabel{(PE3)}{snk:PE3}] $\forall{x_1} \pE(x_1) \to \left( \neg\pT_0(x_1) \to \forall{x_2}\left[ \pH(x_1x_2) \to \alpha(x_2) \right] \right)$.
    \item[\desclabel{(PE4)}{snk:PE4}] $\forall{x_1} \pE(x_1) \to \forall{x_2}\left[ \pV(x_1x_2) \to \pE(x_2) \right]$.
    \item[\desclabel{(PE5)}{snk:PE5}] $\forall{x_1} \pE(x_1) \to \left( \pT_0(x_1) \to \forall{x_2}\left[ \pV(x_1x_2) {\to} \pT_1(x_2) \right] \right)$.
    \item[\desclabel{(PE6)}{snk:PE6}] $\forall{x_1} \pE(x_1) \to \left( \pT_1(x_1) \to \forall{x_2}\left[ \pV(x_1x_2) {\to} \pT_2(x_2) \right] \right)$.
    \item[\desclabel{(PE7)}{snk:PE7}] $\forall{x_1} \pE(x_1) \to \left( \pT_2(x_1) \to \forall{x_2}\left[ \pV(x_1x_2) {\to} \alpha(x_2) \right] \right)$.
    \item[\desclabel{(PE8)}{snk:PE8}] $\forall{x_1} \pE(x_1) \to \left( \alpha(x_1) \to \forall{x_2}\left[ \pV(x_1x_2) {\to} \alpha(x_2) \right] \right)$.
    \item[\desclabel{(PE9)}{snk:PE9}] $\forall{x_1} \pE(x_1) \to \neg \exists{x_2}\left[ \pV(x_1x_2) {\land} \pT_1(x_2) {\land} \pB(x_2) \right]$.

    where $\alpha(x) \deff \textstyle\bigwedge_{i=0}^{2} \neg \pT_i(x)$.
  \end{description}
  \noindent 2. Propagation of atoms: odd columns. (part II of $\phisnakeprop$)
  \begin{description}[itemsep=0em, leftmargin=*]
    \item[\desclabel{(PO1)}{snk:PO1}] $\forall{x_1}\ \neg\pE(x_1) \to \forall{x_2} \left[ \pH(x_1x_2) \to \pE(x_2) \right]$.
    \item[\desclabel{(PO2)}{snk:PO2}] For all $i \in \{ 0, 1, 2 \}$ we have:
    \item[\desclabel{(PO2.i)}{snk:PO2.i}] $\forall{x_1} \neg\pE(x_1) \to \left( \pT_i(x_1) \to \forall{x_2}\left[ \pH(x_1x_2) {\to} \pT_i(x_2) \right] \right)$.
    \item[\desclabel{(PO3)}{snk:PO3}] $\forall{x_1} \neg\pE(x_1) \to \forall{x_2}\left[ \pV(x_1x_2) \to \neg\pE(x_2) \right]$.
    \item[\desclabel{(PO4)}{snk:PO4}] $\forall{x_1} \neg\pE(x_1) \to \left( \pT_2(x_1) \to \forall{x_2}\left[ \pV(x_1x_2) {\to} \pT_1(x_2) \right] \right).$
    \item[\desclabel{(PO5)}{snk:PO5}] $\forall{x_1} \neg\pE(x_1) \to \left( \pT_1(x_1) \to \forall{x_2}\left[ \pV(x_1x_2) {\to} \pT_0(x_2) \right] \right).$
    \item[\desclabel{(PO6)}{snk:PO6}] $\forall{x_1} \neg\pE(x_1) \to \left( \alpha(x_1) \to \forall{x_2} [\pV(x_1x_2) \to \beta(x_2)] \right).$
    \item[\desclabel{(PO7)}{snk:PO7}] $\forall{x_1} \neg\pE(x_1) \to \forall{x_2} [\pV(x_1x_2) \to \neg\pB(x_2)]$.
    \item[\desclabel{(PO8)}{snk:PO8}] $\forall{x_1} \neg\pE(x_1) \to \left( \alpha(x_1) \to \forall{x_2}\left[ \pH(x_1x_2) \to \alpha(x_2) \right] \right)$.
    where $\alpha$ is as defined above and $\beta(x) \deff \neg\pT_0(x) \land \neg \pT_1(x)$.

  \end{description}
  \noindent 3. Propagation: bottom and last. (part III of $\phisnakeprop$)
  \begin{description}[itemsep=0em, leftmargin=*]
    \item[\desclabel{(PB1)}{snk:PB1}] $\forall{x_1} \pB(x_1) \to \forall{x_2} \left[ \pH(x_1x_2) \to \pB(x_2) \right]$.
    \item[\desclabel{(PB2)}{snk:PB2}] $\forall{x_1} \neg\pB(x_1) \to \forall{x_2} \left[ \pH(x_1x_2) \to \neg\pB(x_2) \right]$.
    \item[\desclabel{(PL)}{snk:PL}] $\forall{x_1} \pL(x_1) \to \forall{x_2} \left[ \pV(x_1x_2) \to \pL(x_2) \right]$.
  \end{description}

  \noindent 4. Some basic consistency. ($\phisnakecons$)
  \begin{description}[itemsep=0em, leftmargin=*]
  \item[\desclabel{(C1)}{snk:C1}] $\textstyle\bigwedge_{i=0}^{2}\forall{x_1}\ \pT_i(x_1) \to \left( \textstyle\bigwedge_{j \in \{ 0, 1, 2 \} \setminus \{ i \}} \neg \pT_i(x_1) \right)$.
  \item[\desclabel{(C2)}{snk:C2}] $\forall{x_1x_2}\ \pP(x_1x_2) \to \left( \pH(x_1x_2) \leftrightarrow \neg\pHnot(x_1x_2)] \right)$.

The above formula can rewritten into a conjunction of $\forall{x_1x_2}\ [\pH(x_1x_2) \land \pHnot(x_1x_2)] \to \bot$ (expressing disjointness of $\pH$ and $\pHnot$) and $\forall{x_1x_2}\ \pP(x_1x_2) \to [\pH(x_1x_2) \vee \pHnot(x_1x_2)]$.

  \end{description}

  \noindent 5. Grid building formul{\ae}. ($\phisnakeinit$)
  \begin{description}[itemsep=0em, leftmargin=*]
    \item[\desclabel{(GVO1)}{snk:GVO1}] $\forall{x_1}\ \neg\pE(x_1) {\to} (\neg\pT_0(x_1) {\to} \exists{x_2}[\pP(x_1x_2) {\land} \pV(x_1x_2)])$.
    \item[\desclabel{(GVO2)}{snk:GVO2}] $\forall{x_1}\ \neg\pE(x_1) \to \left[ \pT_0(x_1) \to \neg\exists{x_2}\ \pV(x_1x_2) \right]$.

    \item[\desclabel{(GVE1)}{snk:GVE1}] $\forall{x_1}\ \pE(x_1) {\to} (\neg\pB(x_1) {\to} \exists{x_2}[\pP(x_1x_2) {\land} \pV(x_1x_2)])$.
    \item[\desclabel{(GVE2)}{snk:GVE2}] $\forall{x_1}\ \pE(x_1) \to \left[ \pB(x_1) \to \neg\exists{x_2}\ \pV(x_1x_2) \right]$.

    \item[\desclabel{(GH1)}{snk:GH1}] $\forall{x_1}\ \neg\pL(x_1) \to \exists{x_2}\ [ \pP(x_1x_2) \land \pH(x_1x_2) ]$.
    \item[\desclabel{(GH2)}{snk:GH2}] $\forall{x_1}\ \pL(x_1) \to \neg\exists{x_2}\ \pH(x_1x_2)$.
  \end{description}

  \noindent 6. Encoding $\strSnk_0$. ($\phisnakeinit$)
  \begin{description}[itemsep=0em, leftmargin=*]
        \item[\desclabel{(I)}{snk:I}] $\exists{x_1}\ [\pB(x_1) \land \pT_0(x_1) \land \pE(x_1) \land \neg \pL(x_1)]$.
  \end{description}

  \noindent 7. Grid ``closing'' conjunctive query. ($\querysnake$)
  \begin{description}[itemsep=0em, leftmargin=*]
    \item[\desclabel{(Q)}{snk:Q}] $\exists{x}\exists{y}\exists{z}\exists{t}\, \pV(xy) \land \pH(yz) \land \pV(zt)     \land \pHnot(xt)$.
  \end{description}

  \noindent 8. The transitivity assertion concerning $\pP$. (implicit)
  \begin{description}[itemsep=0em, leftmargin=*]
    \item[\desclabel{(T)}{snk:T}] $\forall{xyz}\ [\pP(xy) \land \pP(yz)] \to \pP(xz)$.
  \end{description}

  \noindent All but the last two formul{\ae} belong to the fluted two-variable guarded fragment (with a sole transitive guard $\pP$), the second to last one is a {\CQ}, and the last one asserts transitivity~of~$\pP$.~\myqed
\end{definition}

\subsection{Inductive proof for encoding grid lemma.}\label{app:inductive_proof_for_encoding_grid_lemma}


\begin{proof}
  Let $\homog \deff \homof \cup \{ ((\rmN', \rmM'), \ela) \}$, for an $\ela \in A$ below.
  \begin{itemize}[itemsep=0em, leftmargin=*]
    \item Suppose $\rmN$ is odd. There are two possible cases:
    \begin{itemize}[itemsep=0em, leftmargin=*]
      \item $(\rmN', \rmM') = (\rmN{+}1, \rmN{+}1)$.
        As $\homof(\rmN, \rmM) \not\in \pL^{\strA}$ we can take any $\ela \in A$ with $\strA \models \mathrm{\ref{snk:GH1}}[\homof(\rmN, \rmM), \ela]$.~Thus:
    \end{itemize}
    $\CogOOne{\colon} \strA \models \pP[\homog(\rmN,\rmM),\homog(\rmN',\rmM')] {\land} \pH[\homog(\rmN,\rmM),\homog(\rmN',\rmM')].$
    \begin{itemize}[itemsep=0em, leftmargin=*]
      \item $(\rmN', \rmM') = (\rmN, \rmM{+}1)$.
      By design, $(\rmN, \rmM)$ does not satisfy both $\pT_0$ and $\pE$ in $\strSnk$, and hence neither $\homof(\rmN, \rmM)$ in $\strA$ (due to the semi-strong homomorphicity of $\homof$).
      Hence, take any $\ela \in A$ with $\strA \models \mathrm{\ref{snk:GVO1}}[\homof(\rmN, \rmM), \ela]$.~Thus:
    \end{itemize}    
    $\CogOTwo{\colon} \strA \models \pP[\homog(\rmN,\rmM),\homog(\rmN',\rmM')] {\land} \pV[\homog(\rmN,\rmM),\homog(\rmN',\rmM')].$
    \item Suppose $\rmN$ is even. There are two possible cases:
    \begin{itemize}[itemsep=0em, leftmargin=*]
      \item $(\rmN', \rmM') = (\rmN{+}1, 0)$.
        As $\homof(\rmN, \rmM) \not\in \pL^{\strA}$ we can take any $\ela \in A$ witnessing $\strA \models \mathrm{\ref{snk:GH1}}[\homof(\rmN, \rmM), \ela]$. Thus:
      \end{itemize}
      $\CogEOne{\colon} \strA \models \pP[\homog(\rmN,\rmM),\homog(\rmN',\rmM')] {\land} \pH[\homog(\rmN,\rmM),\homog(\rmN',\rmM')].$
      \begin{itemize}[itemsep=0em, leftmargin=*]
      \item $(\rmN', \rmM') = (\rmN, \rmM{-}1)$.
      By design, $(\rmN, \rmM)$ satisfies $\pE$ in $\strSnk$ but dissatisfies $\pB$. 
      Thus, due to the semi-strong homomorphicity of $\homof$, the same holds for $\homof(\rmN, \rmM)$ in $\strA$.
      We take an $\ela \in A$ with $\strA \models \mathrm{\ref{snk:GVE1}}[\homof(\rmN, \rmM), \ela]$.~Thus:
    \end{itemize}
    $\CogETwo{\colon} \strA \models \pP[\homog(\rmN,\rmM),\homog(\rmN',\rmM')] {\land} \pV[\homog(\rmN,\rmM),\homog(\rmN',\rmM')].$

  \end{itemize}

  \noindent By $\homof(n,m) {=} \homog(n,m)$ for all $(n,m) \lesssnk (\rmN',\rmM')$ and~by~the (semi-strong) homomorphicity of $\homof$ by analysing the definition of a snake, we infer the following properties~for~all $(n,m), (n',m'), (n{+}1, m), (n,m{-}1), (n,m{+}1)$~$\lesssnk$~$(\rmN', \rmM')$.
  \begin{itemize}[itemsep=0em, leftmargin=*]
    \item $(\heartsuit){\colon}$ $\strA \models \pE[\homog(n, m)]$ if and only if $n$ is even;
    \item $(\clubsuit){\colon}$ $\strA \models \pB[\homog(n, m)]$ if and only if $m=0$;
    \item $(\spadesuit){\colon}$ $\strA \models \pP[\homog(n, m), \homog(n', m')]$ if $(n,m) \lesssnk (n',m')$;
    \item $(\diamondsuit){\colon}$ $\strA \models \pH[\homog(n, m), \homog(n{+}1, m)]$;
    \item $(\triangle){\colon}$ $\strA \models \pV[\homog(n, m), \homog(n, m{+}1)]$ if $n$ is odd;
    \item $(\nabla){\colon}$ $\strA \models \pV[\homog(n, m), \homog(n, m{-}1)]$ if $n$ is even;
    \item $(\oplus_i){\colon}$ $\strA \models \pT_i[\homog(n, m)]$ whenever $i \in \N$ is the minimal non-negative integer satisfying $(n, m{+}i{+}1) \not\in \snk_{\infty}$.
  \end{itemize}
  \noindent To see that $\homog$ is as desired we show the upcoming claims.~\qedhere
  
\end{proof}

When presenting the proofs of the upcoming claims we employ the naming conventions from the proof of \Cref{lemma:injecting-snake-main}. 
We start with a pair of twin claims establishing $\pH$-connections between elements from snakes that are not successors w.r.t.~$\lesssnk$.

\begin{claim}\label{claim:injecting-snake-missing-H}
Consider $\strA$, $\homog$, and $(\rmN', \rmM')$ from the proof of \Cref{lemma:injecting-snake-main} and suppose that $\rmN'$ is positive and even.\\
Then $\strA \models \pH[\homog(\rmN'{-}1, \rmM'), \homog(\rmN', \rmM')]$.~\myqed
\end{claim}

\begin{proof}
  As $\rmN'$ is positive and even, the elements $(\rmN'{-}1,\rmM'{+}1)$, $(\rmN',\rmM'{+}1)$ and $(\rmN'{-}1, \rmM')$ belong to a snake and are smaller (w.r.t. $\lesssnk$) than $(\rmN', \rmM')$.
  In particular, $(\rmN, \rmM) = (\rmN', \rmM'{+}1)$.
    \vspace{-1.25em}
    \begin{figure}[H]
      \begin{center}
        \begin{tikzpicture}[transform shape]
        \draw (0.00,3.00) node[] (Vx) {$\homog(\rmN'{-}1,\rmM'{+}1)$};
        \draw (5.725,3.00) node[] (Vt) {$\homog(\rmN',\rmM'{+}1) = \homog(\rmN, \rmM)$};
        \draw (0.00,0.00) node[] (Vy) {$\homog(\rmN'{-}1, \rmM')$};
        \draw (5.725,0.00) node[] (Vz) {$\homog(\rmN', \rmM')$};

        \path[->] (Vy) edge [blue, -stealth] node[fill=white] {{$\pV \text{\, by \,} (\triangle)$}} (Vx);
        \path[->] (Vt) edge [blue, -stealth] node[fill=white] {{$\pV, \pP \text{\, by \,} \CogETwo$}} (Vz);
        \path[->] (Vy) edge [red, -stealth, densely dashdotted] node[fill=white] {{???}} (Vz);
        \path[->] (Vx) edge [red, -stealth] node[fill=white] {{$\pH \text{\, by \,} (\diamondsuit)$}} (Vt);
        \path[->] (Vy) edge [black] node[fill=white] {{$\pP \text{\, by \,} (\spadesuit)$}} (Vt);

        \draw (0.00,3.50) node[] (Vx) {$y$};
        \draw (5.725,3.50) node[] (Vt) {$z$};
        \draw (0.00,-0.50) node[] (Vy) {$x$};
        \draw (5.725,-0.50) node[] (Vz) {$t$};
        \end{tikzpicture}%
      \end{center}
    \end{figure}
    \vspace{-2em}
    Applying $(\triangle)$, $(\diamondsuit)$, $(\spadesuit)$, and $\CogETwo$ we see that the relations between the elements hold as illustrated above. 
    By the transitivity of $\pP$ we conclude $\strA \models \pP[\homog(\rmN'{-}1, \rmM'), \homog(\rmN', \rmM')]$.
    Hence, by \ref{snk:C2} we either have $\strA \models \pH[\homog(\rmN'{-}1, \rmM'), \homog(\rmN', \rmM')]$ or $\strA \models \pHnot[\homog(\rmN'{-}1, \rmM'), \homog(\rmN', \rmM')]$. 
    The second option, however, results in a match $x \mapsto \homog(\rmN'{-}1, \rmM')$, $y \mapsto \homog(\rmN'{-}1, \rmM'{+}1)$, $z \mapsto \homog(\rmN, \rmM)$, $t \mapsto \homog(\rmN', \rmM')$ to the query $\querysnake$, as depicted above. 
    Thus, $\strA \models \pH[\homog(\rmN'{-}1, \rmM'), \homog(\rmN', \rmM')]$, as desired.\qedhere 
\end{proof}

\begin{claim}\label{claim:injecting-snake-missing-H-bis}
Consider $\strA$, $\homog$, and $(\rmN', \rmM')$ from the proof of \Cref{lemma:injecting-snake-main} and suppose that $\rmN' \geq 3$ and that $\rmN'$ is odd.\\
Then $\strA \models \pH[\homog(\rmN'{-}1, \rmM'), \homog(\rmN', \rmM')]$.~\myqed
\end{claim}

\begin{proof}
  As $\rmN' \geq 3$ and $\rmN'$ is odd, the elements $(\rmN'{-}1,\rmM')$, $(\rmN'{-}1, \rmM'{-}1)$ and $(\rmN', \rmM'{-}1)$ are smaller (w.r.t. $\lesssnk$) than $(\rmN', \rmM')$.
  Moreover, $(\rmN, \rmM) = (\rmN', \rmM'{-}1)$.
    \vspace{-1.25em}
    \begin{figure}[H]
      \begin{center}
        \begin{tikzpicture}[transform shape]
        \draw (0.00,3.00) node[] (Vx) {$\homog(\rmN'{-}1,\rmM')$};
        \draw (5.725,3.00) node[] (Vt) {$\homog(\rmN',\rmM')$};
        \draw (0.00,0.00) node[] (Vy) {$\homog(\rmN'{-}1, \rmM'{-}1)$};
        \draw (5.725,0.00) node[] (Vz) {$\homog(\rmN', \rmM'{-}1) = \homog(\rmN, \rmM)$};

        \path[->] (Vx) edge [blue, -stealth] node[fill=white] {{$\pV \text{\, by \,} (\nabla)$}} (Vy);
        \path[->] (Vz) edge [blue, -stealth] node[fill=white] {{$\pV, \pP \text{\, by \,} \CogOTwo$}} (Vt);
        \path[->] (Vx) edge [red, -stealth, densely dashdotted] node[fill=white] {{???}} (Vt);
        \path[->] (Vy) edge [red, -stealth] node[fill=white] {{$\pH \text{\, by \,} (\diamondsuit)$}} (Vz);
        \path[->] (Vx) edge [black] node[fill=white] {{$\pP \text{\, by \,} (\spadesuit)$}} (Vz);

        \draw (0.00,3.50) node[] (Vx) {$x$};
        \draw (5.725,3.50) node[] (Vt) {$t$};
        \draw (0.00,-0.50) node[] (Vy) {$y$};
        \draw (5.725,-0.50) node[] (Vz) {$z$};
        \end{tikzpicture}%
      \end{center}
    \end{figure}
    \vspace{-2em}
    Applying $(\nabla)$, $(\diamondsuit)$, $(\spadesuit)$, and $\CogOTwo$ we see that the relations between the elements hold as illustrated above. 
    By the transitivity of $\pP$ we conclude $\strA \models \pP[\homog(\rmN'{-}1, \rmM'), \homog(\rmN', \rmM')]$.
    Hence, by \ref{snk:C2} we either have $\strA \models \pH[\homog(\rmN'{-}1, \rmM'), \homog(\rmN', \rmM')]$ or $\strA \models \pHnot[\homog(\rmN'{-}1, \rmM'), \homog(\rmN', \rmM')]$. 
    However, the second option results in a match $x \mapsto \homog(\rmN'{-}1, \rmM')$, $y \mapsto \homog(\rmN'{-}1, \rmM'{-}1)$, $z \mapsto \homog(\rmN, \rmM)$, $t \mapsto \homog(\rmN', \rmM')$ to the query $\querysnake$, as depicted above. 
    Thus $\strA \models \pH[\homog(\rmN'{-}1, \rmM'), \homog(\rmN', \rmM')]$ as desired.\qedhere 
\end{proof}

We are now ready to show that $\homog$ is a (semi-strong) homomorphism. 
The first part of the proof concerns unary relations. 
\begin{claim}\label{claim:injecting-snake-unary}
The map $\homog$ from the proof of \Cref{lemma:injecting-snake-main} preserves satisfaction and dissatisfaction of unary atoms made from~$\sigma$.~\myqed
\end{claim}

\begin{proof}
For brevity, let $\strSnk$ denote $\strSnk_{(\rmN', \rmM')}$.
Consider any pair $(n,m)$ from $\strSnk$.
For all unary predicates $\alpha \in \sigma$ we prove that $\strSnk \models \alpha[(n,m)]$ if and only if $\strA \models \alpha[\homog(n,m)]$.
If $(n,m) \lesssnk (\rmN', \rmM')$ we have $\homof(n,m) = \homog(n,m)$, so we are done by the (semi-strong) homomorphicity of $\homof$. 
Consider any unary $\alpha \in \sigma$:
\begin{itemize}[itemsep=0em, leftmargin=*]

\item Let $\alpha = \pE$ and suppose $\strSnk \models \alpha[(\rmN', \rmM')]$.
  By design of $\strSnk$ we have that $\rmN'$ is even. By $\lesssnk$ there are two possible cases:
  \begin{itemize}
    \item $(\rmN', \rmM') = (\rmN, \rmM{-}1)$. Apply $(\heartsuit)$, $\CogETwo$ and \ref{snk:PE4}. 
    \item $(\rmN', \rmM') = (\rmN{+}1, \rmN{+}1)$. Apply $(\heartsuit)$, $\CogOOne$ and \ref{snk:PO1}. 
  \end{itemize}
\item Let $\alpha = \pE$ and suppose $\strSnk \not\models \alpha[(\rmN', \rmM')]$.
  By design of $\strSnk$ we have that $\rmN'$ is odd. By $\lesssnk$ there are two possible cases:
  \begin{itemize}
    \item $(\rmN', \rmM') = (\rmN, \rmM{+}1)$. Apply $(\heartsuit)$, $\CogOTwo$ and \ref{snk:PO3}. 
    \item $(\rmN', \rmM') = (\rmN{+}1, 0)$. Apply $(\heartsuit)$, $\CogEOne$ and \ref{snk:PE1}. 
  \end{itemize}
\item Let $\alpha = \pB$ and suppose $\strSnk \models \alpha[(\rmN', \rmM')]$.
  By design of $\strSnk$ we have that $\rmM' = 0$. By $\lesssnk$ there are two possible cases:
  \begin{itemize}
    \item $\rmN'$ is odd. Then $\rmM = 0$ and $(\rmN', \rmM') = (\rmN{+}1, 0)$.
    Apply both $(\heartsuit)$ and $(\clubsuit)$ to $(\rmN, \rmM)$, then invoke $\CogOOne$ and \ref{snk:PB1}.
    \item $\rmN'$ is even. Then $(\rmN, \rmM) = (\rmN, 1)$ and $(\rmN', \rmM') = (\rmN, 0)$.
    Then $\pH[\homog(\rmN{-}1, 0), \homog(\rmN, 0)]$ holds by \Cref{claim:injecting-snake-missing-H}.
    By $(\clubsuit)$ applied to $(\rmN{-}1, 0)$ and \ref{snk:PB1} we are done.
  \end{itemize}
\item Let $\alpha = \pB$ and suppose $\strSnk \not\models \alpha[(\rmN', \rmM')]$. 
  By design of $\strSnk$ we have that $\rmM' > 0$. By $\lesssnk$ there are two possible cases:
  \begin{itemize}
    \item $\rmN$ is odd. 
    If $(\rmN', \rmM') = (\rmN{+}1, \rmN{+}1)$ we apply $(\heartsuit)$, $\CogOOne$ and \ref{snk:PB2}.
    Otherwise we apply $(\heartsuit)$, $\CogOTwo$ and \ref{snk:PO7}.
    \item $\rmN$ is even. Then $(\rmN', \rmM') = (\rmN, \rmM{-}1)$ and $(\rmM{-}1) > 0$.
    Thus $\pH[\homog(\rmN{-}1, \rmM{-}1), \homog(\rmN, \rmM{-}1)]$ holds by \Cref{claim:injecting-snake-missing-H}.
    By $(\clubsuit)$ applied to $(\rmN{-}1, \rmM{-}1)$ and \ref{snk:PB2} we are done.
  \end{itemize}
\item Let $\alpha = \pT_0$ and suppose $\strSnk \models \alpha[(\rmN', \rmM')]$. 
  By design of~$\strSnk$, we see that $\rmN$ is odd and one of the options holds:
  \begin{itemize}
    \item  $(\rmN', \rmM') = (\rmN{+}1, \rmN{+}1)$. 
    Simply apply $(\heartsuit)$ and $(\oplus_0)$ to $(\rmN, \rmM)$, then invoke $\CogOOne$, and \ref{snk:PO2.i} (for $i{=}0$).
    \item $(\rmN', \rmM') = (\rmN, \rmN{+}1)$.
    Simply apply $(\heartsuit)$ and $(\oplus_1)$ to $(\rmN, \rmM)$, then invoke $\CogOTwo$, and \ref{snk:PO5}.
  \end{itemize}
\item Let $\alpha = \pT_0$ and suppose $\strSnk \not\models \alpha[(\rmN', \rmM')]$. 
  By design of~$\strSnk$, we have to consider the following cases:
  \begin{itemize}
    \item $\rmN$ is odd, then $(\rmN', \rmM') = (\rmN, \rmM{+}1)$, and $\rmM{+}1 \leq \rmN$.\\ 
    If $\rmM = \rmN{-}1$ we apply $(\heartsuit)$ and $(\oplus_2)$ to $(\rmN, \rmM)$, and then use both $\CogOTwo$ and \ref{snk:PO4} to conclude $\strA \models \pT_1[(\rmN',\rmM')]$. By \ref{snk:C1} we are done.
    Otherwise, we apply $(\heartsuit)$ and $(\oplus_i)$ for all $i \in \{ 0, 1, 2 \}$ to $(\rmN, \rmM)$, concluding that $\homog(\rmN,\rmM)$ satisfies none of $\pT_i$. With $\CogOTwo$ and \ref{snk:PO6} we are done.

    \item $\rmN$ is even. One option is that $(\rmN', \rmM') = (\rmN, \rmM{-}1)$, and $\rmM{-}1 < \rmN$.
    By $(\heartsuit)$ we have $\strA \models \pE[\homog(\rmN,\rmM)]$.~Observe:
    \begin{itemize}
    \item If $(\rmN', \rmM') {=} (\rmN, \rmN{-}1)$, use $(\oplus_0)$, $\CogETwo$, \ref{snk:PE5}, \ref{snk:C1}.
    \item If $(\rmN', \rmM') {=} (\rmN, \rmN{-}2)$, use $(\oplus_1)$, $\CogETwo$, \ref{snk:PE6}, \ref{snk:C1}.
    \item If $(\rmN', \rmM') {=} (\rmN, \rmN{-}3)$, use $(\oplus_2)$, $\CogETwo$, and \ref{snk:PE7}.
    \item Finally, if $\rmM' < \rmN{-}3$ we apply $(\oplus_i)$ (for all indices $i \in \{ 0, 1, 2 \}$) to $(\rmN,\rmM)$, and invoke $\CogETwo$ and \ref{snk:PE8}.
    \end{itemize}
    Otherwise, $(\rmN, \rmM) = (\rmN, 0)$ and $(\rmN', \rmM') = (\rmN{+}1, 0)$. 
    \begin{itemize}
    \item If $\rmN = 0$. We first apply both $(\oplus_0)$ and $\CogEOne$ to $(\rmN,\rmM)$. We then invoke \ref{snk:PE2} and \ref{snk:C1}.
    \item If $\rmN > 0$, apply $(\oplus_0)$, $\CogEOne$ to $(\rmN,\rmM)$. Invoke \ref{snk:PE3}.
    \end{itemize}

  \end{itemize}
\item Let $\alpha = \pT_1$ and suppose $\strSnk \models \alpha[(\rmN', \rmM')]$. 
  By $\lesssnk$ there are two possible cases:
  \begin{itemize}
    \item $\rmN'$ is odd, $(\rmN, \rmM) = (\rmN, \rmN{-}1)$, and $(\rmN', \rmM') = (\rmN, \rmN)$.
    First, apply $(\oplus_2)$ and $(\heartsuit)$ to $(\rmN, \rmM)$. We then conclude by~$\CogOTwo$ and \ref{snk:PO4}.
    \item $\rmN'$ is even, $(\rmN, \rmM) = (\rmN, \rmN)$, and $(\rmN', \rmM') = (\rmN, \rmN{-}1)$.
    First, apply $(\oplus_0)$ and $(\heartsuit)$ to $(\rmN, \rmM)$. We then conclude by~$\CogETwo$ and \ref{snk:PE5}.
  \end{itemize}
\item Let $\alpha = \pT_1$ and suppose $\strSnk \not\models \alpha[(\rmN', \rmM')]$. 
  By $\lesssnk$ there one of the following cases holds:
  \begin{itemize}
    \item $\rmN'$ is even, $(\rmN, \rmM) = (\rmN, \rmN{-}1)$, $(\rmN', \rmM') = (\rmN, \rmN{-}2)$.\\
    First, apply $(\oplus_1)$ and $(\heartsuit)$ to $(\rmN, \rmM)$. With $\CogETwo$ and \ref{snk:PE6} we have $\strA \models \pT_2[\homog(\rmN', \rmM')]$. By \ref{snk:C1} we are done.
    \item $\rmN'$ is even, $(\rmN, \rmM) = (\rmN, \rmN{-}2)$, $(\rmN', \rmM') = (\rmN, \rmN{-}3)$.\\
    Apply $(\oplus_2)$, $(\heartsuit)$ to $(\rmN, \rmM)$. Then use $\CogETwo$ and \ref{snk:PE7}.
    \item $\rmN'$ is even, $(\rmN', \rmM') = (\rmN, \rmN{-}1)$, $\rmM < \rmN{-}2$.\\
    Apply $(\heartsuit)$ and $(\oplus_i)$ for all $i \in \{ 0, 1, 2 \}$ to $(\rmN, \rmM)$. 
    We~can now conclude with $\CogETwo$ and \ref{snk:PE8}.
    \item $\rmN'$ is even, $(\rmN, \rmM) {=} (\rmN, \rmN{+}1)$, $(\rmN', \rmM') {=} (\rmN{+}1, \rmN{+}1)$.\\
    Note that $\strSnk \models \pT_0[(\rmN',\rmM')]$ by design of $\strSnk$.
    We already know that $\strA \models \pT_0[\homog(\rmN',\rmM')]$. Thus by \ref{snk:C1} we are done.
    \item $\rmN'$ is odd, $(\rmN, \rmM) {=} (\rmN, \rmN)$, $(\rmN', \rmM') {=} (\rmN, \rmN{+}1)$.\\
    First, apply $(\oplus_1)$ and $(\heartsuit)$ to $(\rmN, \rmM)$. With $\CogOTwo$ and \ref{snk:PO5} we have $\strA \models \pT_0[\homog(\rmN', \rmM')]$. By \ref{snk:C1} we are done.
    \item $\rmN'$ is odd, $(\rmN', \rmM') {=} (\rmN, \rmM{+}1)$, and $\rmM < \rmN$.\\
    Apply $(\heartsuit)$ and $(\oplus_i)$ for all $i \in \{ 0, 1, 2 \}$ to $(\rmN, \rmM)$. 
    We~can now conclude with $\CogOTwo$ and \ref{snk:PO6}.
    \item $\rmN'$ is odd, $(\rmN, \rmM) {=} (\rmN, 0)$, $(\rmN', \rmM') {=} (\rmN{+}1, 0)$.\\
    If $\rmN = 0$ then we use $(\oplus_0)$ and $(\heartsuit)$ to $(\rmN, \rmM)$. By~$\CogEOne$ and \ref{snk:PE2} we have $\strA \models \pT_2[\homog(\rmN', \rmM')]$. By \ref{snk:C1} we are done.
    Otherwise $\rmN > 0$. Apply $(\heartsuit)$ and $(\oplus_i)$ for all $i \in \{ 0, 1, 2 \}$ to $(\rmN, \rmM)$. 
    By $\CogEOne$ and \ref{snk:PE8} we are done.
  \end{itemize}
\item Let $\alpha = \pT_2$ and suppose $\strSnk \models \alpha[(\rmN', \rmM')]$. 
  By $\lesssnk$ there one of the following cases holds:
  \begin{itemize}
    \item $\rmN' = 1$. 
    Then $(\rmN, \rmM) = (0,0)$ and $(\rmN', \rmM') = (1,0)$.
    Firstly, we apply $(\heartsuit)$ and $(\oplus_0)$ to infer $\strA \models \pT_0[\homog(0,0)]$. 
    Next, we conclude by $\CogEOne$ and \ref{snk:PE2}.
    \item $\rmN'$ is odd and $\rmN' \geq 3$.
    By $(\heartsuit)$ and $(\oplus_0)$ we have $\strA \models \pT_0[\homog(\rmN'{-}1,\rmM')]$. 
    Next $\strA \models \pH[\homog(\rmN'{-}1,\rmM'), \homog(\rmN',\rmM')]$ by \Cref{claim:injecting-snake-missing-H-bis}. 
    By~\ref{snk:PE2}~we~are~done.
    \item $\rmN'$ is even. Then $(\rmN', \rmM') = (\rmN, \rmM{-}1)$.
    We apply $(\heartsuit)$ and $(\oplus_0)$ to $(\rmN, \rmM)$. We finish by $\CogETwo$ and \ref{snk:PE6}.

  \end{itemize}
\item Let $\alpha = \pT_2$ and suppose $\strSnk \not\models \alpha[(\rmN', \rmM')]$. 
  If $(\rmN', \rmM')$ satisfies either $\pT_0$ or $\pT_1$ in $\strSnk$ then by the previous developments we know that $\homog(\rmN', \rmM')$ satisfies $\pT_0$ or $\pT_1$ in $\strA$. Hence, by \ref{snk:C1} we conclude $\strA \not\models \pT_2[(\rmN', \rmM')]$. 
  Otherwise, by design of $\strSnk$, it suffices to consider the following cases:
  \begin{itemize}
    \item $\rmN'$ is even, $(\rmN', \rmM') {=} (\rmN, \rmM{-}1)$, $\rmM' < \rmN{-}1$.\\
    If $\rmM=\rmN{-}2$ then we apply $(\heartsuit)$ and $(\oplus_2)$ to $(\rmN, \rmM)$. By $\CogETwo$ and \ref{snk:PE7} we have $\strA \not\models \pT_2[\homog(\rmN',\rmM')]$ as desired.
    Otherwise, we apply  $(\heartsuit)$ and $(\oplus_i)$ for all $i \in \{ 0, 1, 2 \}$ to $(\rmN, \rmM)$. Now it is sufficient to employ $\CogETwo$ and \ref{snk:PE8}.

    \item $\rmN'$ is odd. Then $\rmN' \geq 3$ and $\rmM' < \rmN'{-}2$.
    By \Cref{claim:injecting-snake-missing-H-bis} we have $\strA \models \pH[\homog(\rmN'{-}1, \rmM'), \homog(\rmN', \rmM')]$.
    Applying $(\heartsuit)$ and $(\oplus_0)$ to $(\rmN'{-}1, \rmM')$ we infer that $\homog(\rmN'{-}1, \rmM')$ in $\strA$ satisfies $\pE$ and dissatisfies $\pT_0$.
    By \ref{snk:PE3} we are done.\qedhere
  \end{itemize}
\end{itemize}

\end{proof}

\noindent The second part of the proof concerns binary relations. 
\begin{claim}\label{claim:injecting-snake-binary}
The map $\homog$ from the proof of \Cref{lemma:injecting-snake-main} preserves satisfaction of binary atoms made from~$\sigma$.~\myqed
\end{claim}

\begin{proof}
Let $\strSnk$ denote $\strSnk_{(\rmN', \rmM')}$.
By the design of $\strSnk$ and the fact that $\homog$ restricted to $\strSnk_{(\rmN, \rmM)}$ is a homomorphism, it suffices to consider any pair $(n,m)$ from $\strSnk$ and any binary predicate~$\beta \in \sigma$ for which $\strSnk \models \beta[(n,m), (\rmN',\rmM')]$, and prove $\strA \models \beta[\homog(n,m), \homog(\rmN',\rmM')]$.
Suppose $(\homog(n,m),\homog(\rmN',\rmM'))$ satisfies some $\beta \in \sigma$ in $\strA$.
If $(n,m) = (\rmN, \rmM)$ then we are done by the construction of $\homog$. Hence, assume $(n,m) \lesssnk (\rmN',\rmM')$.
By the design of $\strSnk$ we have that $\beta$ is either $\pP$ or $\pH$. 
If $\beta = \pP$ we have $\strA \models \beta[\homog(n,m),\homog(\rmN,\rmM)]$ (by homomorphicity of~$\homog$) and $\strA \models \beta[\homog(\rmN,\rmM),\homog(\rmN',\rmM')]$ (by the construction of~$\homog$).
By transitivity of $\pP$, we are done.
Otherwise $\beta = \pH$. 
By design of $\strSnk$, we have $(n,m) = (\rmN'{-}1,\rmM')$. 
Thus, $\strA \models \beta[\homog(n,m),\homog(\rmN',\rmM')]$ follows from \Cref{claim:injecting-snake-missing-H} or \Cref{claim:injecting-snake-missing-H-bis}, depending on the parity of $\rmN'$ (the extra conditions mentioned in the statements of the claims are fulfilled due to $(n,m) \neq (\rmN, \rmM)$).
\end{proof}

\noindent The third and the last part of the proof concerns injectivity.
\begin{claim}\label{claim:injecting-snake-injectivity}
The map $\homog$ from the proof of \Cref{lemma:injecting-snake-main} is injective. Hence, $\homog$ is a semi-strong homomorphism.~\myqed
\end{claim}

\begin{proof}
  \emph{Ad absurdum}, suppose $\homog$ is not injective. 
  By injectivity of $\homof$, there is $(n,m) \lesssnk (\rmN, \rmM')$ with $\homog(\rmN', \rmM') = \homog(n,m)$.
  Consider cases based on the construction of $(\rmN',\rmM')$ w.r.t. $\lesssnk$:
  \begin{itemize}[itemsep=0em, leftmargin=*]
    \item $\rmN$ is even, $(\rmN, \rmM) = (\rmN, 0)$, $(\rmN',\rmM') = (\rmN{+}1, 0)$.\\
      By our previous developments we have $\strA \models \neg\pE[\homog(\rmN{+}1, 0)]$. Thus by $(\heartsuit)$ we know that $n$ is odd. 
      Similarly, we already know that $\strA \models \pB[\homog(\rmN{+}1, 0)]$, thus by $(\clubsuit)$ we have~$m{=}0$.
      By~induction, we prove that $(\homog(\rmN,i), \homog(n, i))$ satisfies both~$\pP$ and $\pH$ in~$\strA$, for all indices $0 \leq i \leq n{+}1$.  Indeed:
      \vspace{-1em}
      \begin{figure}[H]
        \begin{center}
          \begin{tikzpicture}[transform shape]
          \draw (0.00,3.00) node[] (Vx) {$\homog(\rmN,i{+}1)$};
          \draw (5.725,3.00) node[] (Vt) {$\homog(n,i{+}1)$};
          \draw (0.00,0.00) node[] (Vy) {$\homog(\rmN,i)$};
          \draw (5.725,0.00) node[] (Vz) {$\homog(n, i)$};

          \path[->] (Vx) edge [blue, -stealth] node[fill=white] {{$\pV, \pP \text{\, by \,} (\nabla), (\spadesuit)$}} (Vy);
          \path[->] (Vz) edge [blue, -stealth] node[fill=white] {{$\pV, \pP \text{\, by \,} (\triangle), (\spadesuit)$}} (Vt);
          \path[->] (Vx) edge [red, -stealth, densely dashdotted] node[fill=white] {{???}} (Vt);
          \path[->] (Vy) edge [red, -stealth] node[fill=white] {{$\pH, \pP \text{\, by assumption}$}} (Vz);
          \path[->] (Vx) edge [black, bend right] node[fill=white] {{$\pP \text{\, by transitivity}$}} (Vt);

          \draw (0.00,3.50) node[] (Vx) {$x$};
          \draw (5.725,3.50) node[] (Vt) {$t$};
          \draw (0.00,-0.50) node[] (Vy) {$y$};
          \draw (5.725,-0.50) node[] (Vz) {$z$};
          \end{tikzpicture}%
        \end{center}
      \end{figure}
      \vspace{-2em}
      Observe that $\strA \models \pH[\homog(\rmN, i), \homog(n,i)]$. Indeed, if $i = 0$ then it follows by $\homog(n, 0) = \homog(\rmN{+}1,0)$ and $\CogEOne$. Otherwise, this follows from the inductive assumption.
      Then, applying $(\triangle)$, $(\nabla)$, $(\spadesuit)$ and the transitivity of $\pP$, we see that the relations between the elements hold as illustrated above. 
      Applying \ref{snk:C2}, we either have $\strA \models \pH[\homog(\rmN,i{+}1), \homog(n, i{+}1)]$ or $\strA \models \pHnot[\homog(\rmN,i{+}1), \homog(n, i{+}1)]$. 
      The second option is not possible as it would result in (a depicted) match for the query $\querysnake$.
      This finishes the induction. We thus conclude $\strA \models \pH[\homog(\rmN, n{+}1), \homog(n,n{+}1)]$.
      With $(\oplus_0)$ we infer $\strA \models \pT_0[\homog(n,n{+}1)]$ and $\strA \not\models \pT_0[\homog(\rmN,n{+}1)]$.
      Applying \ref{snk:PE3} to $\homog(n,n{+}1)$ we arrive at a contradiction. \text{\faBolt}

    \item $\rmN$ is odd, $(\rmN, \rmM) = (\rmN, \rmN{+}1)$, $(\rmN',\rmM') = (\rmN{+}1, \rmN{+}1)$.
      By our previous developments we have $\strA \models \pE[\homog(\rmN{+}1, \rmN{+}1)]$. 
      Note that $n$ is even by $(\heartsuit)$.
      Analogously, we proved $\strA \models \pT_0[\homog(\rmN{+}1, \rmN{+}1)]$, so $m{=}n$ by $(\clubsuit)$.
      By~induction, we prove that $(\homog(\rmN,\rmN{+}1{-}i), \homog(n, n{-}i))$ satisfies both~$\pP$ and $\pH$ in~$\strA$, for all indices $0 \leq i \leq n{+}1$. 
      We illustrate the proof with:
      \vspace{-1em}
      \begin{figure}[H]
        \begin{center}
          \begin{tikzpicture}[transform shape]
          \draw (0.00,3.00) node[] (Vx) {$\homog(\rmN, \rmN{+}1{-}i)$};
          \draw (5.725,3.00) node[] (Vt) {$\homog(n, n{-}i)$};
          \draw (0.00,0.00) node[] (Vy) {$\homog(\rmN, \rmN{-}i)$};
          \draw (5.725,0.00) node[] (Vz) {$\homog(n, n{-}i{-}1)$};

          \path[->] (Vy) edge [blue, -stealth] node[fill=white] {{$\pV, \pP \text{\, by \,} (\triangle), (\spadesuit)$}} (Vx);
          \path[->] (Vt) edge [blue, -stealth] node[fill=white] {{$\pV, \pP \text{\, by \,} (\nabla), (\spadesuit)$}} (Vz);
          \path[->] (Vy) edge [red, -stealth, densely dashdotted] node[fill=white] {{???}} (Vz);
          \path[->] (Vx) edge [red, -stealth] node[fill=white] {{$\pH, \pP \text{\, by assumption}$}} (Vt);
          \path[->] (Vy) edge [black, bend left] node[fill=white] {{$\pP \text{\, by transitivity}$}} (Vz);

          \draw (0.00,3.50) node[] (Vx) {$y$};
          \draw (5.725,3.50) node[] (Vt) {$z$};
          \draw (0.00,-0.50) node[] (Vy) {$x$};
          \draw (5.725,-0.50) node[] (Vz) {$t$};
          \end{tikzpicture}%
        \end{center}
      \end{figure}
      \vspace{-1.5em}
      Note that $\strA \models \pH[\homog(\rmN, \rmN{+}1{-}i), \homog(n,n{-}i)]$. Indeed, if~$i {=} 0$ we use $\homog(n, n) = \homog(\rmN{+}1, \rmN{+}1)$ and $\CogOOne$. Otherwise, this follows from the inductive assumption.
      Then, applying $(\triangle)$, $(\nabla)$, $(\spadesuit)$ and the transitivity of $\pP$, we see that the relations between the elements hold as illustrated above. 
      By~\ref{snk:C2}, we either have $\strA \models \pH[\homog(\rmN,\rmN{-}i), \homog(n, n{-}i{-}1)]$ or $\strA \models \pHnot[\homog(\rmN,\rmN{-}i), \homog(n, n{-}i{-}1)]$. 
      The second option is not possible as it would result in (a depicted) match for the query $\querysnake$.
      Hence, $\strA \models \pH[\homog(\rmN, \rmN{-}n) \homog(n,0)]$ by induction.
      By $(\clubsuit)$ we have $\strA \models \pB[\homog(n,0)]$ and $\strA \not\models \pB[\homog(\rmN,\rmN{-}n)]$.
      Applying \ref{snk:PB1} to $\homog(n,0)$ we arrive at a contradiction. \text{\faBolt}
      
      \item $\rmN$ is even and $(\rmN',\rmM') = (\rmN, \rmM{-}1)$.\\
      We first resolve the case of $(n,m) = (\rmN, \rmM{+}k)$ for some non-negative $k$.
      We show by induction that for all indices $0 \leq i \leq \rmN{-}\rmM{-}k$ the pair $(\homog(\rmN{-}1,\rmM{-}1{+}i), \homog(\rmN, \rmM{+}k{+}i))$ satisfies $\pH$ in~$\strA$.
      The inductive step is depicted below. 
      \vspace{-1.25em}
      \begin{figure}[H]
        \begin{center}
          \begin{tikzpicture}[transform shape]
          \draw (0.00,3.00) node[] (Vx) {$\homog(\rmN{-}1,\rmM{+}i)$};
          \draw (5.85,3.00) node[] (Vt) {$\homog(\rmN, \rmM{+}k{+}i{+}1)$};
          \draw (0.00,0.00) node[] (Vy) {$\homog(\rmN{-}1,\rmM{-}1{+}i)$};
          \draw (5.85,0.00) node[] (Vz) {$\homog(\rmN, \rmM{+}k{+}i)$};

          \path[->] (Vy) edge [blue, -stealth] node[fill=white] {{$\pV \text{\, by \,} (\triangle)$}} (Vx);
          \path[->] (Vt) edge [blue, -stealth] node[fill=white] {{$\pV \text{\, by \,} (\nabla)$}} (Vz);
          \path[->] (Vx) edge [red, -stealth, densely dashdotted] node[fill=white] {{???}} (Vt);
          \path[->] (Vy) edge [red, -stealth] node[fill=white] {{$\pH \text{\, by assumption}$}} (Vz);
          \path[->] (Vx) edge [black, bend right] node[fill=white] {{$\pP \text{\, by \,} (\spadesuit)$}} (Vt);

          \draw (0.00,3.50) node[] (Vx) {$x$};
          \draw (5.725,3.50) node[] (Vt) {$t$};
          \draw (0.00,-0.50) node[] (Vy) {$y$};
          \draw (5.725,-0.50) node[] (Vz) {$z$};
          \end{tikzpicture}%
        \end{center}
      \end{figure}
      \vspace{-2em}
      For $i{=}0$ this follows from the homomorphicity of $\homoh$ and the equality of $(n,m)$ and $(\rmN, \rmM{-}1)$.
      For the inductive step, we apply $(\triangle)$, $(\nabla)$, $(\spadesuit)$ to see that the relations between the elements hold as illustrated above. 
      By~\ref{snk:C2}, we have that the pair $(\homog(\rmN{-}1,\rmM{-}1{+}i), \homog(\rmN, \rmM{+}k{+}i))$  in $\strA$ satisfies either $\pH$ or~$\pHnot$.
      The second option is not possible as it would result in (a depicted) match for the query $\querysnake$. 
      Now, apply the induction hypothesis for $i \deff (\rmN{-}\rmM{-}k)$ to infer $\strA \models \pH[\homog(\rmN{-}1, \rmN{-}k{-}1), \homog(\rmN,\rmN)]$.
      Consider the cases:
      \begin{itemize}\itemsep0em
        \item If $k \leq 1$ we apply $(\oplus_{k{+}1})$ and \ref{snk:PO2} to $(\rmN{-}1, \rmN{-}k{-}1)$ and conclude that $\strA \models \pT_{k{+}1}[\homog(\rmN,\rmN)]$. 
        This, by $(\oplus_{0})$, yields a contradiction with $\strA \models \pT_0[\homog(\rmN,\rmN)]$.
        \item Otherwise, we have $k > 1$. We invoke $(\oplus_{i})$ (for all indices $0 \leq i \leq 2$) and \ref{snk:PO8} to $(\rmN{-}1, \rmN{-}k{-}1)$ and infer $\strA \models \neg \pT_0[\homog(\rmN,\rmN)]$. 
        A contradiction with $(\oplus_{0})$.
      \end{itemize}
      Otherwise, $n < \rmN$.
      If $n = 0$ then $(n,m) = (0,0)$. 
      By $(\clubsuit)$ and $(\oplus_0)$ we see that $\homog(0,0)$ satisfies both $\pB$ and $\pT_0$ in~$\strA$. 
      By equality $\homog(\rmN',\rmM') = \homog(0,0)$ we infer $\strA \models \pH[\homog(\rmN,\rmM), \homog(0,0)]$, contradicting \ref{snk:PE9}. So assume $n > 0$.
      We already know $\strA \models \pE[\homog(\rmN, \rmM{-}1)]$, and so $n$ is even by $(\heartsuit)$.
      By induction, we show that for all indices $0 \leq i \leq \min{(\rmM{+}1, m)}$~the~pair $(\homog(\rmN{-}1, \rmM{-}1{-}i), \homog(n,m{-}i))$ satisfies both $\pH$ and $\pP$ in~$\strA$.
      The base case follows from the homomorphicity of $\homog$ together with $\homog(n,m) = \homog(\rmN, \rmM{-}1)$.
      \vspace{-1.05em}
      \begin{figure}[H]
        \begin{center}
          \begin{tikzpicture}[transform shape]
          \draw (0.00,3.0) node[] (Vx) {$\homog(\rmN{-}1, \rmM{-}1{-}i)$};
          \draw (5.725,3.0) node[] (Vt) {$\homog(n, m{-}i)$};
          \draw (0.00,0.00) node[] (Vy) {$\homog(\rmN{-}1, \rmM{-}1{-}i{-}1)$};
          \draw (5.725,0.00) node[] (Vz) {$\homog(n, m{-}i{-}1)$};

          \path[->] (Vy) edge [blue, -stealth] node[fill=white] {{$\pV, \pP \text{\, by \,} (\triangle), (\spadesuit)$}} (Vx);
          \path[->] (Vt) edge [blue, -stealth] node[fill=white] {{$\pV, \pP \text{\, by \,} (\nabla), (\spadesuit)$}} (Vz);
          \path[->] (Vy) edge [red, -stealth, densely dashdotted] node[fill=white] {{???}} (Vz);
          \path[->] (Vx) edge [red, -stealth] node[fill=white] {{$\pH, \pP \text{\, by assumption}$}} (Vt);
          \path[->] (Vy) edge [black, bend left] node[fill=white] {{$\pP \text{\, by transitivity}$}} (Vz);

          \draw (0.00,3.5) node[] (Vx) {$y$};
          \draw (5.725,3.5) node[] (Vt) {$z$};
          \draw (0.00,-0.50) node[] (Vy) {$x$};
          \draw (5.725,-0.50) node[] (Vz) {$t$};
          \end{tikzpicture}%
        \end{center}
      \end{figure}
      \vspace{-1.8em}
      The inductive step once more follows the already well-known route (the above picture hints the proof).
      Consider cases: 
      \begin{itemize}\itemsep0em
        \item $m < \rmM{+}1$. In particular, this means that $\rmM{-}1{-}m$ is positive.
        By putting $i \deff m$ in our induction, we infer that $\strA \models \pH[\homog(\rmN{-}1, \rmM{-}1{-}m), \homog(n,0)]$.
        By $(\clubsuit)$ we have that  $\strA \models \neg\pB[\homog(\rmN{-}1, \rmM{-}1{-}m)]$ and $\strA \models \pB[(\homog(n,0)]$.
        A~contradiction with \ref{snk:PB2}.

        \item $\rmM{+}1 < m$. In particular, this means that $m{-}\rmM{-}1$ is positive.
        By putting $i \deff \rmM{+}1$ in our induction, we infer that $\strA \models \pH[\homog(\rmN{-}1, 0), \homog(n,m{-}\rmM{-}1)]$.
        By $(\clubsuit)$ we have that  $\strA \models \pB[\homog(\rmN{-}1, 0)]$ and $\strA \models \neg\pB[\homog(n,m{-}\rmM{-}1)]$.
        A~contradiction with \ref{snk:PB1}.
      \end{itemize}
      Otherwise $m = \rmM{+}1$. By putting $i \deff m$ in our inductive hypothesis, we infer $\strA \models \pH[\homog(\rmN{-}1, 0), \homog(n, 0)]$. 
      We employ yet another induction and show that for all indices $0 \leq j \leq n$ the pair $(\homog(\rmN{-}1, j), \homog(n, j))$ satisfies both $\pP$ and $\pH$ in $\strA$. The inductive base is already established.
      We depict the proof~as:
      \vspace{-1.5em}
      \begin{figure}[H]
        \begin{center}
          \begin{tikzpicture}[transform shape]
          \draw (0.00,3.00) node[] (Vx) {$\homog(\rmN{-}1, j{+}1)$};
          \draw (5.85,3.00) node[] (Vt) {$\homog(n, j{+}1)$};
          \draw (0.00,0.00) node[] (Vy) {$\homog(\rmN{-}1, j)$};
          \draw (5.85,0.00) node[] (Vz) {$\homog(n, j)$};

          \path[->] (Vy) edge [blue, -stealth] node[fill=white] {{$\pV, \pH \text{\, by \,} (\triangle), (\spadesuit)$}} (Vx);
          \path[->] (Vt) edge [blue, -stealth] node[fill=white] {{$\pV, \pH \text{\, by \,} (\nabla), (\spadesuit)$}} (Vz);
          \path[->] (Vx) edge [red, -stealth, densely dashdotted] node[fill=white] {{???}} (Vt);
          \path[->] (Vy) edge [red, -stealth] node[fill=white] {{$\pH, \pP \text{\, by assumption}$}} (Vz);
          \path[->] (Vx) edge [black, bend right] node[fill=white] {{$\pP \text{\, by transitivity\,}$}} (Vt);
          \path[->] (Vz) edge [black, bend right] node[fill=white] {{$\pP \text{\, by \,} (\spadesuit) \text{\, due to \,} n < \rmN{-}1$}} (Vy);

          \draw (0.00,3.50) node[] (Vx) {$x$};
          \draw (5.725,3.50) node[] (Vt) {$t$};
          \draw (0.00,-0.50) node[] (Vy) {$y$};
          \draw (5.725,-0.50) node[] (Vz) {$z$};
          \end{tikzpicture}%
        \end{center}
      \end{figure}
      \vspace{-1.5em}
      By repeating a similar reasoning to the one presented before (see the picture for a proof hint) we conclude the induction.
      We thus have $\strA \models \pH[\homog(\rmN{-}1, n), \homog(n, n)]$.
      By $(\oplus_0)$ and the fact that $n$ is even we have $\strA \models \pT_0[\homog(n, n)]$.
      Similarly, by $(\heartsuit)$ and the fact that $\rmN$ is even, we have $\strA \models \neg\pE[\homog(\rmN{-}1, n)]$.
      Suppose that $\strA \models \neg\pT_0[\homog(\rmN{-}1, n)]$.
      Then we obtain a contradiction with either \ref{snk:PO2} or \ref{snk:PO8} and \ref{snk:C1}.
      Hence, $\strA \models \pT_0[\homog(\rmN{-}1, n)]$. 
      But $(\oplus_0)$ yields $n = \rmN$, contradicting the initial assumption $n < \rmN$.
      This concludes the third~case.

    \item $\rmN$ is odd and $(\rmN',\rmM') = (\rmN, \rmM{+}1)$.
     This case can be treated analogously to the previous ones. \qedhere 
  \end{itemize}
\end{proof}

\subsection{Proof of \Cref{lemma:injecting-snake-summary}}\label{app:lemma:injecting-snake-summary}

\begin{proof}
  To establish (A), take an $\rmN$-snake $\strSnk$ and expand it by interpreting $\pHnot$ as $\pP^{\strSnk} \setminus \pH^{\strSnk}$ and $\pL$ as $\{ (\rmN,m) \in \snkN \}$.
  Hence, based on Definitions \ref{def:snake}--\ref{def:phisnake}, it can be readily verified that the expanded $\strSnk$ indeed satisfies~$\phisnake$. 
  Take any $\strA \models \phisnake$ and define an auxiliary sequence of maps $\{\homof_{(n,m)}\}_{(n,m) \in \snk_{\infty}}$ as:
  \begin{itemize}[itemsep=0em, leftmargin=*]
  \item Let $\homof_{(0,0)} \deff \{ ((0,0), \ela)\}$ for any fixed $\ela \in A$ witnessing $\strA \models \phisnakeinit[\ela]$; We stress that $\homof_{(0,0)}$ witnesses $\strSnk_0 \sinjto_{\homof_{(0,0)}} \strA$.
  \item Suppose that $\homof_{(m,n)}$ is already defined. 
  If $\homof_{(m,n)}$ is non-empty and it is not the case that $(m,n)$ is simultaneously final and $\homof_{(m,n)}(n,m) \in \pL^{\strA}$, we define $\homof_{(m',n')}$ for the successor $(m',n')$ w.r.t. $\lesssnk$ of $(m,n)$ as any fixed extension of $\homof_{(m,n)}$ guaranteed by \Cref{lemma:injecting-snake-main}.
  Otherwise, we put $\homof_{(m',n')} \deff \emptyset$. 
  \end{itemize}
  For the proof of (B) suppose $\strA \models \forall{x_1}\neg\pL(x_1)$. Then for all indices $(n,m) \in \snk_{\infty}$ the maps $\homof_{(n,m)}$ are non-empty. By design $\{\homof_{(n,m)}\}_{(n,m) \in \snk_{\infty}}$ is a growing w.r.t. $\subseteq$ sequence of semi-strong homomorphisms. Hence, $\homof \deff \textstyle\bigcup_{(n,m) \in \snk_{\infty}} \homof_{(n,m)}$ is the desired semi-strong homomorphism witnessing $\strSnk_{\infty} \injto_{\homof} \strA$.
  To establish (C) suppose that $\strA$ is finite. 
  Note that there is an index $(n,m)$ such that $\homof_{(n,m)}$ is non-empty while $\homof_{(n', m')}$ is empty for the successor w.r.t. $\lesssnk$ of $(n,m)$.
  Indeed, if all the maps $\homof_{(n,m)}$ were non-empty, then $\strA$ would be infinite by their injectivity.
  Take the smallest w.r.t. $\lesssnk$ such an index $(\rmN,\rmM)$. By design $(\rmN,\rmM)$ is final and $\homof_{(\rmN,\rmM)}(\rmN,\rmM) \in \pL^{\strA}$. Thus the number $\rmN$ and a map $\homof_{(\rmN,\rmM)}$ are (by design) as desired.
\end{proof}

\subsection{Reduction from Tiling Systems}\label{app:reduction_from_the_tiling_systems}

We assume the reader is familiar with the octant tiling problem as described in the main body of the paper. Before we proceed with the reduction, we emphasize that, in the case of the \emph{finite} octant tiling problem, a minor correction to the standard formulation is required—otherwise, each domino tiling system trivially covers a singleton octant. 
Specifically, we assume that each domino tiling system $\tilingsys$ includes two distinguished tiles: $\pT_{\mathrm{fst}}$ and $\pT_{\mathrm{lst}}$, which are required to cover the first and last cells of the octant, respectively, under the usual lexicographic ordering.
Unfortunately, we could not locate a definitive reference for the undecidability of this problem. The result is often attributed to Appendix~A of \emph{The Classical Decision Problem} by Börger, Grädel, and Gurevich\footnote{See the discussion on TCS Stack Exchange: \url{https://cstheory.stackexchange.com/questions/54748/reference-request-undecidability-of-tiling-problem-of-finite-octant}} but the is not contained there. However, a suitable reduction from the halting problem for Minsky machines is rather easy: each ``row'' of the octant can represent one of the possible configurations, and counter values are encoded by using special tiles placed at appropriate heights. Transition consistency can be enforced by the $\tilingH$ component of $\tilingsys$, while configuration consistency is enforced by the $\tilingV$ component. We plan to provide further details on this graduate-level construction in the journal version of this paper.

Let $\tilingsys \deff (\tiles, \tilingV, \tilingH, \pT_{\mathrm{fst}}, \pT_{\mathrm{lst}})$ be (an extended) tiling system. As discussed in the main paper, we aim to construct a formula $\phitiling$ such that $\phisnake \land \phitiling$ has a (finite) model if and only if $\tilingsys$ covers some (finite) octant.
For every $\tile \in \tiles$, we introduce a fresh unary predicate~$\pT_{\tile}$ and define $\phitiling$ to ensure that all elements of the snake (except the topmost elements in odd columns, which are irrelevant for the reduction) are labelled with exactly one $\pT_{\tile}$, and that $\rmH$- and $\rmV$-connected elements respect the relations in $\tilingH$ and $\tilingV$.
For convenience, define $\mathrm{irrel}(x_1) \deff \pT_0(x_1) \land \neg \pE(x_1)$. The formula $\phitiling$ is then the conjunction of the following clauses:

\begin{itemize}[itemsep=0em, leftmargin=*]
\item Every element is either irrelevant or labelled with exactly one tile:
\[
  \forall{x_1} \mathrm{irrel}(x_1) \veedot \textstyle\bigveedot_{\tile \in \tiles} \pT_{\tile}(x_1)
\]

\item Consistency with $\tilingH$ (ignoring irrelevant elements):
\[
  \textstyle\bigwedge_{\tile \in \tiles} \forall{x_1} \pT_{\tile}(x_1) \to \forall{x_2} \pH(x_1x_2) \to \textstyle\bigvee_{(\tile,\tile') \in \tilingH} \pT_{\tile'}(x_2)
\]

\item Consistency with $\tilingV$ for odd columns:
\[
  \bigwedge_{\tile \in \tiles} \forall{x_1} \pT_{\tile}(x_1) {\to} \neg\pE(x_1) {\to} \forall{x_2} \pV(x_1x_2) {\to} \bigvee_{(\tile,\tile') \in \tilingV} \pT_{\tile'}(x_2)
\]

\item Consistency with $\tilingV$ for even columns (note the reversed order):
\[
  \bigwedge_{\tile \in \tiles} \forall{x_1} \pT_{\tile}(x_1) \to \pE(x_1) \to \forall{x_2} \pV(x_1x_2) \to \bigvee_{(\tile',\tile) \in \tilingV} \pT_{\tile'}(x_2)
\]

\item In the general (infinite) tiling problem, we add the clause $\forall{x_1}\neg\pL(x_1)$ to $\phitiling$ to enforce an infinite grid.

\item In the case of the finite tiling, we enforce placement of the special~tiles:
\begin{align*}
  \forall{x_1} \pL(x_1) \to \pT_0(x_1) \to \pT_{\tile_{\mathrm{lst}}}(x_1)\\
  \forall{x_1} \pB(x_1) \to \pT_0(x_1) \to \pT_{\tile_{\mathrm{fst}}}(x_1)
\end{align*}
\end{itemize}

This concludes the definition of $\phitiling$. We next establish correctness of the reduction.

\begin{lemma}
The guarded fluted two-variable sentence $\phisnake \land \phitiling$ has a (finite) model iff $\tilingsys$ covers some (finite) octant.\myqed
\end{lemma}
\begin{proof}
Assume $\tilingsys$ covers some $\rmN$-octant, as witnessed by a map $\tilesmap$.
By part A of \Cref{lemma:injecting-snake-summary}, the $\rmN$-snake $\strSnk_{\rmN}$ extends to a (finite if $\rmN$ is finite) model of $\phisnake$.
We interpret each predicate $\pT_{\tile}$ as the set of positions $(n,m)$ such that $\tilesmap(n,m) = \tile$. Note that the irrelevant positions are not in the domain of~$\tilesmap$, so no tile predicate $\pT_{\tile}$ is assigned to them. 
Since $\tilesmap$ is a valid solution to the tiling problem, all conjuncts of $\phitiling$ are satisfied, yielding a (finite) model of $\phisnake \land \phitiling$.

\noindent Conversely, suppose $\strA$ is a model of $\phisnake \land \phitiling$.

\begin{itemize}[itemsep=0em, leftmargin=*]
\item If $\strA$ is infinite, then by part B of \Cref{lemma:injecting-snake-summary}, we have $\strSnk_{\infty} \sinjto \strA$ via a homomorphism $\homof$. Our goal will be to construct a map $\tilesmap\colon \octant_{\infty} \to \tiles$.

\item If $\strA$ is finite, then by part C of \Cref{lemma:injecting-snake-summary}, we have $\strSnk_{\rmN} \sinjto_{\homof} \strA$ for some $\rmN$, with $\homof(\rmN, \rmM) \in \pL^{\strA} \cap \pT_0^{\strA}$. Our goal will be to construct a map $\tilesmap\colon \octant_{\rmN} \to \tiles$.
\end{itemize}

\noindent In either case, define $\tilesmap(n,m)$ as the unique tile $\tile$ such that $\strA \models \pT_{\tile}[\homof(n,m)]$. This is well-defined by the first conjunct of~$\phitiling$. Furthermore, $\tilesmap$ satisfies all horizontal and vertical adjacency constraints by the second, third, and fourth conjuncts. 
Finally, in the finite case, we additionally have $\tilesmap(0,0) = \tile_{\mathrm{fst}}$ and $\tilesmap(\rmN,\rmM) = \tile_{\mathrm{lst}}$ due to the last conjunct of~$\phitiling$.
Hence, $\tilesmap$ is a correct tiling of the octant, completing the proof.
\end{proof}

\section{Appendix to \Cref{sec:expspace}} \label{appendix:expspace}


\begin{figure*}
  \centering
      \begin{tikzpicture}[transform shape]

\draw (-11,0) node[minirond] (A0) {\tiny{0}};
\draw (-10.25,0) node[minirond] (A1) {\tiny{1}};
\draw (-9.5,0) node[minirond] (A2) {\tiny{2}};
\draw (-8.75,0) node[minirond] (A3) {\tiny{3}};

\draw (-8,-0.75) node[minirond] (B0) {\tiny{0}};
\draw (-7.25,-0.75) node[minirond] (B1) {\tiny{1}};
\draw (-6.5,-0.75) node[minirond] (B2) {\tiny{2}};
\draw (-5.75,-0.75) node[minirond] (B3) {\tiny{3}};

\draw (-4.75,-0.75) node[minirond] (D0) {\tiny{0}};
\draw (-4,-0.75) node[minirond] (D1) {\tiny{1}};
\draw (-3.25,-0.75) node[minirond] (D2) {\tiny{2}};
\draw (-2.5,-0.75) node[minirond] (D3) {\tiny{3}};

\draw (-1.5,-1.5) node[minirond] (F0) {\tiny{0}};
\draw (-0.75,-1.5) node[minirond] (F1) {\tiny{1}};
\draw (0,-1.5) node[minirond] (F2) {\tiny{2}};
\draw (0.75,-1.5) node[minirond] (F3) {\tiny{3}};

\draw (-1.5,-0.75) node[minirond] (G0) {\tiny{0}};
\draw (-0.75,-0.75) node[minirond] (G1) {\tiny{1}};
\draw (0,-0.75) node[minirond] (G2) {\tiny{2}};
\draw (0.75,-0.75) node[minirond] (G3) {\tiny{3}};

\draw (-8,0.75) node[minirond] (C0) {\tiny{0}};
\draw (-7.25,0.75) node[minirond] (C1) {\tiny{1}};
\draw (-6.5,0.75) node[minirond] (C2) {\tiny{2}};
\draw (-5.75,0.75) node[minirond] (C3) {\tiny{3}};

\draw (-4.75,0.75) node[minirond] (E0) {\tiny{0}};
\draw (-4,0.75) node[minirond] (E1) {\tiny{1}};
\draw (-3.25,0.75) node[minirond] (E2) {\tiny{2}};
\draw (-2.5,0.75) node[minirond] (E3) {\tiny{3}};

\draw (-1.5,1.5) node[minirond] (H0) {\tiny{0}};
\draw (-0.75,1.5) node[minirond] (H1) {\tiny{1}};
\draw (0,1.5) node[minirond] (H2) {\tiny{2}};
\draw (0.75,1.5) node[minirond] (H3) {\tiny{3}};

\draw (-1.5,0.75) node[minirond] (I0) {\tiny{0}};
\draw (-0.75,0.75) node[minirond] (I1) {\tiny{1}};
\draw (0,0.75) node[minirond] (I2) {\tiny{2}};
\draw (0.75,0.75) node[minirond] (I3) {\tiny{3}};

      \path[->] (A0) edge [blue, ->] node[yshift=0.5em] {\scriptsize{$\pG_0$}} (A1);
      \path[->] (A1) edge [blue, ->] node[yshift=0.5em] {\scriptsize{$\pG_0$}} (A2);
      \path[->] (A2) edge [blue, ->] node[yshift=0.5em] {\scriptsize{$\pG_0$}} (A3);

      \path[->] (A3) edge [red, ->] node[yshift=0.25em] {\scriptsize{$\pG_1$}} (B0);
      \path[->] (B0) edge [red, ->] node[yshift=0.5em] {\scriptsize{$\pG_1$}} (B1);
      \path[->] (B1) edge [red, ->] node[yshift=0.5em] {\scriptsize{$\pG_1$}} (B2);
      \path[->] (B2) edge [red, ->] node[yshift=0.5em] {\scriptsize{$\pG_1$}} (B3);

      \path[->] (A3) edge [red, ->] node[yshift=0.25em] {\scriptsize{$\pG_1$}} (C0);
      \path[->] (C0) edge [red, ->] node[yshift=0.5em] {\scriptsize{$\pG_1$}} (C1);
      \path[->] (C1) edge [red, ->] node[yshift=0.5em] {\scriptsize{$\pG_1$}} (C2);
      \path[->] (C2) edge [red, ->] node[yshift=0.5em] {\scriptsize{$\pG_1$}} (C3);

      \path[->] (B3) edge [black, ->] node[yshift=0.5em] {\scriptsize{$\pG_2$}} (D0);
      \path[->] (D0) edge [black, ->] node[yshift=0.5em] {\scriptsize{$\pG_2$}} (D1);
      \path[->] (D1) edge [black, ->] node[yshift=0.5em] {\scriptsize{$\pG_2$}} (D2);
      \path[->] (D2) edge [black, ->] node[yshift=0.5em] {\scriptsize{$\pG_2$}} (D3);

      \path[->] (C3) edge [black, ->] node[yshift=0.5em] {\scriptsize{$\pG_2$}} (E0);
      \path[->] (E0) edge [black, ->] node[yshift=0.5em] {\scriptsize{$\pG_2$}} (E1);
      \path[->] (E1) edge [black, ->] node[yshift=0.5em] {\scriptsize{$\pG_2$}} (E2);
      \path[->] (E2) edge [black, ->] node[yshift=0.5em] {\scriptsize{$\pG_2$}} (E3);

      \path[->] (D3) edge [blue, ->] node[yshift=0.5em] {\scriptsize{$\pG_0$}} (G0);
      \path[->] (G0) edge [blue, ->] node[yshift=0.5em] {\scriptsize{$\pG_0$}} (G1);
      \path[->] (G1) edge [blue, ->] node[yshift=0.5em] {\scriptsize{$\pG_0$}} (G2);
      \path[->] (G2) edge [blue, ->] node[yshift=0.5em] {\scriptsize{$\pG_0$}} (G3);

      \path[->] (D3) edge [blue, ->] node[yshift=0.5em] {\scriptsize{$\pG_0$}} (F0);
      \path[->] (F0) edge [blue, ->] node[yshift=0.5em] {\scriptsize{$\pG_0$}} (F1);
      \path[->] (F1) edge [blue, ->] node[yshift=0.5em] {\scriptsize{$\pG_0$}} (F2);
      \path[->] (F2) edge [blue, ->] node[yshift=0.5em] {\scriptsize{$\pG_0$}} (F3);      

      \path[->] (E3) edge [blue, ->] node[yshift=0.5em] {\scriptsize{$\pG_0$}} (I0);
      \path[->] (I0) edge [blue, ->] node[yshift=0.5em] {\scriptsize{$\pG_0$}} (I1);
      \path[->] (I1) edge [blue, ->] node[yshift=0.5em] {\scriptsize{$\pG_0$}} (I2);
      \path[->] (I2) edge [blue, ->] node[yshift=0.5em] {\scriptsize{$\pG_0$}} (I3);

      \path[->] (E3) edge [blue, ->] node[yshift=0.5em] {\scriptsize{$\pG_0$}} (H0);
      \path[->] (H0) edge [blue, ->] node[yshift=0.5em] {\scriptsize{$\pG_0$}} (H1);
      \path[->] (H1) edge [blue, ->] node[yshift=0.5em] {\scriptsize{$\pG_0$}} (H2);
      \path[->] (H2) edge [blue, ->, thick] node[yshift=0.5em] {\scriptsize{$\pG_0$}} (H3);  

      \draw (1.5,1.75) node (dots1) {{$\ldots$}};
      \draw (1.5,1.5) node (dots1) {{$\ldots$}};
      \draw (1.5,1) node (dots1) {{$\ldots$}};
      \draw (1.5,0.75) node (dots1) {{$\ldots$}};
      \draw (1.5,-1.65) node (dots1) {{$\ldots$}};
      \draw (1.5,-1.4) node (dots1) {{$\ldots$}};
      \draw (1.5,-0.5) node (dots1) {{$\ldots$}};
      \draw (1.5,-0.75) node (dots1) {{$\ldots$}};

      \draw[decorate,decoration={brace,mirror},thick]
  ([yshift=-0.3cm]A0.south) -- ([yshift=-0.3cm]A3.south) 
  node[midway,yshift=-0.4cm] {\scriptsize example configuration};
    \end{tikzpicture}%
  \vspace{-1em}
  \caption{A skeleton for encodings. In the case of $\cup$ and $\circ$, each $\pG_i$ denotes $\pR_i$ for $i \in \{ 1,2,3\}$. Otherwise, each $\pG_i$ denotes $\pR_i \cap \pR_{i{+}1 \bmod 3}$.}

\label{f:twoexplower}
\end{figure*}
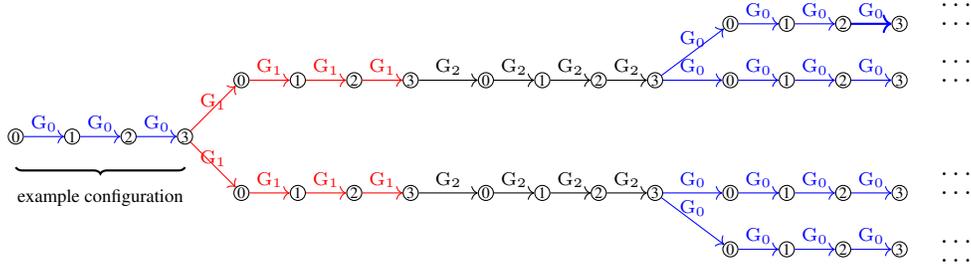


\subsection{Hardness Results}

We first establish the inclusion of any of the operators $\circ$, $\cup$, or~$\cap$ (even with $\cap$ restricted to atomic formul{\ae}) increases the complexity of $\SAT(\FRGFt[\cdot^{+}])$ to $\TwoExpTime$. We focus on the operator $\cdot^+$, because the proof for $\cdot^*$ is almost identical.

\begin{thm}
Both finite and general satisfiability problems for
$\FRGFt[\cdot^{+}, \circ]$,
$\FRGFt[\cdot^{+}, \cup]$, and
$\FRGFt[\cdot^{+}, \cap]$
are $\TwoExpTime$-hard.\myqed
\end{thm}

\newcommand{\turingM}{\texttt{M}}
\newcommand{\inputw}{\texttt{w}}
\newcommand{\phiM}{\varphi_{\turingM}}

\noindent We follow a well-established route for proving $\TwoExpTime$-hardness via encodings of alternating Turing machines operating in exponential space. We focus only on the details specific to our construction; for a more detailed exposition, consult works of Kieroński~\shortcite[Sec.~4.1]{Kieronski06}.
All of our cases are based on similar encodings.
Suppose an alternating Turing machine $\turingM$ and its input $\inputw$ of length $n$ are given. 
W.l.o.g., we may assume that $\turingM$ works in space bounded by $2^n$, that every non-final configuration 
has two possible moves, and that $\turingM$ accepts or rejects precisely in $2^{2^n}$-th step.
We construct a formula $\phiM$ such that $\turingM$ accepts $\inputw$ iff $\phiM$ is satisfiable.
The shape of the intended tree-shaped models~is~given~in~Fig.~\ref{f:twoexplower}.

\noindent Our trees are partitioned into vertical segments of length~$2^n$, each corresponding to a configuration of $\turingM$. 
These segments are linked via a binary relation (depending on a variant), and each element within a segment represents a single~tape~cell. Configurations are encoded using unary predicates to mark tape symbols, states, and the head position.
Tape cell positions are encoded in binary using unary predicates $\pP_0, \ldots, \pP_{n{-}1}$.
A predicate $\pL$ distinguishes the $\pL$eft-successor configurations from the others.
Universal configurations have two successors---one satisfying $\pL$, the other $\neg\pL$---while existential ones have a single successor.
It is routine to write formul{\ae} defining the skeleton of the encoding—essentially stating that each element has the correct successor within its configuration, and that the last element has one or two successors as appropriate. This is almost verbatim from Kieroński's formalization~\shortcite[Thm.~5]{Kieronski06}.
The non-trivial part is identifying, for a given element, those occupying the same or neighbouring tape positions in the successor configurations. 
This step is essential to ensure that configuration updates follow $\turingM$'s transition function.
To achieve this, we use three binary regular relations $\pR_0$, $\pR_1$, and $\pR_2$ linking consecutive elements of a tree. Fig.~\ref{f:twoexplower} shows the skeleton for our encoding.
For example, the formula{\ae} stating that cells not observed by the head retain their letter in the successor configuration, may have the form
$
\textstyle\bigwedge_{i \in \{0,1,2 \}} \forall{x_1x_2}\, \rmG_i(x_1x_2) {\to} \psi(x_1x_2) {\to} \phi(x_1x_2),
$
where $\psi(x_1x_2)$ asserts that $x_1$ and $x_2$ encode the same cell (represent the same numbers with predicates $\pP$) and $x_1$ is not scanned by the head, while $\phi(x_1x_2)$ ensures they both carry the same letter. 
The relation $\pG_i$ depends then on the operator: for composition $\circ$ we let $\pG_i \deff \pR_i^+ \circ \pR_{i+1}^+$; for union $\cup$ we let $\pG_i \deff (\pR_i \cup \pR_{i+1})^+$; and for intersection $\cap$ we let $\pG_i \deff \pR_i^+$ (with all subscripts taken $\bmod\, 3$) but ensuring first that successors in trees follow precisely $\pR_i \cap \pR_{i{+}1}$.
Similarly, we can say that successor configuration of an existential configuration is obtained by applying one of the two possible transitions of $\turingM$,
and that the left and right successor configurations of an universal configuration are obtained by applying the first, and respectively, the second $\turingM$'s transition. 
Finally, we say that there is no rejecting state in the model.


\subsection{Towards an $\ExpSpace$-Complete Sublogic}

As the culminating result of the paper, we demonstrate an $\ExpSpace$ upper bound for $\FRGF[\cdot^{+}, \cdot^*, ?]$.
Similarly to the case of full $\RGF$, it suffices to focus only on its two-variable fragment; the general case is handled~via~the~fusion. 
We  actually prove a slightly stronger result, aiming to generalize $\GFt$ with one-way transitive guards of Kieroński~\shortcite{Kieronski06}.
\noindent The lower bound is due to Kieroński~\shortcite[Thm.~5]{Kieronski06}, while the upper bound adapts and non-trivially generalizes his technique for handling $\GFt$ with one-way transitive guards.  
The key lemma shows that satisfiable sentences of our logic admit regular tree-shaped models verifiable in alternating exponential time, and thus in $\ExpSpace$.  
The main challenge, unlike the transitive-guards case, is that witnesses for $\existsreg$-conjuncts are not necessarily direct successors in tree-like models but may require exponentially long witness paths.~We~prove:

\begin{thm}\label{thm:frgft}
$\SAT(\logicL)$ is $\ExpSpace$-complete for a logic~$\logicL$ obtained from $\FRGFt[\cdot^{+}, \cdot^*, ?]$ by  
\emph{extending} the set of allowed variable sequences in atoms by $x_1x_1$, $x_2x_2$, and~$x_2x_1$, where the last sequence is forbidden in regular guards.~\myqed
\end{thm}

\noindent In the absence of $\circ$ and $\cap$, the test operator ? is not so~useful. 
Indeed, one of the main motivations for the extended syntax of our logic is that we can internalize all 
formul{\ae} of the form $\exists{x_2} (\varphi?)(x_1x_2) \land \psi(x_1x_2)$ with $\varphi(x_1) \land \psi(x_1)$. Thus, from now on, we assume that the input formul{\ae}~are~test-free.

\vspace{-1em}
\paragraph*{Tree-shaped Models with Short Witnessing Paths.}

Fix a formula $\varphi$ in normal form (with components as in Def.~\ref{def:normal-form-RGF}) over (the extended) $\FRGFt[\cdot^{+}, \cdot^*, ?]$, and suppose $\strA \models \varphi$.  
A~sequence $\rho \deff (\ela_0, \ela_1, \ldots, \ela_k)$ in $\strA$ is a \emph{witnessing path} for $\ela_0$ and the $i$-th $\existsreg$-conjunct of $\varphi$ whenever $\strA \models \piiex[\ela_0\ela_k]$, $\strA \models \phiiex[\ela_0\ela_k]$, and $\rho$ \emph{matches} $\phiiex$. 
This last notion is defined as expected. 
For instance, $\rho$ \emph{matches} $\textstyle\pR^+$ if any two consecutive elements of $\rho$ are $\pR$-connected (other cases are handled similarly).
Analogously, $\rho \deff (\ela_0,\ela_1)$ is a \emph{witnessing path} for $\ela_0$ and the $i$-th $\existsfo$-conjunct 
if $\strA \models \varthetaiex[\ela_0 \ela_1]$ and $\strA \models \psiiex[\ela_0 \ela_1]$. 
The second elements in the witnessing paths are called \emph{primal}.
A structure $\str{A}$ is a \emph{proper tree-shaped} model of $\varphi$ if its domain can be arranged into a tree such~that:\\  
\noindent $\bullet$ if $\strA \models \pT[\ela\elb]$ for $\pT \in \sigreg$ then $\elb$ is a child of $\ela$ or $\ela = \elb$.\\
\noindent $\bullet$ For any distinct $\pS, \pR \in \sigreg$ and $(\ela,\elb)$ in $\pS^{\strA} \cap \pR^{\strA}$ we have $\ela = \elb$ (no two regular relations link a node with~its~child).\\
\noindent $\bullet$ One can assign witnessing paths for all $\ela \in A$ and all $\existsreg$- or $\existsfo$-conjuncts of $\varphi$ in a way that every edge from $\strA$ lies on one of such path, each path contains at most $|\AAA_\varphi|{+}1$ elements, and each element has at most $|\varphi|$ witnessing paths and, except for their primal elements,~has~at~most~one~extra~child.

\vspace{0.5em}
\noindent \textbf{Claim A:}~If~satisfiable,~$\varphi$~has~a~proper~tree-shaped~model.~\myqed
\vspace{-0.5em}
\begin{proof}[Sketch.]
We employ a formula-driven unravelling. Starting from an $\strA \models \varphi$, we choose an initial element $\ela_0 \in A$. We unravel $\strA$ into a new structure $\strA'$ from $\ela_0$: for each required witnessing path of $\ela_0$, we shorten it to include at most one realization of each $1$-type and simplify its $2$-types so that each of them contains at most one regular symbol (the one used in the relevant guard); this is safe as regular symbols appear only in guards. 
We then add copies of these simplified witnessing paths to $\strA'$.
Each newly added element $\ela'$ is associated with a \emph{pattern element} $\homof(\ela')$ in $\strA$, and we recursively build the subtree of $\ela'$ by unraveling $\strA$ from $\homof(\ela')$. 
Finally, we define the $2$-types between elements $\ela', \elb'$ from different paths using $\tp{\str{A}}{2}(\homof(\ela'), \homof(\elb'))$, replacing each atom $\pT(x_1x_2)$ and $\pT(x_2x_1)$ for a regular $\pT \in \sigreg$ with its negation.
Note that this ensures no violation of any $\forallfo$-conjuncts. 
If $\ela'$ and $\elb'$ are not parent and child (in which case their $2$-type is already fixed), then they are not connected by any direct edge. If they are connected by $\pT^+$, then by design $\homof(\ela')$ and $\homof(\elb')$ are as well, preserving the satisfaction of $\forallreg$-conjuncts.
\end{proof}

\vspace{-2em}
\paragraph*{Regular Tree-Shaped Models.}
For a proper tree-shaped model $\strA \models \varphi$ we next construct an auxiliary structure $\strA^*$, enhanced with additional pointers from some nodes of $\strA$ to their ancestors (we call them \emph{links} and \emph{half-links}), in a similar way to how \emph{blocking} works in tableaux methods. 
Their presence is useful to unravel $\strA^*$ into a \emph{regular} (in the sense of regular trees) proper
tree-shaped model $\strA' \models \varphi$.
A \emph{profile} of a node $\ela$ in $\strA$ is the triple $\prof{\strA}(\ela) \deff (\tp{\strA}{1}(\ela), \pT_\ela, \AAA_\ela)$, where $\pT_\ela \in \sigreg \cup \{ \varepsilon \}$ is the unique regular predicate $\pT$ (if any) connecting $\ela$ to its parent~$\elp$, and $\AAA_\ela \subseteq \AAA_\varphi$ collects all $1$-types on the path from the root to $\elp$ of elements $\pT$-connected to $\ela$ (if there is no such $\pT$ we put $\pT_{\ela} \deff \varepsilon$ and $\AAA_{\ela} \deff \emptyset$).

\noindent
The clearest way to describe $\strA^*$ is via its algorithmic construction. We traverse $\strA$ in a breadth-first manner and consider each visited node $\ela$. If there exists a node $\elb$ on the path from the root to $\ela$'s parent $\elp$ with the same profile as~$\ela$, we remove all subtrees rooted at $\ela$'s children (except its non-primal child, if any) and add a \emph{link} from $\ela$ to $\elb$. If no such $\elb$ exists but there is a node on the path with the same $1$-type as $\ela$, we remove all subtrees rooted at children of $\ela$ that are not $\pT_{\ela}$-connected to $\ela$ (or simply all subtrees if $\pT_\ela = \varepsilon$) and add a \emph{half-link} from $\ela$ to $\elb$. Otherwise, we continue the traversal.

\noindent Let $\strA^*$ be the resulting structure.
We unravel $\strA^*$ into a tree-shaped model of $\varphi$ by following links and half-links. If an element $\ela$ sends a link to $\elb$, we copy each subtree rooted at a child $\elc$ of $\elb$, attach its root $\elc'$ as a child of $\ela$, and replicate the connections from $\elb$ to $\elc$ and its subtree. This ensures that $\ela$ inherits all necessary witnesses from $\elb$, which are preserved during link creation—non-primal children (those continuing some witnessing paths started above their parents) are never removed. Links and half-links are copied to the new elements with their targets unchanged, and the process continues recursively. If $\ela$ sends a half-link to $\elb$, we proceed similarly but below $\ela$ we only copy the subtrees rooted at the children of $\elb$ to which $\elb$ is not joined by $\pT_{\ela}$. 
This finishes the construction.

\noindent Note that the unravelling of $\strA^*$ may leave some $2$-types undefined. 
Fortunately, our linking strategy allows us to ``complete'' them while preserving the satisfaction of $\varphi$. 
Specifically, even if two elements $\ela^*$ and $\elb^*$ in the unravelling are $\pT^+$-reachable for some regular $\pT$, the use of profiles ensures that we can find corresponding elements $\ela, \elb$ in $\strA$ with the same $1$-types and also $\pT^+$-reachable. 
We can then simply assign the $2$-type of $(\ela, \elb)$ to $(\ela^*, \elb^*)$. This way we can show:

\vspace{0.25em}
\noindent \textbf{Claim B:} The unravelling of $\strA^*$ can be completed by defining the missing $2$-types to obtain a tree-shaped $\strA' \models \varphi$.\myqed

\paragraph*{The Algorithm.}
Our decision procedure searches for a regular proper tree-shaped model of $\varphi$ as described above. 
Since $\strA^*$ compactly represents the intended model $\strA'$ of $\varphi$, it suffices to look for such a compact representation. We show:

\vspace{0.5em}
\noindent \textbf{Claim C:} Every root-to-leaf path in $\strA^*$ has length bounded exponentially in $|\varphi|$, more precisely by $(|\AAA_{\varphi}|{+}1)^4$.~\myqed

\vspace{-1em}
\begin{proof}
Consider any root-to-leaf path $\rho$ in $\strA^*$, and let $\AAA_0$ be the set of $1$-types realized along $\rho$. An element $\ela \in \rho$ is called a \emph{splitting point} if it is not connected to its successor on $\rho$ by $\pT_\ela$ (the predicate from its profile). The number of splitting points is at most $|\AAA_0|$: otherwise, two splitting points $\elb$, $\ela$ with the same $1$-type (where $\elb$ is an ancestor of $\ela$) would trigger a (half-)link from $\ela$ to $\elb$, preventing $\ela$ from having a child not connected via $\pT_\ela$.
Consider a maximal subpath $\hat{\rho}$ of $\rho$ starting at a splitting point $\ela$ and containing no other splitting points (we will shortly see that $\hat{\rho}$ is finite). Let $\hat{\rho}_-$ denote $\hat{\rho}$ without its first element, and let $p \in \N$ be the total number of distinct profiles on $\hat{\rho}_-$. Note that $p \leq |\AAA_0|^2$: all elements on $\hat{\rho}_-$ share the same second profile component (by definition of splitting points), while there are at most $|\AAA_0|$ options each for the first and third components. The latter holds because the sets $\AAA_{\ela}$ in the third component can only grow or remain the same along $\hat{\rho}$, as all edges in $\hat{\rho}$ contain the same regular predicate $\pT$, meaning each element is $\pT^+$-connected to its descendants on $\hat{\rho}$.
Next, observe that the number of primal elements on $\hat{\rho}$ is at most $p$, since any two elements with primal successors must have distinct profiles. Indeed, if~$\elb$ is an ancestor of $\ela$ sharing the same profile, a link from~$\ela$ to $\elb$ would be created, preventing $\ela$ from having a primal successor.
Finally, any subpath $\tilde{\rho} \subseteq \hat{\rho}$ consisting solely of non-primal elements has length at most $|\AAA_0|$, since~$\tilde{\rho}$ must lie within some witnessing path, itself bounded by $|\AAA_0|$. Moreover, the last element $\ela_k$ of such a path cannot have a non-primal $\pT_{\ela_k}$-successor in $\strA^*$.
Wrapping up, the length of $\rho$ is roughly bounded by $(|\AAA_0| {+} 1)^4$, so exponentially in~$|\varphi|$: there are at most $|\AAA_0|^2 {+} 1$ segments $\hat{\rho}$ between splitting points, each divided into at most $|\AAA_0|{+}1$ subsegments $\tilde{\rho}$ with primal endpoints, each of length $\leq |\AAA_0|$.~\qedhere
\end{proof}

\vspace{-0.5em}
\noindent Our algorithm works as follows. It attempts to construct~$\strA^*$, a compact representation of a tree-shaped model of $\varphi$. It begins by creating the root, guessing its $1$-type, and marking it as \emph{unserved}. Then, in a loop, it universally selects an unserved element $\ela$ and checks whether $\ela$ can be linked or half-linked to one of its previously constructed ancestors.
By a \emph{witnessing component} of $\ela$ we mean the structure induced by $\ela$ and its required (exponentially long) witnessing paths. If $\ela$ can be linked to a previous element, we accept--its witnesses will be ensured by unravelling the link. If $\ela$ can only be half-linked, we guess a partial witnessing component consisting of paths for conjuncts involving $\pT_\ela$; the remaining witnesses will be provided by the half-link. If neither link is possible, we guess the full witnessing component.
We then check whether the guessed component (or its guessed fragment) provides all required witnesses for $\ela$, and whether the $2$-types for all pairs in the current model fragment can be set without violating the $\forallfo$- and $\forallreg$-conjuncts of $\varphi$. 
Finally, we mark all newly added elements as \emph{unserved} and the old ones as \emph{served}.

By Claim C, the algorithm runs in $\AExpTime$ and thus it yields an $\ExpSpace$ decision procedure. For correctness, note that if $\varphi$ has a model, then by Claim A it has a proper tree-shaped model, from which $\strA^*$ can be constructed. The algorithm makes all its guesses in line with such a structure.
Conversely, any accepting run of the algorithm produces a structure $\strA^*$ with (half-)links. The properties verified by the algorithm ensure that $\strA^*$ can be unravelled into a regular tree-shaped model of $\varphi$, by applying a strategy from Claim~B.

\vspace{-1em}
\paragraph*{The Multi-Variable Case.}
To extend Thm.~\ref{thm:frgft} to the higher-arity case, we rely on the fusion from Section~\ref{sub:the_case_of_many_variables}. As $\FRGFt[\cdot^{+}, \cdot^*, ?]$ and~$\FGF$ are decidable in $\ExpSpace$, their fusion yields the desired algorithm.
Some extra care is needed, however. The formul{\ae} $\varphifofull$ and $\varphiregfull$ used in the reduction must express $1$-types $\alpha$ and $2$-types $\betaminus$, which is not possible in the forward logic: $\alpha$ may include atoms like $\pm\pT(x_1x_1)$ for a regular $\pT \in \sigreg$, while $\betaminus$ involves $\sigfo$-atoms over arbitrary sequences composed of $x_1$ and $x_2$.
To circumvent this issue, we relax the forwardness requirement by extending the syntax of both~$\FGF$ and $\FRGFt[\cdot^{+}, \cdot^*, ?]$ to allow $\sigfo$-atoms over arbitrary sequences of $x_1$, $x_2$, and $\sigreg$-atoms with a single free variable. These generalizations remain decidable in $\ExpSpace$. 
To see this it suffices to adapt Lemma~\ref{lemma:RGF2-small-number-of-2-types} and verify that the proofs for $\FGF$ by Bednarczyk~\shortcite{Bednarczyk21} and from our procedure for $\FRGFt[\cdot^{+}, \cdot^*, ?]$ remain valid under the extension of the appropriate notion of a type. As such modification are minor we simply conclude:

\begin{corollary}
Assuming minor modifications of existing algorithms, the satisfiability problem for $\FRGF[\cdot^{+}, \cdot^*, ?]$ is complete for $\ExpSpace$.~\myqed
\end{corollary}

\vspace{-0.5em}

\subsection{Proof of Claim A} \label{sec:propertree}

For a $2$-type $\beta$ we denote by $\beta \restr_\FO$ the result of replacing in $\beta$ all literals $\pT(x_1x_2)$, $\pT(x_2x_1)$, for all regular $\pT \in \sigreg$ by their negations $\neg \pT(x_1x_2)$, $\neg\pT(x_2x_1)$.
For a regular $\pT \in \sigreg$ let $\beta \restr_{\vec{\pT}}$ denote the result of replacing 
all literals $\pT'(x_1x_2), \pT'(x_2x_1)$ in  $\beta$, for $\pT' \neq \pT$ with regular $\pT' \in \sigreg$, by $\neg \pT'(x_1x_2), \neg \pT'(x_2x_1)$, and further replacing $\pT(x_2x_1)$ by~$\neg \pT(x_2x_1)$.
We are now ready to jump into the proof. The current version of the proof is rather dry but it will be improved in the revised version of the paper.

\begin{proof}
Let $\strA \models \varphi$ (given in NF) and consider $\ela \in A$.
We construct a small \emph{witnessing component} $\strC_{\ela}$ for $\ela$, together with a
\emph{pattern function} $\homof \colon C_\ela \rightarrow A$ as follows. 
First we put $\ela$ to $C_{\ela}$ and set $\tp{\str{C_\ela}}{1}(\ela) \deff \tp{\str{A}}{1}(\ela)$ and $\homof(\ela) \deff \ela$.
For every $\existsreg$-conjunct, if $\strA \models \gammaiex[\ela]$ then choose a witness $\elb$ for $\ela$ ($\strA \models \piiex[\ela\elb] \wedge \phiiex[\ela\elb]$). Let $\pT \in \sigreg$ be the regular predicate used in $\piiex$, that is $\piiex$ is one of $\pT^+(x_1x_2), \pT^*(x_1x_2), \pT(x_1x_2)$. Let $\ela, \ela_1, \ldots, \ela_k ({=}\elb)$ be an appropriate witnessing path.
Add pairwise distinct fresh elements $\ela'_1, \ldots, \ela_k'$ to $\str{C}_{\ela}$. For all $i \in \{1,\ldots,k\}$ we  set  $\tp{\str{C}_\ela}{1}(\ela_i')=\tp{\str{A}}{1}(\ela_i)$. 
Denoting $\ela'_0 = \ela_0 = \ela$, for all $i \in \{ 0, \ldots, k-1\}$ we set $\tp{\str{C}_\ela}{2}(\ela_i', \ela_{i+1}') \deff \tp{\str{A}}{2}(\ela_i, \ela_{i+1}) \restr_{\vec{\pT}}$.
If $k>1$ then set $\tp{\str{C}_\ela}{2}(\ela, \ela_k') \deff \tp{\str{A}}{2}(\ela, \ela_k) \restr_\FO$.
Shorten the added path in steps $l \in \{ 1, \ldots, k-2 \}$ as follows. 
In the $l$-th step, whenever $\ela_l'$ is still in $C_\ela$: we let $j$ be the maximal index such that $l<j < k$ and $\tp{\str{A}}{1}(\ela_l)=\tp{\str{A}}{1}(\ela_j)$.  
Remove the elements $\ela_{l}', \ldots, \ela_{j-1}'$ from $C_\ela$, and
set $\tp{\str{C}_\ela}{2}(\ela_{l-1}', \ela_{j}'') \deff \tp{\str{C}_\ela}{2}(\ela_{l-1}', \ela_{l}')$. 
Note that this procedure protects $\ela$, $\ela_k$ and does not change their $2$-type.
We also add the required $\existsfo$-witnesses for $\ela$ in $\str{C}_\ela$: if $\str{A} \models \etaiex[\ela]$ then we find a witness $\elb \in A$ such that $\str{A} \models \varthetaiex[\ela\elb] \land \psiiex[\ela\elb]$ and add to $\str{C}_\ela$
a fresh $\elb'$, set $\tp{\str{C}_\ela}{1}(\elb')=\tp{\str{\str{A}}}{1}(\elb)$, $\tp{\str{C_\ela}}{2}(\ela,\elb')=\tp{\str{\str{A}}}{2}(\ela,\elb)\restr_\FO$,
and $\homof(\elb')\deff\elb$. The other $2$-types in $\str{C}_\ela$ are left undefined.
 
Now we construct a tree-shaped $\str{A}'$ together with a \emph{pattern function} $\homof \colon A' \rightarrow A$. 
During the process we always define $1$-types of elements added to $\str{A}'$ and we define $2$-types of some pairs of 
elements (parent-child, element-its witness). The remaining $2$-types will be completed at the end of the construction.
We first choose any $\ela \in A$ and put a fresh $\ela'$ to $A'$, setting $\tp{\strA'}{1}(\ela') \deff \tp{\strA}{1}(\ela)$ and $\homof(\ela') \deff \ela$.
The status of $\ela'$ is \emph{unserved}.
We then repeat in a breadth-first manner: take an unserved element $\ela' \in A'$. Construct a witnessing component $\str{C}_{\homof(\ela')}$ and, identifying its root  with $\ela'$, attach it to $\str{A}'$ as a subtree of $\ela'$. We import the $1$-types, $2$-types and values of $\homof$ for the newly added elements from $\str{C}_{\homof(\ela)}$.
All newly added elements get status \emph{unserved}. Change the status of $\ela'$ to \emph{served}.
Let $\str{A}'$ be the limit of the above process, with all elements \emph{served}. It remains to complete the assignment of $2$-types. For any pair of elements $\ela',\elb' \in A'$ if their $2$-type is not yet defined, set  $\tp{\strA'}{2}(\ela', \elb') \deff \tp{\strA}{2}(\homof(\ela'), \homof(\elb')) \restr_{\FO}$ (it may happen that $\homof(\ela') = \homof(\elb')$).

Let us explain why $\str{A}'$ is a model of $\varphi$. 
We have $\str{A}' \models \forall x \lambda(x)$ since the $1$-types of all elements in $\str{A}'$ are realized in the original model $\str{A}$. We explicitly took care for witnesses.
Consider a $\forallreg$-conjunct $\forall{x_1}\forall{x_2}\, \piifa(x_1x_2) \to \phiifa(x_1x_2)$ and assume that $\str{A}' \models \piifa[\ela'\elb']$.
If $\ela'=\elb'$  then $\str{A}' \models \phiifa[\ela',\elb']$, by repeating the argument on $1$-types. 
Assume that $\ela' \not= \elb'$, and $\piifa = \pT^+[\ela'\elb']$. In this case there is a $\pT^+$-path from $\ela'$ to $\elb'$. If this path is just a direct edge from $\ela'$ to $\elb'$
then also $\str{A} \models \pT[\homof(\ela')\homof(\elb')]$ and hence $\str{A} \models \phiifa[\homof(\ela')\homof(\elb')]$ and  by our construction $\tp{\str{A}'}{2}(\ela',\elb')=\tp{\str{A}}{2}(\homof(\ela'),\homof(\elb'))\restr_{\vec{\pT}}$.
It follows that $\str{A}' \models \phiifa[\ela'\elb']$, as $\phiifa$ does not speak about the $\sigreg$-relations. If the $\pT^+$-path contains more than one edge then by our
construction a similar $\pT^+$-path connects $\homof(\ela')$ and $\homof(\elb')$ in $\str{A}$, so $\str{A} \models \phiifa[\homof(\ela')\homof(\elb')]$. 
By our construction,
$\tp{\str{A}'}{2}[\ela',\elb']=\tp{\str{A}}{2}(\homof(\ela'),\homof(\elb'))\restr_\FO$, which again guarantees that $\str{A}' \models \phiifa[\ela'\elb']$.
If $\piifa = \pT^*(x_1x_2)$ then the reasoning is similar. If $\piifa=\pT(x_1x_2)$ and in case of $\forallfo$-conjuncts the reasoning is also similar but simpler---we need to consider path with at most one edge.
It is readily verified that $\str{A}'$ has all the properties required by the definition of proper tree-shaped model, which concludes the construction.
\end{proof}

\subsection{Proof of Claim B} \label{section:proofclaimcompletingAstar}
\begin{proof}
It is routine to verify that $\strA'$ satisfies $\forall{x_1}\, \lambda(x_1)$, and that the $\forallfo$- and $\forallreg$-conjuncts are respected by the already defined $2$-types (since all $1$-types and $2$-types are inherited from $\strA$). Moreover, every element has the required witnesses, ensured by the existing $2$-types, which guarantees the satisfaction of the $\existsfo$- and $\existsreg$-conjuncts.
It remains to complete the undefined $2$-types without violating the universal conjuncts of $\varphi$. Consider a pair of distinct elements $\ela', \elb' \in \strA'$ such that $\tp{\strA'}{2}(\ela', \elb')$ is not yet fully defined.
If $\ela'$ and $\elb'$ are not connected by any $\pT^+$-path (for $\pT \in \sigreg$), we assign them the unique $2$-type consistent with their $1$-types that contains no positive binary literals.
If, on the other hand, $\ela'$ and $\elb'$ are connected by a $\pT^+$-path for some $\pT \in \sigreg$, then by construction, $\elb'$ is a copy of an element in $\strA$ whose profile includes the $1$-type of $\ela'$. This implies that in $\strA$ there exists a pair of elements with the same $1$-types as $\ela'$ and $\elb'$, connected by a $\pT^+$-path. We can thus assign the $\FO$-reduction of their $2$-type to the pair $(\ela', \elb')$ in $\strA'$, thereby completing the $2$-type.
Repeating this process yields a full model $\strA' \models \varphi$.
\end{proof}

\section{Appendix to \Cref{sec:conclusions}}\label{appendix:undecidability-querying}

\subsection{Proof of \Cref{lemma:negative-results-about-the-FMP}}\label{appendix:proof_of_cref_lemma_negative_results_about_the_fmp}

Consider a signature $\sigma$ consisting of a binary predicate $\pB \in \sigfo$ and regular predicates $\pS, \pR \in \sigreg$.  
We define a formula~$\varphi$ that spoils the finite model property (FMP) with:
\[
\forall{x_1}\forall{x_2} \pR^+(x_1,x_2) {\to} \pB(x_1,x_2) \;\land\; \forall{x_1}\exists{x_2} \pR(x_1,x_2) \;\land\; \varphi_{\mathrm{op}},
\]
for each operator $\mathrm{op} \in \{ ?, \cdot^*, \bar{\cdot}, \circ \}$.
Intuitively, the first conjunct asserts that every pair of $\pR$-connected elements is also connected by $\pB$; the second enforces the existence of an infinite $\pR$-path; and the third (operator-dependent) conjunct ensures that $\pR$ remains acyclic.  
Clearly, any model satisfying the first two conjuncts of $\varphi$ must contain an $\pR$-cycle. Moreover, every element on such a cycle must have a $\pB$-self-loop.  
Thus, the formula $\varphi$ has no finite models.

\begin{enumerate}[itemsep=0em, leftmargin=*]
\item In the case of tests $?$ we define:
\[
\varphi_{?} \deff \forall{x_1}\forall{x_2} (\top?)(x_1,x_2) \to \neg \pB(x_1,x_2),
\]
which forbids $\pB$-self-loops and thus rules out finite models.  
However, the structure of natural numbers where both $\pB$ and $\pR$ are interpreted as the successor relation satisfies the formula, yielding an infinite model of $\varphi$.  
Hence, $\FRGF[\cdot^+, ?]$ does not have the FMP.

\item In the case of Kleene star, we apply the same trick.  
It suffices to force $\pS$ to be interpreted as the empty relation and then use $\pS^*$ to simulate $\top?$.  
Taking $\varphi_{\cdot^*}$ as
\begin{align*}
\left( \forall{x_1}\forall{x_2} \pS(x_1,x_2) \to \bot \right) \;\land\;\\ \forall{x_1}\forall{x_2} \pS^*(x_1,x_2) \to \neg \pB(x_1,x_2)
\end{align*}
we conclude that $\FRGF[\cdot^+, \cdot^*]$ does not have the FMP.

\item In the case of the inverse operator $\bar{\cdot}$, we define:
\[
\varphi_{\bar{\cdot}} \deff \forall{x_1}\forall{x_2} \left(\bar{\pR}\right)^+(x_1,x_2) \to \neg \pB(x_1,x_2).
\]
The same infinite model as before satisfies $\varphi$.  
Suppose, for contradiction, that a finite model $\strA$ exists.  
Then some element $\ela$ on an $\pR$-cycle must be self-reachable via $\pR^+$, implying $\strA \models \pB[\ela,\ela]$ by the first conjunct of $\varphi$.  
But since $\ela$ is also reachable via $\left(\bar{\pR}\right)^+$, we obtain $\strA \models \neg\pB[\ela,\ela]$ from the last conjunct.  
This contradiction shows that $\varphi$ has no finite model.  
So, $\FRGF[\cdot^+, \bar{\cdot}]$ does not have~the~FMP.

\item The case of composition $\circ$ is the most intricate. Define:
\begin{align*}
\varphi_{\circ} \deff \forall{x_1}\exists{x_2} \pS(x_1,x_2) \land \pB(x_1,x_2)\\ \;\land\; \forall{x_1}\forall{x_2} \left( \pR^+ \circ \pS \right)(x_1,x_2) \to \neg \pB(x_1,x_2)
\end{align*}
One can verify that the natural numbers, extended by adding a fresh $(\pS \cap \pB)$-successor to each element, satisfy the formula.  
Formally, an example infinite model $\strA$ is defined by:
\[
A \deff \N \times \{ 0, 1\}, \quad \pR^{\strA} \deff \{ ( (n,0), (n{+}1,0) ) \mid n \in \N \}, 
\]
\[
\pS^{\strA} \deff \{ ( (n,0), (n,1) ) \mid n \in \N \}, \quad \pB^{\strA} \deff \pR^{\strA} \cup \pS^{\strA}.
\]
Suppose for contradiction that a finite model $\strA$ of $\varphi$ exists.  
Then, an element $\ela$ on an $\pR$-cycle must have a $\pS$-successor $\elb$, which, by the first conjunct of $\varphi_{\circ}$, satisfies $\strA \models \pB[\ela,\elb]$.  
However, since $\ela$ is $\pR^+$-self-reachable, it follows that $\elb$ is reachable from $\ela$ via $\pR^+ \circ \pS$, contradicting the second conjunct of $\varphi_{\circ}$.  
Hence, $\FRGF[\cdot^+, \circ]$ does not have~the~FMP.
\end{enumerate}

\noindent We emphasize that all the formul{\ae} presented are not only forward but actually \emph{fluted}, making our results slightly stronger.
\end{document}